\documentclass[12pt]{article}

\usepackage{fontspec}

\usepackage[english]{babel}

\usepackage{csquotes}

\usepackage{url}

\usepackage[
    a4paper,
    left=1.5cm,
    right=1.5cm,
    top=2.5cm,
    bottom=2.5cm
]{geometry}

\usepackage{mathtools}
\usepackage{amsfonts}
\usepackage{amssymb}
\usepackage{amsthm}
\usepackage{mathrsfs}
\usepackage{esint}

\usepackage{xcolor}

\definecolor{myurlcolor}{rgb}{0,0,0.4}
\definecolor{mycitecolor}{rgb}{0,0.5,0}
\definecolor{myrefcolor}{rgb}{0.5,0,0}

\usepackage{graphicx}
\usepackage{tikz}
\usepackage{tikz-cd}

\usepackage{etaremune}

\usepackage[normalem]{ulem}

\usepackage{comment}

\usepackage{makeidx}

\usepackage[shortlabels]{enumitem}

\usepackage{caption}

\usepackage{microtype}

\usepackage{hyperref}

\hypersetup{
    colorlinks=true,
    linkcolor=myrefcolor,
    citecolor=mycitecolor,
    urlcolor=myurlcolor
}

\usepackage[capitalize]{cleveref}

\usepackage[
    backend=biber,
    style=numeric-comp,
    sorting=nyt,
    sortcites=true,
    giveninits=true,
    url=false,
    backref=true,
    maxbibnames=99
]{biblatex}

\DefineBibliographyStrings{english}{
    backrefpage  = {$\downarrow$\,p.},
    backrefpages = {$\downarrow$\,pp.}
}

\AtEveryBibitem{%
    \clearfield{note}%
    \clearfield{isbn}%
    \clearfield{issn}%
    \clearfield{abstract}%
    \clearfield{keywords}%
    \clearfield{urlyear}%
    \clearfield{urlmonth}%
    \clearfield{urlday}%
    \clearfield{urldate}%
}

\theoremstyle{plain}

\newtheorem{theorem}{Theorem}[section]
\newtheorem{proposition}[theorem]{Proposition}
\newtheorem{lemma}[theorem]{Lemma}

\theoremstyle{definition}

\newtheorem{definition}[theorem]{Definition}

\usepackage{etoolbox}

\newtheoremstyle{customremark}
    {6pt}          % Space above
    {6pt}          % Space below
    {\normalfont}  % Body font
    {}             % Indentation
    {\bfseries}    % Heading font
    {.}            % Punctuation after heading
    {.5em}         % Space after heading
    {}

\theoremstyle{customremark}
\newtheorem{remark}[theorem]{Remark}

\AtEndEnvironment{remark}{%
    \unskip\nobreak\hfill\(\triangleleft\)%
}

\title{%
    Fields of covariances on non-commutative probability spaces
    in finite dimensions%
}

\author{%
    F. M. Ciaglia$^{1,5}$
    \href{https://orcid.org/0000-0002-8987-1181}
    {\textsuperscript{\textsf{ORCID}}},
    F. Di Cosmo$^{2, 6}$
    \href{https://orcid.org/0000-0003-0256-5913}
    {\textsuperscript{\textsf{ORCID}}},
    L. González-Bravo$^{3,4,7}$
    \href{https://orcid.org/0000-0002-4382-7978}
    {\textsuperscript{\textsf{ORCID}}}%
}

\begin{document}

\maketitle 
\vspace{-0.5cm}
\noindent
{\footnotesize $^{1}$  Universidad Carlos III de Madrid, ROR: \href{https://ror.org/03ths8210}{03ths8210}, Departamento de Matemáticas, Avenida de la Universidad, 30 (edificio Sabatini), 28911 Leganés (Madrid), España.}  \\
{\footnotesize $^{2}$ Universidad de Alcalá, ROR: \href{https://ror.org/04pmn0e78}{04pmn0e78}, Departamento de Física y Matemáticas, Ctra Madrid-Barcelona, km.33, 600, 28805 Alcalá de Henares, Madrid, España.\\
{\footnotesize $^{3}$ Instituto de Ciencias Matem\'{a}ticas ICMAT (CSIC-UAM-UC3M-UCM), ROR: \href{https://ror.org/05e9bn444}{05e9bn444}, Campus de Cantoblanco UAM, Calle Nicolás Cabrera, 13-15, 28049 Madrid, España.} \\}
{\footnotesize $^{4}$ Universidad Complutense de Madrid, ROR: \href{ https://ror.org/02p0gd045}{02p0gd045}. Departamento de Álgebra, Geometría y Topología, Facultad de Ciencias Matemáticas, Pl. de las Ciencias, 3, Moncloa-Aravaca, 28040 Madrid, España}

%\bigskip
\noindent
{\footnotesize $^{5}$\texttt{fciaglia[at]math.uc3m.es}  $^{6}$\texttt{fabio.di[at]uah.es}   $^{7}$\texttt{lauraego[at]ucm.es}}

\begin{abstract}
We introduce the notion of a field of covariances, a contravariant functor from non-commutative probability spaces to Hilbert spaces, as a categorical analogue of statistical covariance. 
%%%%%%%%%%
In the finite-dimensional setting, we obtain a complete characterization of the fields of covariances
satisfying an additional continuity property in terms of a strictly positive constant $\alpha$ and a continuous function $F\colon[0,+\infty)\rightarrow(0,+\infty)$ that is operator monotone on
$(0,+\infty)$. 
%%%%%%%%%
In the commutative case, our result recovers a contravariant formulation of Čencov's characterization of the Fisher--Rao metric tensor, and extends the corresponding uniqueness statement to tracial states on
possibly noncommutative algebras. 
%%%%%%%%%%%%
On faithful states, a subfamily satisfying a symmetry constraint recovers the regular Morozova--Čencov--Petz monotone metric tensors. 
%%%%%%%%%%
More generally, the covariance fields induce positive
contravariant symmetric tensors on the homogeneous manifolds of states generated by the action of the invertible elements of an arbitrary
finite-dimensional $C^*$-algebra. 
%%%%%%%%%
The geometry internal to the support is determined by the positive spectrum of the modular operator, whereas
support-changing directions are governed by $F(0)$. 
%%%%%%%%%%
For $F(t)=(1+t)/2$, and with a suitable normalization convention, the resulting tensors recover the inverse
of the Riemannian metrics induced by the Jordan product. 
%%%%%%%%
In particular, on pure states, every $F$ induces a multiple of the inverse Fubini--Study tensor, with coefficient $2F(0)$ in our normalization, in agreement after inversion with the Petz--Sudár radial extension.
\end{abstract}

{\footnotesize
\tableofcontents
}

\section{Introduction}\label{sec: introduction}

Classical and quantum information geometry in finite dimensions are based on two celebrated classification results. 
%%%%%%%%%%%%%%%%%%%%%%%%
On the classical side, Čencov’s theorem \cite{C1981a} singles out the Fisher--Rao metric tensor as the unique (up to a positive constant) Riemannian metric tensor invariant under congruent embeddings of statistical models on finite outcome spaces. 
%%%%%%%%%%%%%%%%%%%%%%%%
On the quantum side, the Morozova--Čencov--Petz characterization \cite{MC1991,P1996} shows that there is an entire family of Riemannian metric tensors on the manifold of faithful quantum states of a finite-level quantum system that are monotone under the quantum channels of quantum information theory, each determined by an operator monotone function $F$ satisfying a symmetry condition \cite{P1996}.
%%%%%%%%%%%%%%%%%%%%%%%%%%%%%%%%%%%%%%%%%%%%%%% 

There are two shortcomings concerning these foundational results. 
%%%%%%%%%%
First of all, Čencov's theorem is formulated for statistical models of strictly positive
probability vectors on finite outcome spaces, while the
Morozova--Čencov--Petz characterization applies to the manifold of strictly positive density operators on the Hilbert space of a finite-level quantum system. 
%%%%%%%%%
This restriction excludes classical probability distributions with vanishing components and non-faithful quantum states, including pure states.
%%%%%%%%%%%%%%%%%%%%
The state space of a finite-dimensional $C^*$-algebra is naturally partitioned into smooth homogeneous manifolds generated by the action of the
group of invertible elements.
%%%%%%%%%%%%%%%%
In the case of $\mathcal B(\mathcal H)$,
these manifolds are the spaces of density operators of fixed rank, while in the commutative case they are the relative interiors of the faces of the
probability simplex. 
%%%%%%%%%%
Previous constructions based on a
Jordan-algebraic analogue of the Kostant-Kyrillov-Souriau construction provide Riemannian metric tensors on these homogeneous manifolds \cite{CJS2020a,CJS2020,CJS2023}.
%%%%%%%%%%%%%%%%%%%%%%%%%%%%%%%
For pure quantum states, unitary invariance singles out the Fubini--Study Riemannian metric tensor on complex projective space \cite{BZ2006}. 
%%%%%%%%%
Petz and Sudár investigated the behaviour of the
Morozova--Čencov--Petz metric tensors near the pure-state manifold by means of a radial extension from faithful states. 
%%%%%%%%%%%
They showed that the metric tensors associated with regular operator monotone functions satisfying $F(0)>0$ admit a finite and non-degenerate restriction to the pure-state manifold, proportional to the Fubini--Study metric tensor, with proportionality constant equal to $(2F(0))^{-1}$ \cite{PS1996}. 
%%%%%%%%%%%%%%%%%%%%%%%%%%
However, the relation between the Morozova--Čencov--Petz family and boundary geometries for non-faithful and non-pure states has not been clarified.
%%%%%%%%%%%%%%%% 

Secondly, both results are formulated in  separate mathematical contexts: probability/measure theory for the classical case, and Hilbert space theory in the quantum case.
%%%%%%%%%%%%
Nevertheless, the quantum characterization was developed in close analogy with the classical one, as highlighted in \cite{MC1991}, and it strongly depends on the classical case, as it is clear from how Čencov's theorem is invoked in the proof of theorem 5 in \cite{P1996}.
%%%%%%%%%%%%%%%%%%%%%%
Consequently, there is the need for a unified mathematical framework in which both the classical and quantum characterizations can be simultaneously formulated and directly compared, and that helps understanding the origin of the uniqueness and non-uniqueness of the classical and quantum cases, respectively.
%%%%%%%%%%%%%%%%%%%%%%%%%%%%%%%%%%%%%%
\\

%%%%%%%%%%%%%%%%
These shortcomings motivate the present work.
%%%%%%%%%%%%%%%%
We use the language of  $C^{*}$-algebras   \cite{B2006a,BR1987,D1996,T1979,T2003a}, which provides a powerful mathematical framework where classical and quantum theories can be simultaneously formulated,  to unify the classical and quantum characterizations mentioned above in terms of the characterization of a \textit{field of covariances} on the category $\mathsf{NCP}$ of non-commutative probability spaces in finite dimensions (see definition \ref{defn: field of covariances}).
%%%%%%%%%%%%%%%%%%%%%%%%%
In this context, as it is explained in section \ref{sec: fields of covariances}, the formalism of operator algebras through the so-called \textup{GNS} construction suggests the shift of focus from \textit{covariant} objects like the Riemannian metric tensors in \cite{C1981a,MC1991,P1996} to \textit{contravariant} ones that generalize statistical covariance and its quantum counterpart \cite{GHP2009}.
%%%%%%%%%%%%%%%%%%%%%%%%

Statistical covariance plays a fundamental role in classical probability and statistics.  
%%%%%%%%%%%%%%%%%%%%%%%%%%%5
Given a probability space $(\Omega,\mu)$, the covariance between two complex-valued random variables with finite $\mu$-variance can be expressed in terms of the inner product $\mathcal{L}^{2}(\Omega,\mu)$.
%%%%%%%%%%%%%%%%%
Indeed, denoting with  
\begin{equation}
 	\mathbb{E}_{\mu}(\cdot)=\int_{\Omega}(\cdot)\mathrm{d}\mu
\end{equation} 
the expectation value of $(\cdot)$ with respect to $\mu$, the statistical covariance between $X$ and $Y$ reads 
\begin{equation}\label{eqn: statistical covariance}
 	\mathrm{Cov}_{\mu}(X,Y):=\mathbb{E}_{\mu}\left(\overline{(X-\mathbb{E}_{\mu}(X))}(Y-\mathbb{E}_{\mu}(Y))\right)=\langle \mathbf{P}(X)\mid\mathbf{P}(Y)\rangle_{\mu},
\end{equation}
where $\mathbf{P}$ is the orthogonal projection on the orthogonal complement of the vector subspace generated by the identity function with respect to the Hilbert product of $\mathcal{L}^{2}(\Omega,\mu)$.
%%%%%%%%%%%%%%%%%%%%%%%%%%%%%%%%%%%%%%%%%%%%%%%%%%
The change of variable formula for probability measures implies that the statistical covariance is invariant in the sense that
\begin{equation}\label{eqn: invariance of statistical covariance under *-homomorphisms}
	\mathrm{Cov}_{\sigma}(X\circ\phi,Y\circ\phi)=\mathrm{Cov}_{\mu}(X,Y),
\end{equation}
where $\phi\colon \Lambda\rightarrow\Omega$ is an invertible measurable map such that $\phi_{*}\sigma=\mu$, with $\sigma$ a probability measure on $\Lambda$ and $\phi_{*}$ the pushforward operation between measures.
%%%%%%%%%%%%%%%%%%%%%%%%%%%%%%%%%%%%%%%%%%%%%%%%%%%%
Note that equation \eqref{eqn: invariance of statistical covariance under *-homomorphisms} reads as the invariance of what would be a \textit{contravariant} tensor in differential geometry because the ``transformed point'' $\mu=\phi_{*}\sigma$ and the ``transformed vector'' $X\circ\phi$ are on different sides of the equality.
%%%%%%%%%%%%%%%%%%%%%%%%%%
As noted in \cite{AJLS2017,C1981a,H1991,N2024b}, if we focus on probability measures on discrete and finite outcome spaces, the statistical covariance is the inverse of the Fisher--Rao metric tensor.
%%%%%%%%%%%%%%%%%%%
Čencov's theorem on the uniqueness of the Fisher--Rao metric tensor is based on the invariance with respect to arbitrary state-preserving conditional expectations between probability measures \cite{C1981a,C1986b}, of which the push-forward operations in equation \eqref{eqn: invariance of statistical covariance under *-homomorphisms} are a particular case.
%%%%%%%%%%%%%%
Consequently,   the uniqueness of the Fisher--Rao metric tensor can be equivalently formulated in \textit{contravariant} terms by stating that the statistical covariance is the only inner product on the subspace of centered random variables in $\mathcal{L}^{2}(\Omega,\mu)$ which is invariant under state-preserving conditional expectations.
%%%%%%%%%%%%%%%%%%%%%%%%%%%%%%%%%

On the other hand, in the finite-dimensional quantum case where probability measures are replaced by quantum states (\textit{i.e.},  positive semidefinite operators in the non-commutative algebra $\mathcal{B}(\mathcal{H})$ of bounded linear operators on the finite-dimensional complex Hilbert space $\mathcal{H}$ of the system), a quantum covariance $\mathrm{qCov}_{\rho}$ at the strictly positive quantum state $\rho$ is defined as a sesquilinear product on $\mathcal{B}(\mathcal{H})$ that is Hermitean and positive definite  \cite{GHP2009}.
%%%%%%%%%%%%%%%%%%% %%%%%%%%%%%%%%%%%
Admissible quantum covariances are assumed to satisfy the  so-called \textit{monotonicity property} given by
\begin{equation}\label{eqn: monotonicity of quantum covariance}
 \mathrm{qCov}_{\rho}(\Psi_{*}(\mathbf{b}),\Psi_{*}(\mathbf{b}))\leq\mathrm{qCov}_{\sigma}(\mathbf{b},\mathbf{b}),
\end{equation}
where  $\sigma$ is a strictly positive quantum state on $\mathcal{K}$ with $\mathrm{dim}(\mathcal{K})<\infty$, $\mathbf{b}\in\mathcal{B}(\mathcal{K})$, $\Psi$ is a quantum channel such that $\Psi(\rho)=\sigma$, and $\Psi_{*}$ is its pre-dual map.
%%%%%%%%%%%%%%%%%%%%%%%%%%
Analogously to the classical case, the monotonicity in equation \eqref{eqn: monotonicity of quantum covariance} reads as an inequality between \textit{contravariant} objects (the quantum covariances) because the ``transformed point'' $\sigma=\Phi^{*}\rho$ and the ``transformed vector'' $\Phi(\mathbf{b})$ are on different sides of the inequality. 
%%%%%%%%%%%%%%%%%%%%%%%%%%%%%%%%%%%%%
Moreover, the classification of all quantum covariances satisfying the monotonicity property of equation \eqref{eqn: monotonicity of quantum covariance} developed in \cite{GHP2009} amounts precisely to a contravariant formulation of the Morozova--Čencov--Petz monotone metric tensors \cite{MC1991,P1996}, thus showing that  there are infinitely many quantum monotone covariances, and that the family of all such quantum covariances is parametrized by the operator monotone functions appearing in \cite{P1996}.
%%%%%%%%%%%%%%%%%%%%%%%%%%%%%%%%%%%%%
\\

As explained in more detail in subsection \ref{subsec: NCP}, the operator algebraic framework mentioned above allows us to work with classical and quantum systems as different object in the same category $\mathsf{NCP}$ of non-commutative probability spaces, and to look at classical and quantum covariances as inner products on the so-called \textup{GNS} Hilbert space of the state under consideration.
%%%%%%%%%%%
In this context, the invariance and monotonicity of equations \eqref{eqn: invariance of statistical covariance under *-homomorphisms} and \eqref{eqn: monotonicity of quantum covariance} are captured by what we call a \textit{field of covariances} on $\mathsf{NCP}$ in definition \ref{defn: field of covariances}.
%%%%%%%%%%%%
This is a contravariant functor from   $\mathsf{NCP}$  into the category of complex Hilbert spaces and contractions.
%%%%%%%%%%%%%%%%%%%%%%%%%%%%%%%%%%
This novel categorical reformulation of the problem of characterizing classical and quantum covariances adequately fits the modern line of research centered around a categorical formulation of classical and quantum probability \cite{BF2014,F2020,FGPF2023,P2020f,P2021a,P2023}, and connects with  Čencov's original categorical thinking \cite{C1965,C1978b,C1981a,MC1991}.
%%%%%%%%%%%%%%%%%%%%%%%%%%%%%%%%%%%%%%%%%%%%%%%%%%%%%%%%%%%%%%%%%%%%%%%
Our main result provides a complete characterization of continuous fields of covariances according to definition \ref{defn: continuity for fields of covariances} in the finite-dimensional setting that agrees with the conjectures in \cite{K2016,CDG2023} for the generalization of the Morozova--Čencov--Petz metrics to the operator algebraic setting.
%%%%%%%%%%%%%%%%
More explicitly, theorem \ref{thm: full classification of fields of covariances on fNCP} says that every continuous field of covariances is uniquely determined by a pair $(\alpha,F)$, where $\alpha>0$ and $F:[0,\infty)\to(0,\infty)$ is continuous and operator monotone on $(0,\infty)$, and \textit{vice versa}.
%%%%%%%%%%%%%%%%%%%%

As explained in detail in section \ref{sec:applications}, the characterization in theorem \ref{thm: full classification of fields of covariances on fNCP} has the following geometric consequences.
%%%%%%%%%%%%%%%%

\begin{itemize}
\item
For arbitrary states on commutative algebras of the form $\mathcal{L}^\infty(X_n,\#)$, where $X_n$ is a finite set, we recover the contravariant form of Čencov's uniqueness theorem, or equivalently the
uniqueness of statistical covariance \cite{C1981a,N2024b}. 
%%%%%%%%
The result includes probability distributions that
vanish on subsets of $X_n$, for which the induced metric is the Fisher--Rao metric on the relative interior of the corresponding face of
the probability simplex. 
%%%%%%%%%
The same covariance formula holds for tracial states on arbitrary finite-dimensional, possibly noncommutative, $C^*$-algebras. 
%%%%%%%%%%%%%%%%
In particular, this shows that the classical uniqueness phenomenon is controlled by the triviality of the modular operator associated to a state rather than by the commutativity of the algebra.
%%%%%%%%%%%%%%%%

\item
On the manifold of faithful states of an arbitrary finite-dimensional $C^*$-algebra, the real part of a covariance field defines a positive
contravariant symmetric tensor. 
%%%%%%%%%%
When $F$ satisfies the Petz symmetry condition $F(t)=tF(t^{-1})$, its inverse specializes, for
$\mathscr A=\mathcal B(\mathcal H)$, to the regular
Morozova--Čencov--Petz quantum monotone metric tensor associated with $F$ \cite{MC1991,P1996,GHP2009}.

\item At a non-faithful state $\rho$, the field of covariances induces, after fixing a normalization constant $c$ for the real-linear identification of the cotangent space at $\rho$ with a subspace of the GNS Hilbert space at $\rho$, a positive contravariant symmetric tensor on the homogeneous manifold generated by the action of the invertible elements.
%%%%%%%%%%%%%%%%%%%%%
The contribution tangent to the supported algebra is governed by the positive spectrum of the modular operator of $\rho$, whereas support-changing
directions are proportional to the scalar value $F(0)$. 
%%%%%%%%%%%%%%%%%%%%
The dependence on $F(0)$ is intrinsic, while the proportionality constant depends on $c$.
%%%%%%%%%%%%%%%%%%%%
For $F(t)=(1+t)/2$ and $c=\sqrt{2}$, the resulting tensor agrees with the contravariant tensor induced by the Jordan product, and its inverse recovers the Riemannian metrics constructed in \cite{CJS2020,CJS2020a,CJS2023}.
%%%%%%%%%%%%%%%%%%%%%%
For pure states on $\mathcal B(\mathcal H)$  and $c=\sqrt{2}$, every $F$ yields $2F(0)$ times the inverse Fubini--Study tensor. This agrees with the Petz--Sudár radial extension for regular symmetric operator monotone functions \cite{PS1996}.
\end{itemize}

\textbf{Structure of the paper.}
In section \ref{sec: ncp}, we recall the basics of operator algebra theory needed in later sections, and we introduce the category $\mathsf{NCP}$ of non-commutative probability spaces and its finite-dimensional subcategories $\mathsf{fNCP}$ and $\mathsf{fNCT}$ of non-commutative probability spaces and tracial states in finite dimensions, respectively.
%%%%%%%%%%%%%%%%%%%%%%%%%%%
In section \ref{sec: fields of covariances},  we define fields of covariances and their continuity properties.
%%%%%%%%%%%%%%%%%%%%%%%%
In section \ref{sec: classification}, we prove the main theorem \ref{thm: full classification of fields of covariances on fNCP} giving a complete characterization of continuous fields of covariances on $\mathsf{fNCP}$ through the following steps.
%%%%%%%%%%%%%%%%%%%
Proposition \ref{prop: fields of covariances at the tracial state of B(H)} fixes the covariance at the tracial state of a full matrix algebra.
%%%%%%%%%%%%%%%
Propositions
\ref{prop: field of covariances on faithful rational tracial states},
\ref{prop: classification of invariant covariances for any algebra and tracial states}
and theorem \ref{thm: full classification of fields of covariances on fNCT}
extend the result to all finite-dimensional tracial states.
%%%%%%%%%%%%%%%%%%%%%%%%
Propositions
\ref{prop: covariance operator on the centralizer},
\ref{prop: fields of covariances on faithful states in fNCP with non-degenerate modular operator},
and 
\ref{prop: fields of covariances on faithful states in fNCP}
identify the faithful non-tracial case.
%%%%%%%%%%%%%%%%%%%%%
Proposition \ref{prop: fields of covariances on faithful states in fNCP have operator monotone functions}
then proves that the scalar function appearing in the covariance formula is operator monotone, and proposition
\ref{prop: continuous fields of covariances on fNCP}
extends the formula to non-faithful states.
%%%%%%%%%%%%%%%%%%%%%%
Theorem \ref{thm: full classification of fields of covariances on fNCP} then provides a complete characterization of continuous fields of covariances on $\mathsf{fNCP}$ essentially showing the converse of proposition \ref{prop: continuous fields of covariances on fNCP}. 
%%%%%%%%%%%%%%%%%%%%%%%%%%
In section \ref{sec:applications}, we investigate the geometric tensors induced by the continuous fields of covariances characterized in theorem \ref{thm: full classification of fields of covariances on fNCP} in various explicit examples. 
%%%%%%%%%%%%%%%%%%%%%%%%%%%%%%%%%
Finally, section \ref{sec: conclusions} summarizes the results and discusses directions for future work.

\section{Operator algebras and the category of non-commutative probability spaces}\label{sec: ncp}

A $C^*$-algebra $\mathscr{A}$ is a complex Banach algebra endowed with an involution $\dagger$, that is, a bounded anti-linear map $\dagger$ such that $(\mathbf{a}^{\dagger})^{\dagger}=\mathbf{a}$ and $(\mathbf{ab})^{\dagger}=\mathbf{b}^{\dagger}\,\mathbf{a}^{\dagger}$ for all $\mathbf{a},\mathbf{b}\in\mathscr{A}$, and satisfying the so-called $C^*$-property $\Vert\mathbf{a}^{\dagger}\mathbf{a}\Vert = \Vert\mathbf{a}\Vert ^{2}$, where $\Vert \cdot\Vert$ is the Banach norm on $\mathscr{A}$.
%%%%%%%%%%%%%%%%%%%%%%%%%%%%%%%%%%%%%%%%%%%
Typical examples of $C^*$-algebras are the commutative algebra $\mathbb{C}^{n}$ of complex vectors with component-wise multiplication and the standard norm, the non-commutative matrix algebra $\mathbb{M}_{n}(\mathbb{C})$ of square complex matrices with usual algebraic operations and the operator/spectral norm, the commutative algebra $\mathcal{L}^{\infty}(\Omega,\nu)$ of  equivalence classes of $\nu$-absolutely bounded measurable functions on the measure space $(\Omega, \nu)$ with standard algebraic operations and the sup norm, the non-commutative algebra $\mathcal{B}(\mathcal{H})$ of bounded linear operators on the complex Hilbert space $\mathcal{H}$ with the usual algebraic operations among linear operators and the operator norm.
%%%%%%%%%%%%%%%%%%
A bounded linear map $\phi\colon\mathscr{A}\rightarrow\mathscr{B}$ between $C^*$-algebras is called *-preserving if  $\phi(\mathbf{a}^{\dagger})=\phi(\mathbf{a})^{\dagger}$ for all $\mathbf{a}\in \mathscr{A}$, it is a *-homomorphism if it is *-preserving and $\phi(\mathbf{xy})=\phi(\mathbf{x})\phi(\mathbf{y})$ for all $\mathbf{x},\mathbf{y}\in\mathscr{A}$, it is a *-isomorphism if it is a bijective *-homomorphism, and a *-automorphism if it is a *-isomorphism with $\mathscr{B}=\mathscr{A}$. 
%%%%%%%%%%%%%%%%%%%%%%%%%%%%%%%%%%%%%%%
We refer to  \cite{B2006a,BR1987,D1996,T1979,T2003a} for all details on operator algebras that are not discussed here.
%%%%%%%%%%%%%%%%%%%%

\subsection{States and the \textup{GNS} construction}
In the  operator algebraic context, a classical system is associated with the commutative $C^*$-algebra $\mathcal{L}^{\infty}(\Omega,\nu)$, while a quantum system with the non-commutative $C^{*}$-algebra $\mathcal{B}(\mathcal{H})$.
%%%%%%%%%%%%%%%%%%%
Probability measures and quantum states are then different examples of the notion of \textit{state} on a $C^*$-algebra $\mathscr{A}$.
%%%%%%%%%%%%%%%%%
Denoting with $\mathscr{A}^{*}$ the Banach dual of $\mathscr{A}$ and with $\Vert\cdot\Vert_{\mathscr A^*}$ its dual norm, a state $\rho$ on $\mathscr{A}$ is an element in $\mathscr{A}^{*}$ such that $\rho(\mathbf{a}^{\dagger}\mathbf{a})\geq 0$ for all $\mathbf{a}\in \mathscr{A}$  and such that $\Vert \rho\Vert_{\mathscr{A}^{*}}=1$ (or $\rho(\mathbb{I})=1$ whenever $\mathscr{A}$ has an identity element $\mathbb{I}$) \cite{B2006a,BR1987,T1979}.
%%%%%%%%%%%%%%%%%%%%%%%%%%%%%%%%%%%%%%%%%%%%%%
The space of states of $\mathscr{A}$ is denoted by $\mathcal{S}(\mathscr{A})$.
%%%%%%%%%%%%%%%%%%%%%%%%%%%%

The couple $(\mathscr{A},\rho)$, where $\mathscr{A}$ is a $C^*$-algebra and $\rho$ is a state on $\mathscr{A}$, is often called a \textit{non-commutative probability space}, and is at the heart of Voiculescu's free probability theory \cite{S2009,VDN1992,VSW2016}.
%%%%%%%%%%%%%%%%%%%%%%%%%%%%

A state $\rho$ is called \textit{faithful} if the set
\begin{equation}\label{equation: Gelfand ideal}
	\mathscr{N}_{\rho}:=\{\mathbf{a}\in\mathscr{A}\mid\;\rho(\mathbf{a}^{\dagger}\mathbf{a})=0\},
\end{equation} 
only contains the zero element.
%%%%%%%%%%%%%%%%%%%%%%%%%%%%%%%%%%%%%%%%%
A state $\rho$ is called \textit{tracial} if
\begin{equation}\label{eqn: tracial states}
	\rho(\mathbf{ab})=\rho(\mathbf{ba})	
\end{equation}
for all $\mathbf{a},\mathbf{b}\in\mathscr{A}$.
%%%%%%%%%%%%%%%%%%%%%%
Clearly, all states are tracial when $\mathscr{A}$ is Abelian.
%%%%%%%%%%%%%%%%%%%
When $\mathscr{A}$ is finite-dimensional, it holds (see, \textit{e.g.}, \cite{B2006a,D1996} and \cite[thm. 11.2]{T2003a}) 
\begin{equation}\label{eqn: isomorphism of finite-dimensional algebras} 
\mathscr{A}\cong \bigoplus_{j=1}^{N<+\infty}\mathcal{B}(\mathcal{H}_{j}) \cong \bigoplus_{j=1}^{N}\mathbb{M}_{n_{j}}(\mathbb{C}),
\end{equation}
where $\mathrm{dim}(\mathcal{H}_{j})=n_{j}$, and thus there exist a faithful tracial state $\tau$ on $\mathscr{A}$ determined by the trace on $\mathbb{M}_{K}(\mathbb{C})$, where $K=\sum_{j=1}^{N}n_{j}$, in which $\mathscr{A}$ faithfully embeds.
%%%%%%%%%%%%%%%%%%%%%%%%%% 
Every state $\rho$ on $\mathscr{A}$ can then be written as
\begin{equation}\label{eqn: density operator of a state}
\rho(\mathbf{a})=\tau(\varrho\,\mathbf{a}),
\end{equation}
with $\varrho\in\mathscr{A}$ the so-called \textit{density operator} associated with $\rho$.
%%%%%%%%%%%%%%%%%%%%%%
Note that $\varrho$ is uniquely determined because $\tau$ is faithful.
%%%%%%%%%%%%%%%%%%%%
Moreover, $\rho$ is faithful if and only if $\varrho$  is invertible.
%%%%%%%%%%%%%%%%%%%%%%%%%%%%%%%%%%%

Associated with $\rho$ there is a Hilbert space $\mathcal{H}_{\rho}$, its so-called \textup{GNS} Hilbert space.
%%%%%%%%%%%%%%%%%%%%%%%%%%%%%%%%%%%%
Specifically, the set $\mathscr{N}_{\rho}$ in equation \eqref{equation: Gelfand ideal} turns out to be  a left ideal called \textit{Gelfand ideal} of $\rho$, the vector space $\mathscr{A}/\mathscr{N}_{\rho}$ inherits the inner product 
\begin{equation}\label{eqn: gns inner product}
	\langle[\mathbf{a}]\mid[\mathbf{b}]\rangle_{\rho}=\rho(\mathbf{a}^{\dagger}\mathbf{b}),	
\end{equation}
and the \textup{GNS} Hilbert space is defined as
\begin{equation}\label{eqn: gns hilbert space} 
	\mathcal{H}_{\rho}:=\overline{\mathscr{A}}^{\langle\cdot\mid\cdot\rangle_{\rho}},
\end{equation}
that is,  $\mathcal{H}_{\rho}$ is the closure of $\mathscr{A}/\mathscr{N}_{\rho}$ with respect to the norm induced by $\langle\cdot\mid\cdot\rangle_{\rho}$.
%%%%%%%%%%%%%%%%%%%%%%%%%%%%%%%%%%%%%%%%%%%%%%%%%%%%%%%%%%%%%%%%%
The image in $\mathcal{H}_{\rho}$ of  $\mathbf{a}\in \mathscr{A}$ is denoted with $\xi_{\mathbf{a}}^{\rho}$.
%%%%%%%%%%%%%%
There is a natural representation $\pi_{\rho}$ of $\mathscr{A}$ in $\mathcal{B}(\mathcal{H}_{\rho})$ induced by 
\begin{equation}\label{eqn: gns representation}
	\pi_{\rho}(\mathbf{a})\xi_{\mathbf{b}}^{\rho}=\xi_{\mathbf{ab}}^{\rho}.
\end{equation}
%%%%%%%%%%%%%%%%%%%%%%%%%
This representation is called the \textup{GNS} representation associated with $\rho$, and it always admits a cyclic vector that coincides with $\xi_{\mathbb{I}}^{\rho}$ whenever $\mathscr{A}$ is unital \cite{BR1987}.
%%%%%%%%%%%%%%%%%%%%%%%%%%

In the classical case where $\mathscr{A}=\mathcal{L}^{\infty}(\Omega,\nu)$ and $\rho$ is a probability measure on $\Omega$ which is absolutely-continuous with respect to $\nu$, it turns out that $\mathcal{H}_{\rho}$ coincides with the space $\mathcal{L}^{2}(\Omega,\rho)$ endowed with its standard Hilbert product.
%%%%%%%%%%%%%%%%%%%%%%%%%%%%%%%%%%%%%
The \textup{GNS} Hilbert space is thus the space of complex-valued random variables on $\Omega$ having finite-variance with respect to $\rho$, and   we may interpret the elements of the \textup{GNS} Hilbert space of a state on a non-commutative $C^{*}$-algebra as the non-commutative analogues of complex-valued random variables with finite variance.
%%%%%%%%%%%%%%%%%%%%%%%%%%%%%%%%%%
Therefore, equation \eqref{eqn: statistical covariance}  simply states that the statistical covariance $\mathrm{Cov}_{\rho}$ is nothing but the real part of the \textup{GNS} Hilbert product on the orthogonal complement of the subspace generated by the identity function on $\Omega$, thus hinting at a strong connection between the \textup{GNS} construction and the statistical covariance, which is the inverse of the Fisher-Rao metric tensor when $\Omega$ is finite and $\rho$ is faithful \cite{C1981a}.
%%%%%%%%%%%%%%%%%%%%

In the quantum case where $\mathscr{A}=\mathcal{B}(\mathcal{H})$ and $\rho$ is the linear functional $\rho(\mathbf{a})=\mathrm{Tr}(\varrho\,\mathbf{a})$ associated with the positive-semidefinite, trace-class operator $\varrho$ with unit trace, the \textup{GNS} Hilbert space $\mathcal{H}_{\rho}$ is isomorphic to a subspace of the Hilbert-Schmidt operators on $\mathcal{H}$.
%%%%%%%%%%%%%%%%%%%%%%%%%%%%%%%%%%%
In particular, when $\mathrm{dim}(\mathcal{H})<\infty$ and $\varrho$ is invertible (so that $\rho$ is faithful), $\mathcal{H}_{\rho}$ coincides with $\mathcal{B}(\mathcal{H})$ endowed with the Hilbert product $\langle \mathbf{a}\,|\,\mathbf{b}\rangle_{\rho}=\rho(\mathbf{a}^{\dagger}\mathbf{b})$, whose real part coincides with the quantum covariance in \cite{GHP2009} associated with the operator monotone function $f(t)=\frac{1+t}{2}$, thus hinting at a strong connection between the \textup{GNS} construction and quantum covariances, which are the inverse of the quantum monotone metric tensors in \cite{MC1991,P1996}.
%%%%%%%%%%%%%%%%%%%%%%%%%%%%%%%%
%In  and algebraic quantum field theory, it is very difficult to overestimate the importance of the \textup{GNS} Hilbert space and representation (\textit{cf}. \cite{A2009,FR2020,H1996a}).
%%%%%%%%%%%%%%%%%%%%%
%For instance,  it provides the crucial link between the Heisenberg picture (inherent in the algebraic approach) and the Schrödinger picture (operating within the constructed Hilbert space).
%%%%%%%%%%%%%%%%%%%%%%%%%%%%%%%
%Moreover, the explicit \textup{GNS} construction associated with particular states  often carries important physical meaning (\textit{e.g.}, the fact that the irreducibility of the \textup{GNS} representation is equivalent to the state being pure, or the existence of a link between thermal equilibrium and modular theory for \textit{KMS} states).
%%%%%%%%%%%%%%%%%%%%%%%

Let $\mathscr{A}$ be finite-dimensional.
%%%%%%%%%%%%%%%%%%%%%%%%%%%%%
Associated with every state\footnote{In the infinite-dimensional case, only the so-called \textit{normal states} admit support projections.} $\rho$ there is an orthogonal projection\footnote{Recall that an orthogonal projection in $\mathscr{A}$ is a self-adjoint element $\mathbf{p}$ such that $\mathbf{p}^{2}=\mathbf{p}$.} $\mathbf{p}\in \mathscr{A}$ such that 
\begin{equation}\label{eqn: support projection of normal state}
\rho(\mathbf{a})=\rho(\mathbf{pa})=\rho(\mathbf{ap})=\rho(\mathbf{pap})
\end{equation}
for all $\mathbf{a}\in \mathscr{A}$.
%%%%%%%%%%%%%%%%%%%%%%%%%%%%
This projection is referred to as the \textit{support projection} of $\rho$, and we set $\mathbf{q}:=\mathbb{I}-\mathbf{p}$.
%%%%%%%%%%%%%%%%%%%%%
When $\rho$ is faithful, then $\mathbf{p}=\mathbb{I}$.
%%%%%%%%%%%%%%%%%%%%%%%%%%%%%%
If there is risk of confusion, we will write $\mathbf{p}_{\rho}$ to denote the support projection of $\rho$.
%%%%%%%%%%%%%%%%%%%%%

The algebra $\mathscr{A}$ can be decomposed into the direct sum
\begin{equation} 
\mathscr{A}=\mathbf{p}\mathscr{A}\mathbf{p}\oplus\mathbf{q}\mathscr{A}\mathbf{p}\oplus\mathbf{p}\mathscr{A}\mathbf{q}\oplus\mathbf{q}\mathscr{A}\mathbf{q}\equiv \mathscr{A}_{pp}\oplus \mathscr{A}_{qp} \oplus \mathscr{A}_{pq} \oplus \mathscr{A}_{qq}.
\end{equation}
%%%%%%%%%%%%%%%%%%%%%%%%%%
Note that $\mathscr{A}_{pp}$ and $\mathscr{A}_{qq}$ are *-subalgebras, and the restriction $\tilde{\rho}$  of $\rho$ on $\mathscr{A}_{pp}$ is faithful.
%%%%%%%%%%%%%%%%%%%%%%%%%%%%%%%%%
Moreover, the Gelfand ideal reads
\begin{equation}\label{equation: Gelfand ideal normal state}
\mathscr{N}_{\rho}=\mathscr{A}_{pq}\oplus\mathscr{A}_{qq},
\end{equation}
so that the \textup{GNS} Hilbert space is decomposed according to\footnote{Note that, in the infinite-dimensional case,  $\mathcal{H}_{\rho}^{pp}$ and $\mathcal{H}_{\rho}^{qp}$  are isomorphic to the completions of $\mathscr{A}_{pp}$ and $\mathscr{A}_{qp}$ with respect to the inner product in equation \eqref{eqn: gns inner product}, respectively.}
\begin{equation}\label{eqn: gns hilbert space normal state}
\mathcal{H}_{\rho}\cong\mathscr{A}_{pp}\oplus\mathscr{A}_{qp}\equiv \mathcal{H}_{\rho}^{pp}\oplus\mathcal{H}_{\rho}^{qp}.
\end{equation}
%%%%%%%%%%%%%%%%%%%%%%%%%%%%%%%%%%%%%%%%%%%%%%%%%%%%%%%%%%

\subsection{The modular operator of a state}

Let $\mathscr{A}$ be finite-dimensional and let $\rho$ be a state on $\mathscr{A}$ with support projection $\mathbf{p}$. 
%%%%%%%%%%%%%%%%%%%%%%%%%%%
We define a linear operator $\Delta_\rho$ on $\mathcal{H}_\rho$ by setting
%%%%%%%%%%%%%%%%%%%%%%%%%%%
\begin{equation}\label{eqn: modular operator}
	\langle \xi_{\mathbf{b}} | \Delta_{\rho}(\xi_{\mathbf{a}}) \rangle_{\rho} := \langle \xi_{\mathbf{pa}^{\dagger}\mathbf{p}} |  \xi_{\mathbf{pb}^{\dagger}\mathbf{p}} \rangle_{\rho}=\rho\left(\mathbf{papp}\mathbf{b}^{\dagger}\mathbf{p}  \right) \stackrel{\mbox{\eqref{eqn: support projection of normal state}}}{=} \rho(\mathbf{apb}^\dagger).
\end{equation}
%%%%%%%%%%%%%%%
When $\rho$ is faithful, one has $\mathbf{p}= \mathbb{I}$, and the operator in equation \eqref{eqn: modular operator} coincides with the standard modular operator of Tomita-Takesaki modular theory \cite{T1970a, T2003,S1981}.
%%%%%%%%%%%%%%%%%%%
For a non-faithful state, equation \eqref{eqn: modular operator} gives the faithful modular operator of the reduced state on the supported corner $\mathscr{A}_{pp}$ and annihilates the remaining summand, i.e.,   $\Delta_{\rho}(\xi_{\mathbf{a}})=\mathbf{0}$ when $\xi_{\mathbf{a}}\in\mathcal{H}_{qp}$. It also follows that
\begin{equation}\label{eqn: modular operator classical case}
	\langle \xi_{\mathbf{b}} | \Delta_{\rho}(\xi_{\mathbf{a}}) \rangle_{\rho}=\langle \xi_{\mathbf{b}} |\xi_{\mathbf{a}} \rangle_{\rho}
\end{equation}
whenever $\mathscr{A}$ is Abelian or $\rho$ is a tracial state.
%%%%%%%%%%%%%%%%%%%
We will refer to the operator $\Delta_{\rho}$  as the \textit{modular operator} associated with $\rho$.
%%%%%%%%%%%%%%%%%%%%

When $\rho$ is faithful, the one-parameter unitary group $\mathrm{exp}(it\log(\Delta_{\rho}))$ on $\mathcal{H}_{\rho}$ generates a one-parameter group $\Phi_{t}^{\rho}$ of automorphisms of $\mathscr{A}$ that preserves $\rho$ and satisfies
\begin{equation}\label{eqn: modular flow}
\langle\xi|\pi_{\rho}(\Phi_{t}^{\rho}(\mathbf{a}))(\eta)\rangle_{\rho}=\langle\xi| \Delta_{\rho}^{-it}\pi_{\rho}(\mathbf{a}) \Delta_{\rho}^{it}(\eta)\rangle_{\rho},
\end{equation}
where $\pi_{\rho}$ is the \textup{GNS} representation in equation \eqref{eqn: gns representation}.
%%%%%%%%%%%%%%%%%%%%%%%
The one-parameter group in equation \eqref{eqn: modular flow} is called \textit{the modular flow of $\rho$}, and it is  a cornerstone of the modern theory of $W^{*}$-algebras   and of its application to algebraic quantum field theory \cite{B2000b,B2000a}.
%%%%%%%%%%%%%%%%%%%%%%%%%%%%%%%%%%%%%%%%%%%%
\\

For later use, it is convenient to describe the modular operator of a state $\rho$ in terms of the \textup{GNS} representation of a faithful tracial reference state $\tau$ 
on the \textbf{finite-dimensional}\footnote{The following construction does not carry \textit{verbatim} to the infinite-dimensional case.} algebra $\mathscr{A}$.
%%%%%%%%%%%%%%%%%%
Let $\tau$ be a faithful tracial state on the finite-dimensional $C^{*}$-algebra $\mathscr{A}$, 
and let $\varrho$ denote the density operator associated with a state $\rho$, 
as in equation \eqref{eqn: density operator of a state}. 
The modular operator $\Delta_{\rho}$ can be represented on the Hilbert space $\mathcal{H}_{\tau}$ 
by a linear operator $\tilde{\Delta}_{\rho}$, defined through the relation  
\begin{equation}\label{eqn: modular operator wrt tracial state} 
\langle \xi_{\mathbf{b}}^{\tau}, \tilde{\Delta}_{\rho}R_{\rho}(\xi_{\mathbf{a}}^{\tau}) \rangle_{\tau} 
= \langle \xi_{\mathbf{b}}^{\rho}, \Delta_{\rho}(\xi_{\mathbf{a}}^{\rho}) \rangle_{\rho},
\end{equation}
where $R_{\rho}(\xi_{\mathbf{a}}^{\tau}) = \xi_{\mathbf{a}\varrho}^{\tau}$ denotes right multiplication by $\varrho$.
%%%%%%%%%%%%%%%%%%%%%
The operator $R_{\rho}$ has kernel $ \mathscr{N}_{\rho}\subset \mathscr{A}=\mathcal{H}_{\tau}$, 
and it is invertible on its complement 
\begin{equation} 
\mathcal{H}_{\rho}  = \mathscr{A}_{pp}\oplus \mathscr{A}_{qp} \equiv\mathscr{A}p\subseteq\mathscr{A}=\mathcal{H}_{\tau},
\end{equation}
with inverse given by right multiplication with the inverse of $\varrho$ on $\mathscr{A}_{pp}$, denoted by $\varrho^{+}$.  
%%%%%%%%%%%%%%%%%%%%%%%%%%%%%%%%%%%
We define $W_{\rho}\colon \mathcal{H}_{\tau}\to \mathcal{H}_{\tau}$ as the partial inverse of $R_{\rho}$, namely  
\begin{equation} 
W_{\rho}|_{\ker(R_{\rho})}=0, 
\qquad 
W_{\rho}|_{\mathscr{A}p }=\left(R_{\rho}\right)|_{\mathscr{A}p}^{-1}.
\end{equation}
%%%%%%%%%%%%%%%%%%%%%%%%%%%%
Essentially, $W_{\rho}$ vanishes on the kernel of $R_{\rho}$ and coincides with its inverse on $\mathscr{A}p$.  
%%%%%%%%%%%%%%%%
Finally, letting $L_{\rho}(\xi_{\mathbf{a}}^{\tau})=\xi_{\varrho \mathbf{a}}^{\tau}$, a direct computation shows that  
\begin{equation}\label{eqn: modular operator wrt tracial state 2} 
\tilde{\Delta}_{\rho} \;=\; L_{\rho}W_{\rho}.
\end{equation}  
%%%%%%%%%%%%%%%%%
In particular, when $\rho$ is faithful, equation \eqref{eqn: modular operator wrt tracial state 2} reduces to the familiar formula $\tilde{\Delta}_{\rho}=L_{\rho}R_{\rho}^{-1}$ \cite{P1985,P1986a,P2008}. 
%%%%%%%%%%%%%%%%%%%%%%%%%%%%%%%%%%%%%%%

The next technical lemma investigates what happens to the representation of the modular operator on $\mathcal{H}_{\tau}$ when sequences of states are considered.
%%%%%%%%%%%%%%%%%%%%%%%%%%%%%%%%%%%%%%%%%%%%%%

\begin{lemma}\label{lem: convergence of modular operator}
Let $\{\rho_{n}\}_{n\in\mathbb{N}}$ be a sequence of faithful states on the finite-dimensional $C^*$-algebra $\mathscr{A}$ such that $\|\rho_{n} -\rho\|_{\mathscr{A}^{*}}\to 0$, and let $\mathbf{p}$ be the support projection of $\rho$.
%%%%%%%%%%%%%%%%%%%%%%%%%%%%%%%%%%%%%%%%%%%%%%%%%%%%%%%%%%
Let $\tau$ be a fixed faithful tracial state on $\mathscr{A}$, and let $\varrho$ and $\varrho_{n}$, respectively, be the density operators associated with $\rho$ and $\rho_{n}$ as in equation \eqref{eqn: density operator of a state}.
%%%%%%%%%%%%%%%%%%%%%%%%%%%%%%%%%%%%%%%%%%%%%%%%%%%%%%%
Assume moreover that  $[\varrho_n,\mathbf p]=0$ for $n$ sufficiently large.
%%%%%%%%%%%%%%%%%%%%%%%%%%%%%%%%%%%%%%%%%%%%%%%%%%%%%%%
Let $R_{\rho}$, $L_{\rho}$, $W_{\rho}$, $\tilde{\Delta}_{\rho}$ be the linear operators on the \textup{GNS} Hilbert space $\mathcal{H}_{\tau}$ defined as above, and analogously for $\rho_{n}$.
%%%%%%%%%%%%%%%%%%%%%%%%%%%%%%%%%%%%%%%%%%%
Then:
\begin{enumerate}
\item $\|L_{\rho_{n}}- L_{\rho}\|_{\mathcal{B}(\mathcal{H}_{\tau})}\to 0$ and $\|R_{\rho_{n}}- R_{\rho}\|_{\mathcal{B}(\mathcal{H}_{\tau})}\to 0$;
\item if $\rho$ is faithful (i.e.\ $\varrho$ invertible), then $\|W_{\rho_{n}} - W_{\rho}\|_{\mathcal{B}(\mathcal{H}_{\tau})}\to 0$ and $\|\tilde{\Delta}_{\rho_{n}}-\tilde{\Delta}_{\rho}\|_{\mathcal{B}(\mathcal{H}_{\tau})}\to 0$;
\item if $\rho$ is non-faithful with support projection  $\mathbf{p}=\mathbf{p}_{\rho}<\mathbb{I}$, then:
\begin{enumerate}
\item $\|W_{\rho_{n}}\|_{\mathcal{B}(\mathcal{H}_{\tau})}=\|\varrho_{n}^{-1}\|_{\mathcal{B}(\mathcal{H}_{\tau})}\to\infty$; in particular $(W_{\rho_{n}})$ does not converge in operator norm on $\mathcal H_{\tau}$;
\item on $\mathcal{H}_{\rho}\equiv\mathscr A \mathbf p$, $\|\,W_{\rho_{n}}|_{\mathcal{H}_{\rho}}-W_{\rho}|_{\mathcal{H}_{\rho}}\,\|_{\mathcal{B}(\mathcal{H}_{\rho})}\to 0$ and 
$\|\,\tilde{\Delta}_{\rho_{n}}|_{\mathcal{H}_{\rho}}-\tilde{\Delta}_{\rho}|_{\mathcal{H}_{\rho}}\,\|_{\mathcal{B}(\mathcal{H}_{\rho})}\to 0$.
\end{enumerate}
\end{enumerate}
\end{lemma}

\begin{proof}
Without loss of generality, we may identify $\mathscr{A}$ with a subalgebra of a suitably big matrix algebra because $\mathscr{A}$ is finite-dimensional (see  equation \eqref{eqn: isomorphism of finite-dimensional algebras}).
%%%%%%%%%%%%%%%%%%
The linear map
\begin{equation}
\Phi:\mathscr A\to\mathscr A^{*},\qquad \Phi(x)(a)=\tau(xa),
\end{equation}
is a linear isomorphism. 
%%%%%%%%%%%%%%%%%%%%%%%%%%%%%%%%
All norms on finite-dimensional vector spaces are equivalent, hence $\|\rho_{n}-\rho\|_{\mathscr A^{*}}\to 0$ \textit{iff} the corresponding density operators satisfy $\|\varrho_{n}-\varrho\|_{\mathscr A}\to 0$, and we shall freely use this equivalence.
%%%%%%%%%%%%%%%%%%%%%%%%%%%%%%%%%%%%%%%%%%%%%%%%%%%
\begin{enumerate}
\item By definition, $(L_{\rho_{n}}-L_{\rho})(\mathbf{a})=(\varrho_{n}-\varrho)\mathbf{a}$ for all $\mathbf{a}\in\mathscr A$, so that $\|L_{\rho_{n}}-L_{\rho}\|_{\mathcal{B}(\mathcal{H}_{\tau})}\propto\|\varrho_{n}-\varrho\|_{\mathscr{A}}\to 0$, and the same for $R_{\rho_{n}}$.

\item If $\rho$ is faithful, its associated density operator $\varrho$ is invertible.
%%%%%%%%%%%%%%%%%%%
The set of invertible elements of $\mathscr{A}$ is open in the norm topology, and the inversion map is continuous there.
%%%%%%%%%%%%%%%%%%%%%%%%%%%
Therefore, $\|\varrho_{n}-\varrho\|_{\mathscr{A}}\to 0$ implies $\|\varrho_{n}^{-1} -\varrho^{-1}\|_{\mathscr{A}}\to 0$.
%%%%%%%%%%%%%%%%%%%%%%%%%%%%%%%%%%%
Since $W_{\rho_{n}}$ is right multiplication by $\varrho_{n}^{-1}$ and $W_{\rho}$ is right multiplication by $\varrho^{-1}$, we obtain 
\begin{equation} 
\|W_{\rho_{n}}-W_{\rho}\|_{\mathcal{B}(\mathcal{H}_{\tau})}\propto\|\varrho_{n}^{-1}-\varrho^{-1}\|_{\mathscr{A}}\to 0.
\end{equation}
%%%%%%%%%%%%%%%%%%%%%%%%
Consequently, it holds
\begin{equation} 
\|\tilde{\Delta}_{\rho_{n}}-\tilde{\Delta}_{\rho}\|_{\mathcal{B}(\mathcal{H}_{\tau})}
=\|L_{\rho_{n}}W_{\rho_{n}}-L_{\rho}W_{\rho}\|_{\mathcal{B}(\mathcal{H}_{\tau})}
\le \|L_{\rho_{n}}-L_{\rho}\|\|W_{\rho_{n}}\|+\|L_{\rho}\|\,\|W_{\rho_{n}}-W_{\rho}\|\to 0
\end{equation}
since $\|W_{\rho_{n}}\|=\|\varrho_{n}^{-1}\|$ stays bounded near an invertible $\varrho$.
%%%%%%%%%%%%%%%%%%%%%%%%%%%%%%%

\item 
\begin{enumerate}
\item Since $\mathscr{A}$ may be realized as a subalgebra of a suitably big matrix algebra (see equation \eqref{eqn: isomorphism of finite-dimensional algebras}), and since the eigenvalue functions are continuous \cite[ch. III]{B1997}, the smallest eigenvalue $\lambda_{min}(\varrho_{n})$ tends to $0$ because $\|\varrho_{n}-\varrho\|_{\mathscr{A}}\to 0$ and $\|W_{\rho_{n}}\|\propto\|\varrho_{n}^{-1}\|\propto\lambda_{\min}(\varrho_{n})^{-1}\to\infty$.
%%%%%%%%%%%%%%%%%%%%%%%%%%%%%%%%%%%%%%%%%%%%%

\item By assumption, $[\varrho_n,\mathbf p]=0$ for $n$ sufficiently large. Therefore $R_{\rho_n}$ preserves $\mathscr A\mathbf p$, and so $R_{\rho_n}\!\restriction_{\mathscr A\mathbf p}$ may be regarded as an operator on $\mathcal H_\rho\equiv\mathscr A\mathbf p$.
%%%%%%%%%%%%%%%%%%%%
On the subspace $\mathcal{H}_{\rho}\equiv\mathscr A \mathbf p$, the operator $R_{\rho}$ is positive and invertible, with inverse given by right multiplication with the inverse of $\varrho$ on $ \mathscr{A}_{pp}=\mathbf{p}\mathscr{A}\mathbf{p}$, denoted by $\varrho^{+}$.
%%%%%%%%%%%%%%%%%%%%
Hence there exists $\varepsilon>0$ such that
\begin{equation}
R_\rho\!\restriction_{\mathcal{H}_{\rho}}\;\ge\;\varepsilon\,\mathrm{id}_{\mathcal{H}_{\rho}}.
\end{equation}
%%%%%%%%%%%%%%%%%%%%%%%%
Since $\|R_{\rho_n}- R_\rho\|_{\mathcal{B}(\mathcal{H}_{\tau})}\to 0$, for $n$ sufficiently large and $\xi_{\mathbf{a}}^{\tau}$ in $\mathcal{H}_{\rho}\equiv\mathscr{A}\mathbf p$ we also have
\begin{equation}
\begin{split}
\langle \xi_{\mathbf{a}}^{\tau},R_{\rho_{n}}\xi_{\mathbf{a}}^{\tau}\rangle_{\tau} &=\langle \xi_{\mathbf{a}}^{\tau},R_\rho \xi_{\mathbf{a}}^{\tau}\rangle_{\tau}+\langle \xi_{\mathbf{a}}^{\tau},(R_{\rho_{n}}-R_\rho)\xi_{\mathbf{a}}^{\tau}\rangle_{\tau}
\;\ge \\ & \\ 
\geq &\langle \xi_{\mathbf{a}}^{\tau},R_\rho \xi_{\mathbf{a}}^{\tau}\rangle_{\tau} - \mid\!\langle \xi_{\mathbf{a}}^{\tau},(R_{\rho_{n}}-R_\rho)\xi_{\mathbf{a}}^{\tau}\rangle_{\tau}
\!\mid\;\ge \\&\\
\geq&\;\varepsilon\|\xi_{\mathbf{a}}^{\tau}\|_{\mathcal{H}_{\tau}}^{2}-\|R_{\rho_{n}}-R_\rho\|_{\mathcal{B}(\mathcal{H}_{\tau})}\|\xi_{\mathbf{a}}^{\tau}\|_{\mathcal{H}_{\tau}}^{2}
\;\ge\;\tfrac{\varepsilon}{2}\|\xi_{\mathbf{a}}^{\tau}\|_{\mathcal{H}_{\tau}}^{2},
\end{split}
\end{equation}
so that
\begin{equation}
R_{\rho_n}\!\restriction_{\mathscr A \mathbf p}\;\ge\;\tfrac{\varepsilon}{2}\,\mathrm{id}_{\mathscr A \mathbf p},
\end{equation}
and thus $R_{\rho_n}\!\restriction_{\mathcal{H}_{\rho}}$ is invertible with
\begin{equation}
\big\|\big(R_{\rho_n}\!\restriction_{\mathcal{H}_{\rho}}\big)^{-1}\big\|\;\le\;\tfrac{2}{\varepsilon}.
\end{equation}
Moreover, since $R_{\rho_n}$ preserves $\mathscr A\mathbf p$ and $\varrho_n$ is invertible, one has
\begin{equation}
\big(R_{\rho_n}\!\restriction_{\mathcal H_\rho}\big)^{-1}
=
W_{\rho_n}\!\restriction_{\mathcal H_\rho}.
\end{equation}
Likewise, it holds
\begin{equation}
\big(R_{\rho}\!\restriction_{\mathcal H_\rho}\big)^{-1}
=
W_{\rho}\!\restriction_{\mathcal H_\rho}.
\end{equation}
%%%%%%%%%%%%%%%%%%%%%%%%%%%%%
By continuity of the inversion map on the open set of invertible operators on $\mathcal{H}_{\rho}$, we conclude
\begin{equation}
\big\|W_{\rho_n}\!\restriction_{\mathcal{H}_{\rho}}-W_{\rho}\!\restriction_{\mathcal{H}_{\rho}}\big\|_{\mathcal{B}(\mathcal{H}_{\rho})}\;\longrightarrow\;0.
\end{equation}
Finally, since $\|L_{\rho_n}-L_\rho\|_{\mathcal B(\mathcal H_\tau)}\to 0$ and since $\big\|W_{\rho_n}\!\restriction_{\mathcal H_\rho}-W_{\rho}\!\restriction_{\mathcal H_\rho}\big\|_{\mathcal B(\mathcal H_\rho)}\to 0$, we have
\begin{equation}
\begin{split}
\left\|
\tilde{\Delta}_{\rho_{n}}\!\restriction_{\mathcal{H}_{\rho}}
-
\tilde{\Delta}_{\rho}\!\restriction_{\mathcal{H}_{\rho}}
\right\|_{\mathcal{B}(\mathcal{H}_{\rho})}
& =
\left\|
L_{\rho_n}W_{\rho_n}\!\restriction_{\mathcal{H}_{\rho}}
-
L_{\rho}W_{\rho}\!\restriction_{\mathcal{H}_{\rho}}
\right\|_{\mathcal{B}(\mathcal{H}_{\rho})}    \\
&\le
\left\|
L_{\rho_n}-L_\rho
\right\|_{\mathcal B(\mathcal H_\tau)}
\left\|
W_{\rho_n}\!\restriction_{\mathcal H_\rho}
\right\|_{\mathcal B(\mathcal H_\rho)} \\
&\quad +
\left\|
L_\rho
\right\|_{\mathcal B(\mathcal H_\tau)}
\left\|
W_{\rho_n}\!\restriction_{\mathcal H_\rho}
-
W_\rho\!\restriction_{\mathcal H_\rho}
\right\|_{\mathcal B(\mathcal H_\rho)}
\longrightarrow 0 .
\end{split}
\end{equation}
\end{enumerate}
\end{enumerate}
\end{proof}

\begin{remark}\label{rem: necessity of commuting condition}
The commutation assumption $[\varrho_n,\mathbf{p}]=0$ in Lemma
\ref{lem: convergence of modular operator} is essential. 
%%%%%%%%%%%%%%%
Indeed, let
$\mathscr{A}=M_2(\mathbb{C})$ and let $\tau$ be the normalized trace on
$M_2(\mathbb{C})$, namely
\begin{equation}
\tau(\mathbf{a})=\frac{1}{2}\mathrm{Tr}(\mathbf{a}).
\end{equation}
%%%%%%%%%%%%%%%%%%%%%
Let
\begin{equation}
\varrho=
\begin{pmatrix}
2&0\\
0&0
\end{pmatrix},
\qquad
\mathbf{p}=s(\varrho)=
\begin{pmatrix}
1&0\\
0&0
\end{pmatrix}.
\end{equation}
%%%%%%%%%%%%%%%%%%%%%%%%
Then $\rho\in\mathscr{S}(\mathscr{A})$ is the state defined by
\begin{equation}
\rho(\mathbf{a})=\tau(\varrho\mathbf{a}),
\qquad
\mathbf{a}\in M_2(\mathbb{C}).
\end{equation}
%%%%%%%%%%%%%%%%%%%%%%%%%%%%%%%%
For $0<\varepsilon<3/4$, define
\begin{equation}
\varrho_\varepsilon=
\begin{pmatrix}
2-2\varepsilon & \sqrt{\varepsilon}\\[1mm]
\sqrt{\varepsilon} & 2\varepsilon
\end{pmatrix}.
\end{equation}
%%%%%%%%%%%%%%%%%%%%%%%%%%%%%%%%%
Then $\varrho_\varepsilon$ is strictly positive,
\begin{equation}
\tau(\varrho_\varepsilon)=1,
\end{equation}
and $\varrho_\varepsilon\to\varrho$ as $\varepsilon\to 0$. 
%%%%%%%%%%%%%%%%%%%%%
Let
$\rho_\varepsilon\in\mathscr{S}(\mathscr{A})$ be the faithful state defined by
\begin{equation}
\rho_\varepsilon(\mathbf{a})=\tau(\varrho_\varepsilon\mathbf{a}),
\qquad
\mathbf{a}\in M_2(\mathbb{C}).
\end{equation}
However, it is
\begin{equation}
[\varrho_\varepsilon,\mathbf{p}]\neq 0
\end{equation}
for every $\varepsilon>0$.
%%%%%%%%%%%%%%%%%%%%%%%%%%%%%%%%%%%%%%%%%%%
Let $\mathbf{e}_{ij}$ denote the standard matrix units and set
$\mathbf{x}=\mathbf{e}_{21}\in\mathscr{A}\mathbf{p}$. Since
$\rho_\varepsilon$ is faithful, the corresponding modular operator on
$\mathcal{H}_{\tau}$ is given by
\begin{equation}
\widetilde{\Delta}_{\rho_\varepsilon}(\mathbf{x})
=
\varrho_\varepsilon\mathbf{x}\varrho_\varepsilon^{-1}.
\end{equation}
%%%%%%%%%%%%%%%%%%%%%%%%%%%%%%%%%%%%%
A direct computation gives
\begin{equation}
\varrho_\varepsilon \mathbf{e}_{21}\varrho_\varepsilon^{-1}
=
\begin{pmatrix}
\frac{\sqrt{\varepsilon}}{2(3/4-\varepsilon)}
&
-\frac{1}{4(3/4-\varepsilon)}
\\[2mm]
\frac{\varepsilon}{3/4-\varepsilon}
&
-\frac{\sqrt{\varepsilon}}{2(3/4-\varepsilon)}
\end{pmatrix}.
\end{equation}
%%%%%%%%%%%%%%%%%%
Therefore, it is
\begin{equation}
\lim_{\varepsilon\to 0}
\varrho_\varepsilon \mathbf{e}_{21}\varrho_\varepsilon^{-1}
=
-\frac{1}{3}\mathbf{e}_{12}.
\end{equation}
%%%%%%%%%%%%%%%%%%%%%%%%%%%%%%%%
On the other hand, the  modular operator associated with the limiting
non-faithful state satisfies
\begin{equation}
\widetilde{\Delta}_{\rho}(\mathbf{e}_{21})
=
\varrho\,\mathbf{e}_{21}\,\varrho^+
=
0,
\end{equation}
where
\begin{equation}
\varrho^+=
\begin{pmatrix}
\frac{1}{2}&0\\
0&0
\end{pmatrix}
\end{equation}
is the inverse of $\varrho$ on $\mathbf{p}\mathscr{A}\mathbf{p}$, extended by
zero on $(\mathbb{I}-\mathbf{p})\mathscr{A}(\mathbb{I}-\mathbf{p})$.
%%%%%%%%%%%%%%%%%%%%%%%%%%%
Consequently, we have
\begin{equation}
\widetilde{\Delta}_{\rho_\varepsilon}\!\restriction_{\mathscr{A}\mathbf{p}}
\not\longrightarrow
\widetilde{\Delta}_{\rho}\!\restriction_{\mathscr{A}\mathbf{p}}
\end{equation}
in operator norm. 
%%%%%%%%%%%%%%%%%%%%%%%%%%%%%
This shows that convergence of faithful states to a non-faithful state is not sufficient to guarantee convergence of the
 modular operators on the limiting GNS space. 
 %%%%%%%%%%%%%%%%%%%%%
 The additional condition
$[\varrho_n,\mathbf{p}]=0$, or equivalently preservation of the support
decomposition of the limiting state, excludes precisely this phenomenon.
\end{remark}

\subsection{Completely-positive unital maps}

In both the classical and quantum case, a crucial role is played by how the statistical covariance behaves under the relevant transformations of the theory: the  classical Markov kernels, and the quantum channels, respectively.
%%%%%%%%%%%%%%%%%%%%%%
In the operator algebraic context, both these types of transformations are recovered as dual maps of the so-called \textit{completely positive unital (CPU)} maps.
%%%%%%%%%%%%%%%%%
Given two (possibly infinite-dimensional) $C^{*}$-algebras $\mathscr{A}$ and $\mathscr{B}$, a \textit{positive} map between them is a bounded, linear map $\Phi\colon\mathscr{B}\rightarrow\mathscr{A}$ such that $\Phi(\mathbf{bb}^{\dagger})$ is a positive element in $\mathscr{A}$ for all $\mathbf{b}\in\mathscr{B}$.
%%%%%%%%%%%%%%%%%%%%%%%%
Positive maps are automatically $*$-preserving in the sense that $\Phi(\mathbf{b}^{\dagger})=\Phi(\mathbf{b})^{\dagger}$ for all $\mathbf{b}\in\mathscr{B}$.
%%%%%%%%%%%%%%%%%%%%%%%
A completely positive \textit{CP} map is a positive map $\Phi\colon\mathscr{B}\rightarrow\mathscr{A}$ such that $\Phi\otimes\mathrm{id}_{n}\colon\mathscr{B}\otimes\mathbb{M}_{n}(\mathbb{C})\rightarrow\mathscr{A}\otimes\mathbb{M}_{n}(\mathbb{C})$ is a positive map for all $n\in\mathbb{N}$, where $\mathbb{M}_{n}(\mathbb{C})$ is the $C^{*}$-algebra of complex-valued square matrices, $\mathrm{id}_{n}$ is the identity map on $\mathbb{M}_{n}(\mathbb{C})$, and $\otimes$ denotes the tensor product between $C^{*}$-algebras and their bounded linear maps\footnote{Note that the tensor product between two $C^{*}$-algebras is uniquely defined when at least one of them is finite-dimensional as in this case.}.
%%%%%%%%%%%%%%%%%%%%%%
If $\mathscr{A}$ and $\mathscr{B}$ have identity elements denoted as  $\mathbb{I}_{\mathscr{A}}$ and $\mathbb{I}_{\mathscr{B}}$, respectively, a \textit{CP} map is unital (\textit{CPU}) if $\Phi(\mathbb{I}_{\mathscr{B}})=\mathbb{I}_{\mathscr{A}}$.
%%%%%%%%%%%%%%%%%%
If $\mathscr{A}$ is the $C^{*}$-algebra of continuous functions on a compact Hausdorff space, then every positive map from $\mathscr{A}$ to  $\mathscr{B}$ is automatically completely-positive \cite[prop. IX.4.1]{D1996}.
%%%%%%%%%%%%%%%%%%%

The \textup{GNS} Hilbert spaces behave well under the action of \textit{CPU} maps.
%%%%%%%%%%%%%%%%%%%%%%%%%%
Every \textit{CPU} map $\Phi\colon\mathscr{B}\rightarrow\mathscr{A}$ enjoys  Kadison's inequality \cite[proposition II.6.9.14]{B2006a}
\begin{equation}\label{eqn: kadison inequality}
	 \Phi(\mathbf{a}^{\dagger}\mathbf{a})\geq \Phi(\mathbf{a})^{\dagger}\Phi(\mathbf{a})
\end{equation} 
which implies 
\begin{equation}\label{eqn: schwarz inequality for GNS}
	\rho(\Phi(\mathbf{a}^{\dagger}\mathbf{a}))\geq \rho(\Phi(\mathbf{a})^{\dagger}\Phi(\mathbf{a}))	
\end{equation} 
for every state $\rho$ on $\mathscr{A}$.
%%%%%%%%%%%%%%%%%
Now, let $\mathscr{A},\mathscr{B}$ be finite-dimensional, and let $\Phi\colon \mathscr{B}\rightarrow\mathscr{A}$ be a CPU map such that $\Phi^{*}(\rho)=\sigma$, where $\rho$ and $\sigma$ are normal states on $\mathscr{A}$ and $\mathscr{B}$, respectively.
%%%%%%%%%%%%%%%%%
Since $\Phi$ is unital and satisfies  equation \eqref{eqn: schwarz inequality for GNS}, taking $\mathbf{b}\in\mathscr{B}_{\mathbf{q}_{\sigma}\mathbf{p}_{\sigma}}\oplus\mathscr{B}_{\mathbf{q}_{\sigma}\mathbf{q}_{\sigma}}$ implies
\begin{equation}\label{eqn: CP contraction preserves left-right ideal}
0=\sigma(\mathbf{b}\,\mathbf{b}^{\dagger})\geq \rho\left(  \Phi(\mathbf{b})\,\Phi(\mathbf{b})^{\dagger}\right)\geq 0 ,
\end{equation}
which means   that $\Phi(\mathscr{B}_{\mathbf{q}_{\sigma}\mathbf{p}_{\sigma}}\oplus\mathscr{B}_{\mathbf{q}_{\sigma}\mathbf{q}_{\sigma}})\subseteq \mathscr{A}_{\mathbf{q}_{\rho}\mathbf{p}_{\rho}}\oplus\mathscr{A}_{\mathbf{q}_{\rho}\mathbf{q}_{\rho}}$.
%%%%%%%%%%%%%%%%%
Similarly, taking $\mathbf{b}\in\mathscr{B}_{\mathbf{p}_{\sigma}\mathbf{q}_{\sigma}}\oplus\mathscr{B}_{\mathbf{q}_{\sigma}\mathbf{q}_{\sigma}}\equiv\mathscr{N}_{\sigma}$, equation \eqref{eqn: schwarz inequality for GNS} implies 
\begin{equation}\label{eqn: CP contraction preserves left-right ideal 0}
	0=\sigma(\mathbf{b}^{\dagger}\mathbf{b})\geq \rho\left(  \Phi(\mathbf{b})^{\dagger}\,\Phi(\mathbf{b})\right) \geq 0,
\end{equation}
and thus $\Phi(\mathscr{N}_{\sigma})\subseteq \mathscr{N}_{\rho}$. 
%%%%%%%%%%%%%%%%%%%%%
In particular, $\Phi$ determines the linear map $\tilde{\Phi}\colon\mathscr{B}/\mathscr{N}_{\sigma}\cong\mathcal{H}_{\sigma}\rightarrow\mathscr{A}/\mathscr{N}_{\rho}\cong\mathcal{H}_{\rho}$ given by
\begin{equation}\label{eqn: from CPU to linear contractions} 
\tilde{\Phi}(\xi_{\mathbf{b}}^{\sigma}):=\xi_{\Phi(\mathbf{b})}^{\rho}.
\end{equation}
%%%%%%%%%%%%%%%%%%%%%%%%%%%%%%%
Here again, $\xi_{\mathbf{b}}^{\sigma}$ is the \textup{GNS} vector represented by $\mathbf{b}$ at the state $\sigma$, while $\xi_{\Phi(\mathbf{b})}^{\rho}$ is the \textup{GNS} vector represented by the transformed observable $\Phi(\mathbf{b})$ at the state $\rho$.
%%%%%%%%%%%%%%%%%%%%
which is well defined because of equation \eqref{eqn: CP contraction preserves left-right ideal} and turns out to be a contraction\footnote{In the infinite-dimensional case, $\Phi$ determines a densely-defined map through equation \eqref{eqn: from CPU to linear contractions}  that is a contractive and thus can be extended to a bounded contraction $\tilde{\Phi}\colon\mathcal{H}_{\sigma}\rightarrow\mathcal{H}_{\rho}$, with an evident abuse of notation} in the sense that
\begin{equation} 
\langle\tilde{\Phi}(\xi_{\mathbf{b}})\mid\tilde{\Phi}(\xi_{\mathbf{b}})\rangle_{\rho}\leq\langle \xi_{\mathbf{b}} \mid \xi_{\mathbf{b}}\rangle_{\sigma}	
\end{equation}
for all $\mathbf{b}\in\mathscr{B}$.
%%%%%%%%%%%%%%%%
Finally, we prove an important inequality involving the contraction $\tilde{\Phi}$ and the modular operators $\Delta_{\rho}$ and $\Delta_{\sigma}$.
%%%%%%%%%%%%%%%%%%%%

\begin{lemma}\label{lem: monotonicity of modular operator}
Let $\Phi\colon \mathscr{B}\rightarrow\mathscr{A}$ be a \textit{CPU} map such that $\Phi^{*}(\rho)=\sigma$, where $\Phi^{*}\colon \mathscr{A}^{*}\rightarrow\mathscr{B}^{*}$ is the Banach dual of $\Phi$, and $\rho$ and $\sigma$ are states on the finite-dimensional $C^{*}$-algebras $\mathscr{A}$ and $\mathscr{B}$, respectively.
%%%%%%%%%%%%%%%%%%%%%%
Let $\Delta_{\rho}$ and $\Delta_{\sigma}$ be the modular operators of $\rho$ and $\sigma$, respectively, defined according to equation \eqref{eqn: modular operator}, and let $\tilde{\Phi}$ as in equation \eqref{eqn: from CPU to linear contractions}.
%%%%%%%%%%%%%%%%%%%%%%
We denote by $\tilde{\Phi}^{*}$ the Hilbert-space adjoint of the contraction $\tilde{\Phi}\colon\mathcal{H}_{\sigma}\to\mathcal{H}_{\rho}$.
%%%%
It follows that
\begin{equation}\label{eqn: monotonicity of modular operator FINAL}
\tilde{\Phi}^{*} \,\Delta_{\rho}\,\tilde{\Phi} \leq    \Delta_{\sigma}.
\end{equation} 
%%%%%%%%%%%%%%%%%%%%
\end{lemma}

\begin{proof}
First, note that
\begin{equation}\label{eqn: monotonicity of modular operator}
\begin{split}
\langle \tilde{\Phi}(\xi_{\mathbf{b}}^{\sigma})  | \Delta_{\rho}\tilde{\Phi}(\xi_{\mathbf{b}}^{\sigma}) \rangle_{\rho} & \stackrel{\mbox{\eqref{eqn: from CPU to linear contractions}}}{=} \langle \xi_{\Phi(\mathbf{b})}^{\rho}  | \Delta_{\rho}(\xi_{\Phi(\mathbf{b})}^{\rho}) \rangle_{\rho} \stackrel{\mbox{\eqref{eqn: modular operator}}}{=}\rho \left( \mathbf{p}_{\rho}\Phi(\mathbf{b})\mathbf{p}_{\rho}\,\mathbf{p}_{\rho}\Phi(\mathbf{b})^{\dagger}\mathbf{p}_{\rho} \right)= \\
&\stackrel{\mbox{\eqref{eqn: CP contraction preserves left-right ideal}}}{=}\rho \left( \mathbf{p}_{\rho}\Phi(\mathbf{b}_{0})\mathbf{p}_{\rho}\,\mathbf{p}_{\rho}\Phi(\mathbf{b}_{0})^{\dagger}\mathbf{p}_{\rho} \right),
\end{split}
\end{equation}
where $\mathbf{b}_{0}=\mathbf{p}_{\sigma}\mathbf{b}\mathbf{p}_{\sigma}$.
%%%%%%%%%%%%%%%%%%%%%%
Then, note that 
\begin{equation} 
\mathbf{b}\mapsto \Psi(\mathbf{b}):=\mathbf{p}_{\rho}\Phi(\mathbf{b})\mathbf{p}_{\rho}	
\end{equation}
is a completely-positive contraction, so that  
\begin{equation}\label{eqn: monotonicity of modular operator 2}
\begin{split}
\langle \tilde{\Phi}(\xi_{\mathbf{b}}^{\sigma})  | \Delta_{\rho}\tilde{\Phi}(\xi_{\mathbf{b}}^{\sigma}) \rangle_{\rho} &\stackrel{\mbox{\eqref{eqn: monotonicity of modular operator} }}{=}\rho \left(  \Psi(\mathbf{b}_{0})\, \Psi(\mathbf{b}_{0})^{\dagger}  \right)\stackrel{\mbox{\eqref{eqn: kadison inequality} }}{\leq} \rho\left(\Psi(\mathbf{b}_{0}\mathbf{b}_{0}^\dagger)\right)= \rho\left( \mathbf{p}_{\rho}\Phi(\mathbf{b}_{0}\,\mathbf{b}_{0}^{\dagger})\,\mathbf{p}_{\rho} \right) \\
& \stackrel{\mbox{\eqref{eqn: support projection of normal state}}}{=} \rho\left( \Phi(\mathbf{b}_{0}\,\mathbf{b}_{0}^{\dagger}) \right)  \stackrel{\sigma=\Phi^{*}\rho}{=}\sigma(\mathbf{b_{0}b_{0}}^{\dagger}) \stackrel{\mbox{\eqref{eqn: modular operator}}}{=} \langle  \xi^{\sigma}_{\mathbf{b} }| \Delta_{\sigma}(\xi^{\sigma}_{\mathbf{b}}) \rangle_{\sigma},
\end{split}
\end{equation}
which is equivalent to \eqref{eqn: monotonicity of modular operator FINAL}.
%%%%%%%%%%%%%%%%%%%%%

\end{proof}

\subsection{The category of non-commutative probability spaces}\label{subsec: NCP}

Given a category $\mathsf{C}$, the classes of its objects and morphisms are denoted as $\mathsf{C}_{0}$ and $\mathsf{C}_{1}$, respectively.
%%%%%%%%%%%%%%%%%%%%%%%%%%%%
Then, if $\mathsf{D}$ is another category, a functor $\mathfrak{F}$ from $\mathsf{C}$ to $\mathsf{D}$ will be denoted by a squiggly arrow $\mathfrak{F}\colon\mathsf{C}\rightsquigarrow\mathsf{D}.$
%%%%%%%%%%%%%%%%%%%%%%%%%%%%%%
Moreover,the action of $\mathfrak{F}$ on $c\in\mathsf{C}$ is denoted as $\mathfrak{F}_{0}(c)$, while its action on $f\in\mathsf{C}_{1}$ is denoted by $\mathfrak{F}_{1}(f)$.
%%%%%%%%%%%%%%%%%%%%%%%%%%
Finally, for all the mathematical details on  category theory that are used but not discussed in this work, we refer to \cite{LS1997,P2024,R2016}.
%%%%%%%%%%%%%%%%%%%%%%%%%%%%%%%%

\begin{definition}[The category $\mathsf{NCP}$ of non-commutative probability spaces\footnote{The fact that $\mathsf{NCP}$ as described in definition \ref{defn: NCP} is indeed a category amounts to a routine check.}]\label{defn: NCP}
An object in $\mathsf{NCP}$ is a couple $(\mathscr{A},\rho)$ with $\mathscr{A}$ a $C^{*}$-algebra and $\rho$ a state on $\mathscr{A}$, while a morphism  $\hat{\Phi}$ between $(\mathscr{A},\rho)$ and $(\mathscr{B},\sigma)$ is associated with a \textit{completely-positive and unital} (CPU) map $\Phi\colon\mathscr{B}\rightarrow\mathscr{A}$ such that $\Phi^{*}(\rho)=\sigma$, where $\Phi^{*}\colon \mathscr{A}^{*}\rightarrow\mathscr{B}^{*}$ is the Banach dual of $\Phi$.
%%%%%%%%%%%%%%%%%%%%%%%%%
When $\mathscr{A}$ is constrained to be finite-dimensional, we obtain the  full subcategory $\mathsf{fNCP}$ of finite-dimensional non-commutative probability spaces.
%%%%%%%%%%%%%%%%%%%%%%%%%%%
When $\mathscr{A}$ is constrained to be finite-dimensional and $\rho$ to be a tracial state, we obtain the full subcategory $\mathsf{fNCT}$, which is also a full subcategory of $\mathsf{fNCP}$.
%%%%%%%%%%%%%%%%%%%%%%%%
\end{definition}

Beside the authors' works \cite{CDGIM2023,CDG2025}, the category $\mathsf{NCP}$ and some of its close relatives have already appeared in the literature.
%%%%%%%%%%%%%%%%%%%%%%%%%%%%%
In \cite{BF2014}, the category $\mathsf{FinStat}$ is introduced in order to give a functorial characterization of the Kullback-Leibler relative entropy.
%%%%%%%%%%%%%%%%%%%
The category $\mathsf{FinStat}$ is a sort of variation, at the level of morphisms, of the subcategory $\mathsf{fCP}$ (of $\mathsf{NCP}$) of finite-dimensional classical probability spaces  whose objects are states on finite-dimensional commutative $C^{*}$-algebras (or, equivalently, probability measures on discrete and finite outcome spaces).
%%%%%%%%%%%%%%%%%%%%%%%%%%%%%%
In \cite{P2021a}, the category $\mathsf{NCFinStat}$ is introduced in order to give a functorial characterization of the von Neumann--Umegaki relative entropy.
%%%%%%%%%%%%%%%%%
The category $\mathsf{NCFinStat}$ is a non-commutative version of $\mathsf{FinStat}$, and, in analogy with $\mathsf{FinStat}$, is a sort of variation, at the level of morphisms, of the subcategory $\mathsf{fNCP}\subset\mathsf{NCP}$ of finite-dimensional non-commutative probability spaces.
%%%%%%%%%%%%%%%%%%%
In \cite[definition 3.22 and corollary 3.23]{PR2023}, the opposite category to $\mathsf{NCP}$  is introduced (as well as some of its variations where state-preserving CPU maps are replaced by \textit{a.e.} equivalence classes of 2-positives and positive unital maps), and it is used to investigate the appropriate non-commutative version of classical disintegrations in this categorical context.
%%%%%%%%%%%%%%%%%%%%%%%%%%
In \cite{P1984}, a category whose objects are couples $(\mathscr{V},\varphi)$, where $\mathscr{V}$ is a von Neumann algebra and $\varphi$ is a faithful weight on $\mathscr{V}$, and whose morphisms are associated to suitable normal positive contractions between von Neumann algebras is introduced and used to investigate weight-adapted conditional expectations in a categorical setting.
%%%%%%%%%%%%%%%%%%%%%%
\\

Of particular interest for this work are the so-called \textit{split monomorphisms} in $\mathsf{NCP}$, which provide a possibly non-commutative version of what Čencov originally called \textit{congruent embeddings} in his categorical approach to statistics and decision theory \cite{C1965,C1981a}.
%%%%%%%%%%%%%%%%%%%%%%%%%%%%%%%%%%%%%%%%%%%%%%%%%%%%%%%%%%%%%%%

\begin{definition}[Split monomorphisms]\label{defn: split monomorphisms} 
Let $\mathsf{C}$ be a category.
%%%%%%%%%%%%%%%%%%%%%%%%%%
A \textbf{split monomorphism} in $\mathsf{C}$ is a morphism $f\colon A\rightarrow B$ in $\mathsf{C}$ admitting a left inverse $g\colon B\rightarrow A$ again in $\mathsf{C}$, that is, $g\circ f=\mathrm{id}_{A}$.
%%%%%%%%%%%%%%%%%%%%%%%%
The  morphism $f$ is referred to as a \textbf{section} of $g$, while $g$ is referred to as a \textbf{retraction} of $f$.
%%%%%%%%%%%%%%%%%%%%%%
The object $A$ is referred to as a \textup{retract} of $B$.
\end{definition}

An important example of \textbf{split monomorphism} in $\mathsf{NCP}$ is determined by the so-called \textbf{conditional expectations}.
%%%%%%%%%%%%%%%%%%%%%5
In section \ref{sec: classification}, split monomorphisms will be the ``symmetry transformations'' in $\mathsf{NCP}$ that are used to classify the so-called \textit{fields of covariances}  introduced in section \ref{defn: field of covariances}.
%%%%%%%%%%%%%%%%%%%%%%%%%%%%%%%%%
Recall that, given a $C^{*}$-algebra $\mathscr{A}$ and a $C^{*}$-subalgebra $\mathscr{M}\subseteq \mathscr{A}$, a conditional expectation of $\mathscr{A}$ onto $\mathscr{M}$ is a bounded linear projection $\mathcal{E}\colon\mathscr{A}\rightarrow \mathscr{M}$ of norm $1$.
%%%%%%%%%%%%%%%%%%%%%%
Equivalently \cite[thm. II.6.10.2]{B2006a}, a conditional expectation is a completely-positive contraction  $\mathcal{E}\colon\mathscr{A}\rightarrow\mathscr{M}$ such that $\mathcal{E}(\mathbf{m})=\mathbf{m}$ for all $\mathbf{m}\in\mathscr{M}\subseteq \mathscr{A}$ and such that
\begin{equation}\label{eqn: bimodularity of conditional expectations} 
\mathcal{E}(\mathbf{ma})=\mathbf{m}\mathcal{E}(\mathbf{a}),\qquad \mathcal{E}(\mathbf{am})=\mathcal{E}(\mathbf{a})\mathbf{m}
\end{equation}
for all $\mathbf{a}\in\mathscr{A}$ and all $\mathbf{m}\in\mathscr{M}$.
%%%%%%%%%%%%%%%%%%%%%%%%%

Let $\mathcal{E}\colon \mathscr{A}\rightarrow\mathscr{M}$ be a conditional expectation, and let  $\iota\colon\mathscr{M}\hookrightarrow\mathscr{A}$ be the natural inclusion map, which is a *-homomorphism.
%%%%%%%%%%%%%%%%%%%%%%%%%%
Let $\sigma\in\mathcal{S}(\mathscr{A})$ and $\rho\in\mathcal{S}(\mathscr{M})$ be such that $\mathcal{E}^{*}\rho=\sigma$ and $\iota^{*}\sigma=\rho$.
%%%%%%%%%%%%%%%%%%%%%%%%%
Since $\mathcal{E}\circ\iota=\mathrm{id}_{\mathscr{M}}$, the CPU maps $\mathcal{E}$ and $\iota$ determine morphisms  in $\mathsf{NCP}$ that we denote by $\widehat{\mathcal{E}}\colon(\mathscr{M},\rho)\rightarrow(\mathscr{A},\sigma)$ and $\widehat{\iota}\colon(\mathscr{A},\sigma)\rightarrow(\mathscr{M},\rho)$, respectively.
With this notation, $\widehat{\iota}\circ\widehat{\mathcal{E}}=\mathrm{id}_{(\mathscr{M},\rho)}$, so that $\widehat{\mathcal{E}}$ is a split monomorphism and  $(\mathscr{M},\rho)$ is a retract of $(\mathscr{A},\sigma)$.
If $\mathscr{B}$ is a $C^*$-algebra which is *-isomorphic to $\mathscr{M}$ through $\phi$, the CPU map $\phi\circ\mathcal{E}$ determines an analogous split monomorphism from $(\mathscr{B},\omega)$ to $(\mathscr{A},\sigma)$.
%%%%%%%%%%%%%%%%%%%%

\section{Fields of covariances}\label{sec: fields of covariances}

The observation that is at the heart of this work is that the assignment $\rho\mapsto\mathcal{H}_{\rho}$ may be seen as a contravariant functor $\mathfrak{G}$ from  $\mathsf{NCP}$  to the category $\mathsf{Hilb}$ of complex Hilbert spaces and bounded linear contractions by defining\footnote{The fact that $\mathfrak{G}$ is actually a functor follows from the associativity of the composition of linear maps.} 
\begin{equation}\label{eqn: gns functor}
\begin{split}
\mathfrak{G}&\colon \mathsf{NCP}\rightsquigarrow\mathsf{Hilb} \\
\mathfrak{G}_{0}(\mathscr{A},\rho)&=\mathcal{H}_{\rho}, \quad \mathfrak{G}_{1}(\hat{\Phi})=\tilde{\Phi},
\end{split}	
\end{equation}
where $\tilde{\Phi}$ is as in equation \eqref{eqn: from CPU to linear contractions}.
%%%%%%%%%%%%%%%%%%%%%
We refer to $\mathfrak{G}$ as the \textit{GNS}-functor.
%%%%%%%%%%%%%%%%%%%%%%%%%%%%%%%%%%
Note that the functoriality of $\mathfrak{G}$ entails the monotonicity property
\begin{equation}\label{eqn: monotonicity property GNS functor}
\langle \tilde{\Phi}(\psi)\mid\tilde{\Phi}(\psi)\rangle_{\rho}\leq \langle\psi\mid\psi\rangle_{\sigma}
\end{equation}
because $\tilde{\Phi}$ is a contraction, and equation  \eqref{eqn: monotonicity property GNS functor} coincides with the invariance property of classical statistical covariance in equation \eqref{eqn: invariance of statistical covariance under *-homomorphisms} when $\mathscr{A}=\mathcal{L}^{\infty}(\Omega,\nu)$ and $\Phi$ is invertible, and with  the monotonicity property of the quantum covariance in equation \eqref{eqn: monotonicity of quantum covariance} when $\mathscr{A}=\mathcal{B}(\mathcal{H})$ and its associated operator monotone function \cite{GHP2009} is the constant function $f(t)=1$.
%%%%%%%%%%%%%%%%%%%%%%%%%%%%%%%%
A natural question is then whether all the quantum covariances discussed in \cite{GHP2009} may arise as suitable deformations of the \textit{GNS}-functor given above.
%%%%%%%%%%%%%%%%%%%%%
The \textit{GNS}-functor constructed above fixes the underlying transformation law of non-commutative random variables because a morphism $\hat{\Phi}\colon(\mathscr{A},\rho)\longrightarrow(\mathscr{B},\sigma)$ in $\mathsf{NCP}$ acts on the algebraic \textit{GNS} cores $\mathscr{B}/\mathscr{N}_{\sigma}$ through the linear map defined in equation \eqref{eqn: from CPU to linear contractions}.
%%%%%%%%%%%%%%%%
The quantum covariances appearing in \cite{GHP2009} do not arise by changing this transformation law. 
%%%%%%%%%%%%%%
Rather, they arise by changing the Hilbertian structure on each \textit{GNS} core, replacing the \textit{GNS} inner product with a new positive sesquilinear form.
%%%%%%%%%%%%%%%
Therefore, in order to capture such covariances functorially, we keep the canonical maps $\tilde{\Phi}$ on morphisms, but impose the additional requirement that they be contractive with respect to the new covariance forms. 
%%%%%%%%%%%%%%%%%%%% 
This requirement is not automatic from the \textit{GNS}-contractivity in equation \eqref{eqn: monotonicity property GNS functor}; it is precisely the condition ensuring that the maps $\tilde{\Phi}$ extend continuously to contractions between the corresponding covariance Hilbert spaces. 
%%%%%%%%%%%%%%%%%%%% 
This leads to the following definition.

\begin{definition}[Field of covariances]\label{defn: field of covariances}
Let $\mathsf{D}$ be a (not necessarily proper) subcategory of $\mathsf{NCP}$.
%%%%%%%%%%%%%%%%%%%%%%%%%%%%%%%%%%%%%%%%%%%
A \emph{field of covariances} (or \emph{covariance field}) on $\mathsf{D}$ is a contravariant functor
\begin{equation} 
\mathfrak{C}\colon \mathsf{D}\;\rightsquigarrow\;\mathsf{Hilb}
\end{equation}
satisfying the following properties.
%%%%%%%%%%%%%%%%%%%%%%%%%%%%%%%%%%%%%%%%%%%
For every object $(\mathscr{A},\rho)\in \mathsf{D}_{0}$, the Hilbert space
$\mathfrak{C}_{0}(\mathscr{A},\rho)\equiv\mathcal{H}^{\mathfrak{C}}_{\rho}$ is the completion of
$\mathscr{A}/\mathscr{N}_{\rho}$ with respect to a Hermitian positive definite sesquilinear form
\begin{equation}\label{eqn: covariance form}
\mathfrak{C}_{\rho}\colon
(\mathscr{A}/\mathscr{N}_{\rho})\times(\mathscr{A}/\mathscr{N}_{\rho})
\longrightarrow
\mathbb{C}.
\end{equation}
%%%%%%%%%%%%%%%5
Equivalently, the inner product of $\mathcal{H}^{\mathfrak{C}}_{\rho}$ restricts to
\begin{equation}
\langle \xi_{\mathbf{a}},\xi_{\mathbf{b}}\rangle_{\mathcal{H}^{\mathfrak{C}}_{\rho}}
=
\mathfrak{C}_{\rho}(\xi_{\mathbf{a}},\xi_{\mathbf{b}})
\end{equation}
on $\mathscr{A}/\mathscr{N}_{\rho}$.
%%%%%%%%%%%%%%%%%%%%%%%%%%%%%%%%%%%%%%%%%%%
For every morphism $\hat{\Phi}\colon(\mathscr{A},\rho)\rightarrow(\mathscr{B},\sigma)$ in $\mathsf{D}$, the induced linear map $\tilde{\Phi}\colon
\mathscr{B}/\mathscr{N}_{\sigma} \rightarrow \mathscr{A}/\mathscr{N}_{\rho}$ defined in equation \eqref{eqn: from CPU to linear contractions} is required to be contractive with respect to the covariance forms, namely
\begin{equation}\label{eqn: categorical monotonicity 3}
\mathfrak{C}_{\rho}\big(\tilde{\Phi}(\xi),\tilde{\Phi}(\xi)\big)
\leq
\mathfrak{C}_{\sigma}(\xi,\xi)
\end{equation}
for every $\xi\in\mathscr{B}/\mathscr{N}_{\sigma}$.
%%%%%%%%%%%%%%%%%%%%%%%%%%%%%%%%%%%%%%%%%%%
This condition depends on the chosen covariance forms
$\mathfrak{C}_{\rho}$ and $\mathfrak{C}_{\sigma}$, and is precisely the condition ensuring that
$\tilde{\Phi}$ admits a unique continuous extension as a contraction from $\mathcal{H}^{\mathfrak{C}}_{\sigma}$ to $\mathcal{H}^{\mathfrak{C}}_{\rho}$, denoted again with $\tilde{\Phi}$. 
%%%%%%%%%%%%%%%%
The action of the functor $\mathfrak{C}$ on
morphisms is required to be this extension:
\begin{equation}\label{eqn: covariance on morphisms}
\mathfrak{C}_{1}(\hat{\Phi})
=\tilde{\Phi}.
\end{equation}
%%%%%%%%%%%%%%%%%%%%%%%%%%%%%%%%%%%%%%%%%%%
The sesquilinear form $\mathfrak{C}_{\rho}$ is called the \emph{covariance} at
$(\mathscr{A},\rho)$.
\end{definition}

Throughout the rest of the work, we will restrict our focus to the case when $\mathsf{D}=\mathsf{fNCP}$, the full subcategory of finite-dimensional non-commutative probability spaces.
%%%%%%%%%%%%%%%%
In this case, since $\mathscr{A}$ is finite-dimensional, the completion step is superfluous at the level of vector spaces. Thus, we identify both $\mathcal{H}^{\mathfrak{C}}_{\rho}$ and $\mathcal{H}_{\rho}$ with the vector space $\mathscr{A}/\mathscr{N}_{\rho}$, while keeping track of the fact that they are endowed, in general, with different inner products.
%%%%%%%%%%%%%%%%%%%%%%%%%%%%%%%%%
The covariance form $\mathfrak{C}_{\rho}$ is then represented, with respect to the \textup{GNS} inner product $\langle\cdot,\cdot\rangle_{\rho}$, by a unique positive invertible operator $\mathbf{T}_{\rho}\in\mathcal{B}(\mathcal{H}_{\rho})$ such that
\begin{equation}\label{eqn: covariance operator}
\mathfrak{C}_{\rho}(\xi_{\mathbf{a}},\xi_{\mathbf{b}})
=
\langle \xi_{\mathbf{a}},\,\mathbf{T}_{\rho}(\xi_{\mathbf{b}})\rangle_{\rho}.
\end{equation}
%%%%%%%%%%%%%%%
We call $\mathbf{T}_{\rho}$ the \emph{covariance operator} at $\rho$ associated with the covariance form $\mathfrak{C}_{\rho}$.

\begin{remark}
It is important to stress that the covariance form should be regarded as the primitive datum, while the covariance operator is a derived object\footnote{We thank an anonymous reviewer for pointing out the need to stress this.}.
%%%%%%%%%%%%%%
In finite dimensions, as argued above, every covariance form is represented, with respect to the \textup{GNS} inner product, by a unique bounded positive invertible operator.
%%%%%%%%%%%%%%%%
On the other hand, in the infinite-dimensional case, the existence of a covariance operator requires additional analytic assumptions. 
%%%%%%%%%%%%%%%%%%
If the covariance form $\mathfrak{C}_{\rho}$ is closed as a positive form on the \textup{GNS} Hilbert space $\mathcal{H}_{\rho}$, then Kato's representation theorem \cite[Chapter VI, Theorem 2.23]{K1995} associates with it a unique positive self-adjoint operator $\mathbf{T}_{\rho}$ such that
\begin{equation}
D(\mathfrak{C}_{\rho})=D(\mathbf{T}_{\rho}^{1/2}),
\qquad
\mathfrak{C}_{\rho}(\xi,\eta)
=
\langle \mathbf{T}_{\rho}^{1/2}\xi,\mathbf{T}_{\rho}^{1/2}\eta\rangle_{\rho}.
\end{equation}
%%%%%%%%%%%%%%%%%%%
More generally, if the covariance form is merely closable, one first replaces it by its closure $\overline{\mathfrak{C}}_{\rho}$, and Kato's theorem gives a unique positive self-adjoint operator $\mathbf{T}_{\rho}$ such that
\begin{equation}
D(\overline{\mathfrak{C}}_{\rho})=D(\mathbf{T}_{\rho}^{1/2}),
\qquad
\overline{\mathfrak{C}}_{\rho}(\xi,\eta)
=
\langle \mathbf{T}_{\rho}^{1/2}\xi,\mathbf{T}_{\rho}^{1/2}\eta\rangle_{\rho}.
\end{equation}
%%%%%%%%%%%%%%%%%%%
In either case, the domain of $\mathbf{T}_{\rho}$ need not coincide with the original \textup{GNS} core $\mathscr{A}/\mathscr{N}_{\rho}$.
%%%%%%%%%%%%%%%%%%%%%%%%
\end{remark}

%%%%%%%%%%%%%%%%%%%%%%%%%%%%%%%%%
The fact that $\mathfrak{C} \colon \mathsf{fNCP} \rightsquigarrow\mathsf{Hilb}$ is a functor implies the monotonicity property in equation \eqref{eqn: categorical monotonicity 3} for every morphism $\hat{\Phi}\in\mathsf{fNCP}_{1}$ between $(\mathscr{A},\rho)$ and $(\mathscr{B},\sigma)$,  and for all $\xi\in\mathcal{H}_{\sigma}$.
%%%%%%%%%%%%%%%%%%%%%%
Moreover, if  $(\mathscr{B},\sigma)\in\mathsf{NCP}_{0}$ is the \textit{retract} of $(\mathscr{A},\rho)\in\mathsf{NCP}_{0}$ through $\hat{\Phi}\in\mathsf{NCP}_{1}$ (see definition \ref{defn: split monomorphisms}), the monotonicity property in equation \eqref{eqn: categorical monotonicity 3}  becomes the invariance property
\begin{equation}\label{eqn: categorical invariance condition}
\mathfrak{C}_{\rho}(\widetilde{\Phi}(\xi),\widetilde{\Phi}(\xi))=\mathfrak{C}_{\sigma}(\xi,\xi),
\end{equation}
valid for all $\xi \in\mathcal{H}_{\sigma}$. 
%%%%%%%%%%%%%%%%%%%%%%%%%%%%%%%%%%%%%%%%%%%%%%%%%%%%%%%%%
From the point of view of the \textit{covariance operator} in equation \eqref{eqn: covariance operator}, the invariance condition in equation \eqref{eqn: categorical invariance condition} (coming from the functoriality of $\mathfrak{C}$) implies that
 { \begin{equation}\label{eqn: equivariance of covariance operators} 
	\widetilde{\Phi}^{*}\,\mathbf{T}_{\rho} \, \widetilde{\Phi}= \mathbf{T}_{\sigma}.
\end{equation}} 
%%%%%%%%%%%%%%%%%%%%%%%
In particular, when $\Phi$ is an automorphism of $\mathscr{A}$ preserving $\rho$, equation \eqref{eqn: equivariance of covariance operators} becomes
\begin{equation}\label{eqn: covariance operator commutes with symmetries}
[\mathbf{T}_{\rho},\tilde{\Phi}]=0 .	
\end{equation}

\begin{remark}[Fields of covariances as contravariant metric tensors]\label{rem: universal fNCP model}
To give an intuitive understanding of the role of fields of covariances following the discussion in \cite{CDG2025},  let us note that a Riemannian metric tensor $\mathrm{R}$ on a real, smooth, finite-dimensional manifold $M$ gives rise to a covariant functor $\mathfrak{R}\colon \mathsf{C}(M)\rightsquigarrow \mathsf{Hilb}_{\mathbb{R}}$, where $\mathsf{C}(M)$ is the manifold $M$ itself seen as a trivial category and $\mathsf{Hilb}_{\mathbb{R}}$ the category of real Hilbert spaces and bounded linear contractions, where $\mathfrak{R}_{0}(m) =(T_{m}M,\mathrm{R}_{m})$  and $\mathfrak{R}_{1}(\mathrm{id}_{m})= \mathrm{id}_{T_{m}M}$.
%%%%%%%%%%%%%%%%%%%%%
Moreover, when there is a Lie group $G$ acting smoothly on $M$, a $G$-invariant metric tensor $\mathrm{R}$ on $M$ gives rise to a covariant functor $\mathfrak{R}\colon G\sphericalangle M\rightsquigarrow  \mathsf{Hilb}_{\mathbb{R}}$, where $G\sphericalangle M$ is an action groupoid \cite{M2005b}, where $\mathfrak{R}_{0}(m)=(T_{m}M,\mathrm{R}_{m})$, and $\mathfrak{R}_{1}(\mathrm{g},m)=T_{m}\alpha_{\mathrm{g}}$, where $\alpha_{\mathrm{g}}$ is the action of $\mathrm{g}\in G$ on $M$.
%%%%%%%%%%%%%%%%%%%%%%%%%%%%
If $\mathrm{R}$ is a contravariant Riemannian metric tensor on $M$, then the constructions outlined earlier leads to contravariant functors.
%%%%%%%%%%%%%%%%%%%%%%%%%%
Consequently, interpreting $\mathsf{NCP}$ as a kind of universal model of classical and quantum states in finite dimensions, a field of covariances may be seen as a sort of categorical counterpart of a contravariant Riemannian metric tensor on $\mathsf{NCP}$.
%%%%%%%%%%%%%%%%%%%%

The key insight is that $\mathsf{NCP}$ contains both classical and quantum systems in terms of states on commutative and non-commutative algebras, respectively.
%%%%%%%%%%%%%%%%%%
This point of view allows to recover the Fisher-Rao metric tensor on classical probability measures on discrete and finite spaces and the Morozova-Čencov-Petz monotone metric tensors on faithful quantum states on finite-dimensional Hilbert spaces as different instances of fields of covariances in a sense that is made precise in section \ref{sec:applications}.
%%%%%%%%%%%%%%%%%%%%%%%%%%%%%%%%%%
This is the mechanism by which the categorical formulation unifies the two perspectives: the same functorial contraction condition becomes classical invariance in the commutative case and quantum monotonicity in the non-commutative case.
%%%%%%%%%%%%%%%%%%%%
A more detailed discussion is given in section \ref{sec:applications}.
%%%%%%%%%%%%%%%%%%%
\end{remark}

We now introduce a notion of continuity for a field of covariances on $\mathsf{fNCP}$ that is crucial in the development of section \ref{sec: classification}.
%%%%%%%%%%%%%%%%%%%%%%%%%%%%%%%%%%%%%%%%%
To motivate the continuity conditions we impose on covariance fields on $\mathsf{fNCP}$, we recall that Čencov's result on the uniqueness of the Fisher--Rao metric tensor \cite{C1981a} relies on the assumption of continuity of the Riemannian metric tensors on the smooth manifolds of strictly positive probability measure on finite outcome spaces (that is,  faithful states on commutative algebras of the form $\mathbb{C}^{n}$), while the Morozova-Čencov-Petz characterization of quantum monotone metric tensors \cite{MC1991,P1996} relies on the assumption of continuity of the Riemannian metric tensors on the smooth manifolds of invertible quantum states in finite dimensions (that is, faithful states on non-commutative algebras of the form $\mathcal{B}(\mathcal{H})$ with $\mathrm{dim}(\mathcal{H})<\infty$).
%%%%%%%%%%%%%%%%%%%%%%%%%%%%%%%%%%%%%%%%%%%
In both cases, the topologies come from the norm topology of the dual space of the algebra $\mathscr{A}$  under consideration, namely, $\mathscr{A}=\mathbb{C}^{n}$ in the classical case, and $\mathscr{A}=\mathcal{B}(\mathcal{H})$ in the quantum case.
%%%%%%%%%%%%%%%%%%%%%%%%%%%%%%%%%%%%%%%%%
In particular, the continuity of Riemannian metric tensors at a given faithful state $\rho$ (either in $\mathbb{C}^{n}$ or $\mathcal{B}(\mathcal{H})$) holds for all sequences of faithful states converging to $\rho$.
%%%%%%%%%%%%%%%%%%%%%%%%%%%%%%%%%%%%%%%
In addition, a procedure to extend quantum monotone metric tensors from faithful states on $\mathscr{A}=\mathcal{B}(\mathcal{H})$ (with $\mathcal{H}$ finite-dimensional) to pure states has been introduced in \cite{PS1996}, and is based on sequences of faithful states converging to a given pure state.
%%%%%%%%%%%%%%%%%%%%%%%%%%%
We want to introduce a notion of continuity for fields of covariances on $\mathsf{fNCP}$ that allows us to recover the continuity behaviour mentioned above.
%%%%%%%%%%%%%%%%%%%%%%%%%%

\begin{definition}[Support-compatible sequence for $\rho$]\label{defn: commuting sequence}
A sequence $\{\rho_{n}\}_{n\in\mathbb{N}}$  of faithful states on $\mathscr{A}$ such that\footnote{The choice of the norm topology over  the weak-* topology on $\mathcal{S}(\mathscr{A})$ is not really important in the finite-dimensional case because these topologies are equivalent, but it may  have non-trivial consequences in the infinite-dimensional case.} $\|\rho_{n} -\rho\|_{\mathscr{A}^{*}}\to 0$, is called \textbf{support-compatible for $\rho$} if 
\begin{equation}\label{eqn: support compatibility of sequences}
\Phi_{t}^{\rho_{n}}(\mathbf{p})=\mathbf{p}
\end{equation}
for every $t\in\mathbb{R}$ and all $n$, where $\Phi_{t}^{\rho_{n}}$ denotes the modular flow of $\rho_{n}$ and $\mathbf{p}$ is the support projection of the limit state $\rho$.
%%%%%%%%%%%
\end{definition}

\begin{definition}[Continuous fields of covariances]\label{defn: continuity for fields of covariances}
Let $(\mathscr{A},\rho)\in\mathsf{fNCP}_{0}$, let $\mathbf{p}$ be the support projection of $\rho$.
%%%%%%%%%%%%%%%%%%%%%%%%%%%%%%%%%%%%%%%%%%%%%%%%%%%%%%%%%%%%%%
A field of covariances $\mathfrak{C}$ on  $\mathsf{fNCP}$ is called \textbf{continuous at $\rho$} if, for every sequence $\{\rho_{n}\}_{n\in\mathbb{N}}$ that is support-compatible for $\rho$ according to definition \ref{defn: commuting sequence}, it holds
\begin{equation}\label{eqn: definition of continuity}
\lim_{n\rightarrow\infty}\; \mathfrak{C}_{\rho_{n}}\left( \xi_{\mathbf{ap}}^{\rho_{n}}, \xi_{\mathbf{ap}}^{\rho_{n}}\right)=\mathfrak{C}_{\rho}(\xi_{\mathbf{a}}^{\rho}, \xi_{\mathbf{a}}^{\rho}),
\end{equation}
for every $\mathbf{a}\in\mathscr{A}$.
%%%%%%%%%%%%%%%
The field of covariances $\mathfrak{C}$ is called \textbf{continuous on $\mathsf{fNCP}$} (or simply \textbf{continuous}) if it is continuous at $\rho$ for each $(\mathscr{A},\rho)\in\mathsf{fNCP}_{0}$.
\end{definition}

Note that the support-compatibility condition in definition \ref{defn: commuting sequence} is automatic in several important cases. 
%%%%%%%%%%%%%%%%%%
If the approximating states $\rho_n$ are tracial, then their modular automorphism groups are trivial, and hence they fix the support projection $\mathbf p_\rho$ of the limiting state and thus every sequence of faithful tracial states converging in norm to a tracial state is support-compatible. 
%%%%%%%%%%
In particular, this conclusion holds for every sequence of faithful states on a commutative algebra.
%%%%%%%%%%%%%%%
Moreover, if the limiting state $\rho$ is faithful, then $\mathbf p_\rho=\mathbb I$, so every sequence of faithful states converging to $\rho$ is automatically support-compatible. 
%%%%%%%%%%%%%%%%%%
Consequently, on the faithful state space, as well as in the commutative and tracial settings, definition \ref{defn: continuity for fields of covariances} reduces to the usual notion of continuity considered in classical and quantum information geometry mentioned above.
%%%%%%%%%%%%%%%%

The support-compatibility requirement becomes nontrivial only when $\mathscr A$ is noncommutative and the limiting state $\rho$ is non-faithful. 
%%%%%%%%%%%%%%%%
In this case, the decomposition of $\mathscr A$ determined by the support projection $\mathbf p_\rho$ of the limiting state is preserved by the approximating modular structures. 
%%%%%%%%%%%%%%
This is precisely the hypothesis used in Lemma \ref{lem: convergence of modular operator} to obtain convergence of the partial inverses and of the supported modular operators on $\mathscr A\mathbf p_\rho$. 
%%%%%%%%%%
The restriction is therefore adapted to the modular-convergence argument used below, and we do not claim that it is the strongest possible continuity condition at boundary states.
%%%%%%%%%%%%%%%%%%%%%%%%%%%%%

The occurrence of the support projection $\mathbf p_\rho$ in the vectors $\xi_{\mathbf a\mathbf p_\rho}^{\rho_n}$ in definition \ref{defn: continuity for fields of covariances} has a separate role. 
%%%%%%%%%%%%%%%%%%%%%%%%%%%%%%%%%%
When $\rho$ is non-faithful, so that $\mathbf{p}$ is not the identity, its \textup{GNS} null space is $\mathscr A(\mathbb I-\mathbf p_\rho)$ (see equation \eqref{equation: Gelfand ideal normal state}), and hence its \textup{GNS} Hilbert space is canonically identified with the right ideal $\mathscr A\mathbf p_\rho$ (see equation \eqref{eqn: gns hilbert space normal state}). 
%%%%%%%%%%%%%
The multiplication $\mathbf a\mapsto\mathbf a\mathbf p_\rho$ therefore restricts the comparison to the directions that survive in the limiting \textup{GNS} space. 
%%%%%%%%%%%%%%%%%%%%%%%
Without this restriction, continuity would also constrain directions that vanish at the limiting state but may retain a nonzero covariance along faithful approximants. 
%%%%%%%%%%%%%
As shown in remark \ref{rem: continuity on reduced directions is necessary} in section \ref{sec: classification}, this would impose an additional condition that would exclude the Petz-symmetric covariances associated with the usual quantum monotone metric tensors \cite{P1996,GHP2009}. 
%%%%%%%%%%%%%%%%%%%%%
Further discussion of these two restrictions is given in remarks \ref{rem: continuity holds for all sequences converging to faithful states} and \ref{rem: continuity on reduced directions is necessary}.
%%%%%%%%%%%%%%%%

\section{Characterization of continuous fields of covariances in finite dimensions}\label{sec: classification}

In this section, we obtain a complete characterization of all the \textit{continuous fields of covariances} on $\mathsf{fNCP}$ in the sense of definitions \ref{defn: field of covariances} and \ref{defn: continuity for fields of covariances}.
%%%%%%%%%%%%%%%%%%%%%%%%%%%%%%%%%%%%%%%%%%%%%%%%%

In subsection \ref{subsec: tracial states}, we obtain a characterization of continuous fields of covariances case on the subcategory $\mathsf{fNCT}\subset\mathsf{fNCP}$ of tracial states on finite-dimensional $C^{*}$-algebras.
%%%%%%%%%%%%%%%%%%%%%%%%%%%
This case essentially recovers statistical covariance and the Fisher-Rao metric tensor of classical probability vectors as discussed by Čencov \cite{C1981a}, but in a way that is closer to the recent treatment by Nagaoka \cite{N2024b} because of the contravariant nature of the functor defining fields of covariances, and extends it to tracial states on possibly non-commutative finite-dimensional $C^{*}$-algebras, and to possibly non-faithful tracial states.
%%%%%%%%%%%%%%%%%%%%%%%%%%%%%%%%%%%%%%%%%

In subsection \ref{subsec: faithful states}, we pass to $\mathsf{fNCP}$ by first considering the case of faithful states on finite-dimensional $C^{*}$-algebras.
%%%%%%%%%%%%%%%%%%%%%%%%
This case essentially recovers Čencov-Morozova-Petz's classification of quantum monotone metric  tensor \cite{MC1991,P1996}, but in a way that is closer to the case of quantum covariances \cite{GHP2009}, again because of the contravariant nature of the functor defining fields of covariances, and extends it to faithful states on arbitrary finite-dimensional $C^{*}$-algebras.
%%%%%%%%%%%%%%%%%%%%
Then, we consider the case of non-faithful states on finite-dimensional $C^{*}$-algebras.
%%%%%%%%%%%%%%%%%%%%%%%%%%%%%%%%%%%%%
This case represents a generalized alternative  to  Petz and Sudár's radial procedure \cite{PS1996} that is not constrained to the case of pure states.
%%%%%%%%%%%%%%%%%%%%%%%%%%%%%%%%%%%%%%%%%%

\subsection{Tracial states}\label{subsec: tracial states}

In this section, we give a complete characterization of \textit{continuous fields of covariances}  for the subcategory $\mathsf{fNCT}$ of tracial states on finite-dimensional algebras.
%%%%%%%%%%%%%%%%%%%%%%%%%%%
The strategy of the proof is to first characterize a field of covariance $\mathfrak{C}$ on the unique tracial state $\tau$ on $\mathcal{B}(\mathcal{H})$ with $\mathcal{H}$ finite-dimensional, and then impose continuity as in definition \ref{defn: continuity for fields of covariances} and use the invariance in equation \eqref{eqn: categorical invariance condition} to obtain the covariance at all other states in $\mathsf{fNCT}$.
%%%%%%%%%%%%%%%%%%%%%%%%%%%%%%%%

\begin{proposition}\label{prop: fields of covariances at the tracial state of B(H)}
Let $\mathfrak{C}:\mathsf{fNCT}\rightsquigarrow\mathsf{Hilb}$ be a field of covariances as in definition \ref{defn: field of covariances}. Then, there exist unique strictly positive numbers $\alpha,\beta\in(0,\infty)$ such that, for every finite-dimensional complex Hilbert space $\mathcal{H}$ and for the unique tracial state $\tau_{\mathcal H}$ on $\mathcal{B}(\mathcal{H})$, it holds
\begin{equation}\label{eqn: covariance tracial full matrix algebra}
\mathfrak{C}_{\tau_{\mathcal H}}(\xi,\eta)
=
\beta\langle\xi\mid\eta\rangle_{\tau_{\mathcal H}}
+
(\alpha-\beta)
\langle\xi\mid\xi^{\tau_{\mathcal H}}_{\mathbb{I}_{\mathcal H}}\rangle_{\tau_{\mathcal H}}
\langle\xi^{\tau_{\mathcal H}}_{\mathbb{I}_{\mathcal H}}\mid\eta\rangle_{\tau_{\mathcal H}}
\end{equation}
for all $\xi,\eta\in\mathcal{H}_{\tau_{\mathcal H}}$.
%%%%%%%%%%%%%%%%
\end{proposition}

\begin{proof}
Fix a finite-dimensional complex Hilbert space $\mathcal H$ and write $N=\dim(\mathcal H)$.
%%%%%%%%%%%%%%%%%%%%%
If $N=1$, then $\mathcal B(\mathcal H)\cong\mathbb C$ and $\mathcal H_{\tau_{\mathcal H}}=\mathbb C\xi^{\tau_{\mathcal H}}_{\mathbb{I}}$.
%%%%%%%%%%%%
Therefore, there exists a unique $\alpha_{1}>0$ such that
$\mathfrak{C}_{\tau_{\mathcal H}}(\xi,\eta)=\alpha_{1}\langle\xi\mid\eta\rangle_{\tau_{\mathcal H}}$.
%%%%%%%%%%%%%%%%%%%%%%%%

Suppose now that $N\geq2$. 
%%%%%%%%%%%%%%%%
The \textup{GNS} Hilbert space $\mathcal{H}_{\tau_{\mathcal H}}$ coincides with $\mathscr{A}=\mathcal{B}(\mathcal{H})$ endowed with the Hilbert-Schmidt inner product, and the group of automorphisms of $\mathscr{A}=\mathcal{B}(\mathcal{H})$ preserving $\tau_{\mathcal H}$ coincides with the unitary group $\mathcal{U}(\mathcal{H})$ acting on $\mathcal{H}_{\tau_{\mathcal H}}$  by conjugation $(\mathbf{U},\mathbf{a})\mapsto \mathbf{UaU}^{\dagger}$. 	
%%%%%%%%%%%%%%%%%%%%%%
The subspace $\mathbb{C} \mathbb{I}_{\mathcal H}$ is   invariant under the conjugation action of $\mathcal{U}(\mathcal{H})$ and its Hilbert--Schmidt orthogonal complement 
\begin{equation} 
(\mathbb{C} \mathbb{I_{\mathcal H}})^{\perp}
=\{A\in B(\mathcal{H}) : \operatorname{Tr} A = 0\}
=\mathfrak{sl}(\mathcal{H}),
\end{equation}
is invariant as well. 
%%%%%%%%%%%%%%%%%%%%%%%%%%%%%
Indeed, the representation of $\mathcal{U}(\mathcal{H})$ on $\mathfrak{sl}(\mathcal{H})$ identifies with the complexified adjoint representation of
$\mathfrak{su}(\mathcal{H})$ (since $\mathfrak{sl}(\mathcal{H})=\mathfrak{su}(\mathcal{H})\otimes_{\mathbb{R}}\mathbb{C}$).
%%%%%%%%%%%%%%%%%%%%%%%%%%%%
As $\mathfrak{su}(\mathcal{H})$ is simple, its (complexified) adjoint representation is irreducible; hence the only nonzero proper $\mathcal{U}(\mathcal{H})$-invariant subspaces of $B(\mathcal{H})$ are
$\mathbb{C} \mathbb{I}_{\mathcal H}$ and $\mathfrak{sl}(\mathcal{H})$.
%%%%%%%%%%%%%%%%%%%%%%%%%%%%%%
The invariance condition in equation \eqref{eqn: covariance operator commutes with symmetries} implies that the covariance operator $\mathbf{T}_{\tau_{\mathcal H}}$ must commute with the operators implementing the action of $\mathcal{U}(\mathcal{H})$ by conjugation.
%%%%%%%%%%%%%%%%%%%%%%%%%%%%%%
Then, Schur's lemma implies that $\mathbf{T}_{\tau_{\mathcal H}}$ is  proportional to the identity on $\mathbb{C}\mathbb{I}_{\mathcal H}$ and $\mathbb{C}\mathbb{I}_{\mathcal H}^{\perp}=\mathfrak{sl}(\mathcal{H})$, albeit with possibly different proportionality factors, and it does not contain terms intertwining these two invariant subspaces.
%%%%%%%%%%%%%%%%%%%%%%%%%%%%%%%%
Since  $\mathbf{T}_{\tau_{\mathcal H}}$ must be self-adjoint and strictly  positive, there are $\alpha_{\mathcal{H}},\beta_{\mathcal{H}}>0$ such that 
\begin{equation}\label{eqn: covariance operator on tau}
    \mathbf{T}_{\tau_{\mathcal H}}= \beta_{\mathcal{H}} \mathbb{I} _{\mathcal H}+ (\alpha_{\mathcal{H}} -\beta_{\mathcal{H}}) |\xi_{\mathbb{I}_{\mathcal H}}^{\tau_{\mathcal H}}\rangle\langle \xi_{\mathbb{I}_{\mathcal H}}^{\tau_{\mathcal H}}|,
\end{equation}
and thus equation \eqref{eqn: covariance operator} implies  
\begin{equation}\label{eqn: covariance on tau} 
    \mathfrak{C}_{\tau_{\mathcal H}}(\xi,\eta)=\beta_{\mathcal{H}}\langle\xi\mid\eta\rangle_{\tau_{\mathcal H}} + (\alpha_{\mathcal{H}} -\beta_{\mathcal{H}})\langle\xi\mid \xi_{\mathbb{I}_{\mathcal H}}^{\tau_{\mathcal H}}\rangle_{\tau_{\mathcal{H}}}\langle\xi_{\mathbb{I}_{\mathcal H}}^{\tau_{\mathcal H}}\mid\eta\rangle_{\tau_{\mathcal H}}
\end{equation}
for all $\xi,\eta\in\mathcal{H}_{\tau_{\mathcal H}}$ as required.
%%%%%%%%%%%%%%%%%%

We now prove that these coefficients do not depend on $\mathcal H$. 
%%%%%%%%%%%%%
Let $\mathcal H$ and $\mathcal K$ be finite-dimensional complex Hilbert spaces, and consider the unital $*$-homomorphism
$\iota_{\mathcal H,\mathcal K}:\mathcal B(\mathcal H)\to\mathcal B(\mathcal H\otimes\mathcal K)$ given by
$\iota_{\mathcal H,\mathcal K}(a)=a\otimes \mathbb{I}_{\mathcal K}$, together with the CPU map
$E_{\mathcal H,\mathcal K}=\operatorname{id}_{\mathcal B(\mathcal H)}\otimes\tau_{\mathcal K}$, where $\tau_{\mathcal{K}}$ denotes the tracial state on $\mathcal{B}(\mathcal{K})$.
%%%%%%%%%%%%%%
These maps satisfy
$E_{\mathcal H,\mathcal K}\circ\iota_{\mathcal H,\mathcal K}
=\operatorname{id}_{\mathcal B(\mathcal H)}$,
$\iota_{\mathcal H,\mathcal K}^{*}(\tau_{\mathcal H\otimes\mathcal K})
=\tau_{\mathcal H}$, and
$E_{\mathcal H,\mathcal K}^{*}(\tau_{\mathcal H})
=\tau_{\mathcal H\otimes\mathcal K}$.
%%%%%%%%%%%%%
Consequently, equation \eqref{eqn: categorical invariance condition} implies
\begin{equation}\label{eqn: covariance amplification invariance}
\mathfrak{C}_{\tau_{\mathcal H\otimes\mathcal K}}
\left(
\xi^{\tau_{\mathcal H\otimes\mathcal K}}_{\mathbf{a}\otimes \mathbb{I}_{\mathcal K}},
\xi^{\tau_{\mathcal H\otimes\mathcal K}}_{\mathbf{a}\otimes \mathbb{I}_{\mathcal K}}
\right)
=
\mathfrak{C}_{\tau_{\mathcal H}}
\left(
\xi^{\tau_{\mathcal H}}_{\mathbf{a}},
\xi^{\tau_{\mathcal H}}_{\mathbf{a}}
\right)
\end{equation}
for all $\mathbf{a}\in\mathcal B(\mathcal H)$.
%%%%%%%%%%%%%%%%%%%%
Taking $\mathbf{a}=\mathbb{I}_{\mathcal H}$ gives
$\alpha_{\mathcal H\otimes\mathcal K}=\alpha_{\mathcal H}$. Exchanging the two tensor factors gives
$\alpha_{\mathcal H\otimes\mathcal K}=\alpha_{\mathcal K}$. Therefore, $\alpha_{\mathcal H}$ is independent of $\mathcal H$; we denote its common value by $\alpha$.

Suppose that $\dim(\mathcal H)\geq2$, and choose a nonzero $\mathbf{a}\in\mathcal B(\mathcal H)$ such that $\tau_{\mathcal H}(\mathbf{a})=0$. Then $\mathbf{a}\otimes \mathbb{I}_{\mathcal K}$ is traceless and
$\| \xi^{\tau_{\mathcal H\otimes\mathcal K}}_{\mathbf{a}\otimes \mathbb{I}_{\mathcal K}}\|_{\tau_{\mathcal H\otimes\mathcal K}}^{2}
=
\|\xi^{\tau_{\mathcal H}}_{\mathbf{a}}\|_{\tau_{\mathcal H}}^{2}$.
Equation~\eqref{eqn: covariance amplification invariance} therefore implies
$\beta_{\mathcal H\otimes\mathcal K}=\beta_{\mathcal H}$. If also $\dim(\mathcal K)\geq2$, exchanging the tensor factors gives
$\beta_{\mathcal H\otimes\mathcal K}=\beta_{\mathcal K}$. Hence, $\beta_{\mathcal H}$ is independent of $\mathcal H$ whenever $\dim(\mathcal H)\geq2$; we denote its common value by $\beta$.
%%%%%%%%%%%%%%%%%%%%%
Thus, equation~\eqref{eqn: covariance tracial full matrix algebra} holds whenever $\dim(\mathcal H)\geq2$. 
%%%%%%%%%%%%%%%%%
If $\dim(\mathcal H)=1$, every vector in $\mathcal H_{\tau_{\mathcal H}}$ is proportional to $\xi^{\tau_{\mathcal H}}_{I_{\mathcal H}}$, so the right-hand side of equation \eqref{eqn: covariance tracial full matrix algebra} reduces to
$\alpha\langle\xi\mid\eta\rangle_{\tau_{\mathcal H}}$. Therefore, the same formula also holds in dimension one.
%%%%%%%%%%%%%%%%%
Finally, $\alpha$ is uniquely determined by the value of the covariance on the identity vector, while $\beta$ is uniquely determined by its value on any nonzero traceless vector in $\mathcal B(\mathcal H)$ with $\dim(\mathcal H)\geq2$.
%%%%%%%%%%%%%%%%%%%%%%%
\end{proof}

Once the field of covariance is known at $(\mathcal{B}(\mathcal{H}),\tau)\in\mathsf{fNCT}_{0}$, the invariance property in equation \eqref{eqn: categorical invariance condition} immediately propagates this information to any object $(\mathscr{A},\sigma)\in\mathsf{fNCT}_{0}$ that is a retract of $(\mathcal{B}(\mathcal{H}),\tau)\in\mathsf{fNCT}_{0}$ through some $\hat{\Phi}\in\mathsf{fNCT}_{1}$ according to definition \ref{defn: split monomorphisms}.
%%%%%%%%%%%%%%%%%%%%%%%%%%%%%%
A particularly important and concrete class of such retracts are the faithful \textit{rational tracial states}, namely those states for which  $\mathscr{A}=\bigoplus_{j=1}^{N}\mathcal{B}(\mathcal{H}_{j})$  (see equation \eqref{eqn: isomorphism of finite-dimensional algebras}) and
\begin{equation}\label{eqn: faithful rational tracial states}
\sigma(\mathbf{a}_{1},\cdots,\mathbf{a}_{N}) =\sum_{j=1}^{N}p_{j}\tau_{j}(\textbf{a}_j),\quad \textbf{a}_j \in B(\mathcal{H}_j),  \, p_{j}\in \mathbb{Q}_{>0},\;\sum_{j=1}^{N}p_{j}=1.
\end{equation}
%%%%%%%%%%%%%%%%%%%%%%%%%%%%%%%
In the proposition below, the split monomorphism and its left inverse for faithful rational tracial states are explicitly  built.
%%%%%%%%%%%%%%%%%%%%%%

\begin{proposition}\label{prop: field of covariances on faithful rational tracial states}
Let $ \mathfrak{C}\colon \mathsf{fNCT}\rightsquigarrow\mathsf{Hilb}$ be  a field of covariances as in definition \ref{defn: field of covariances}, and let $\alpha,\beta>0$ be the constants appearing in proposition \ref{prop: fields of covariances at the tracial state of B(H)}.
%%%%%%%%%%%%%
If $(\mathscr{A},\sigma)\in\mathsf{fNCT}_{0}$ with $\sigma$ a faithful rational tracial state as in equation \eqref{eqn: faithful rational tracial states}, then it holds
\begin{equation}\label{eqn: monotone covariance on faithful rational traces}
    \mathfrak{C}_{\sigma}(\xi,\eta)=\beta\langle\xi\mid\eta\rangle_{\sigma} + (\alpha -\beta)\langle\xi\mid \xi_{\mathbb{I}}^{\sigma}\rangle_{\sigma}\langle\xi^{\sigma}_{\mathbb{I}}\mid\eta\rangle_{\sigma}.
\end{equation}
for all $\xi,\eta\in\mathcal{H}_{\sigma}$.
%%%%%%%%%%%%%%%%%%%%%%%%%

\end{proposition}    	

\begin{proof}
%%%%%%%%
Without loss of generality, we take $\mathscr{A}$ to be the direct sum of algebras of bounded linear operators on finite-dimensional complex Hilbert spaces as in equation \eqref{eqn: isomorphism of finite-dimensional algebras}.
%%%%%%%%%%%%%%%%%%%%%%%%%%%%%%%%%%%%%%%%%%%%%%%%%%%%%%%%%%%%%%%%
Let us write $p_{j}=\frac{L_{j}}{M}$, with $L_{j},M\in\mathbb{N}_{>0}$, so that
\begin{equation}\label{eqn: faithful rational trace on finite-dimensional W*-algebras}
    \sigma\equiv \left(\frac{L_{1}}{M}\tau_{1},\cdots,\frac{L_{N}}{M}\tau_{N}\right),
\end{equation}
with $L_{j}$ an integer and $M=\sum_{j=1}^{N}L_{j}$.
%%%%%%%%%%%%%%%%%%%%%%%%%%%%%%%%%%%%%%%%%%%%%%%%%%%%%%%%%%%%%%%%%
Consider the Hilbert space 
\begin{equation}\label{eqn: tensor product factor in rational tracial states}
 \mathcal{K}_{j}=\left(\mathcal{H}_{j}\overbrace{\oplus\cdots\oplus}^{L_{j}\mbox{ times }}\mathcal{H}_{j}\right)\otimes\left(\bigotimes_{k\neq j}^{N}\mathcal{H}_{k}\right),
\end{equation}
whose dimension is $L_{j}D$ with $D=\mathrm{dim}(\otimes_{k=1}^{N}\mathcal{H}_{k})$, define the Hilbert space
\begin{equation}\label{eqn: tensor product factor in rational tracial states 2}
\mathcal{K}=\bigoplus_{j=1}^{N}\mathcal{K}_{j},
\end{equation}
whose dimension is $MD$, and let $\mathbf{P}_{j}$ be the projection onto $\mathcal{K}_{j}$.
%%%%%%%%%%%%%%%%%%%%
Define the map $\phi\colon \mathscr{A}\rightarrow\mathcal{B}(\mathcal{K})$ setting
\begin{equation} 
\mathbf{A}_{\phi}\equiv\phi(\mathbf{a}_{1},\cdots,\mathbf{a}_{N}), \quad \mathbf{P}_{j}\mathbf{A}_{\phi}\mathbf{P}_{k}=\delta_{jk}\left(\mathbf{0},\cdots, \left(\overbrace{\mathbf{a}_{j},\cdots,\mathbf{a}_{j}}^{L_{j}\mbox{ times }}\right)\otimes\mathbb{I}_{j},\cdots\mathbf{0}\right),
\end{equation}
where $\mathbb{I}_{j}$ is the identity operator on $\left(\bigotimes_{k\neq j}^{N}\mathcal{H}_{k}\right)$.
%%%%%%%%%%%%%%%%
A direct check shows that $\phi$ is a unital $*$-homomorphism, so that it is a CPU map.
%%%%%%%%%%%%%%%%%%%%%%
It holds
\begin{equation}\label{eqn: pullback for rational tracial states}
\begin{split}
\phi^{*}\tau(\mathbf{0},\cdots,\mathbf{a}_{j},\cdots,\mathbf{0})&=\tau\left(\mathbf{0},\cdots, \left(\overbrace{\mathbf{a}_{j},\cdots,\mathbf{a}_{j}}^{L_{j}\mbox{ times }}\right)\otimes\mathbb{I}_{j},\cdots\mathbf{0}\right)=\\
&=\frac{\mathrm{dim}(\mathcal{K}_{j})}{\mathrm{dim}(\mathcal{K})}\tau_{L_{j}}\left(\overbrace{\mathbf{a}_{j},\cdots,\mathbf{a}_{j}}^{L_{j}\mbox{ times }}\right)=\frac{L_{j}}{\sum_{r=1}^{N}L_{r}}\tau_{j}(\mathbf{a}_{j}).                                         \end{split}
\end{equation}
%%%%%%%%%%%%%%%%
Note that the tensor product part in equation \eqref{eqn: tensor product factor in rational tracial states} is needed to ensure that $\frac{\mathrm{dim}(\mathcal{K}_{j})}{\mathrm{dim}(\mathcal{K})}=\frac{L_{j}}{M}$ in equation \eqref{eqn: pullback for rational tracial states}.
%%%%%%%%%%%%%%%%
Therefore, recalling that
\begin{equation} 
(\mathbf{a}_{1},\cdots,\mathbf{a}_{N})=\sum_{j=1}^{N}(\mathbf{0},\cdots,\mathbf{a}_{j},\cdots,\mathbf{0})
\end{equation}
and that $\phi$ is linear, we get that
\begin{equation}\label{eqn: rational states are retracts of trace}
\phi^{*}\tau(\mathbf{a}_{1},\cdots,\mathbf{a}_{N})=\sum_{j=1}^{N}\frac{L_{j}}{\sum_{r=1}^{N}L_{r}}\tau_{j}(\mathbf{a}_{j})\stackrel{\mbox{\eqref{eqn: faithful rational trace on finite-dimensional W*-algebras}}}{=}\sigma(\mathbf{a}_{1},\cdots,\mathbf{a}_{N}),
\end{equation}
which means that $\phi^{*}\tau =\sigma$.
%%%%%%%%%%%%%%%%%%%%%%%%%%%%%%%

Since $\tau$ is tracial, its modular automorphism group is trivial.
%%%%%%%%%%%%%%%%%%%%%%%%%%%%%%%
Consequently, the subalgebra $\phi(\mathscr{A})\subseteq\mathcal{B}(\mathcal{K})$ is automatically invariant under the modular automorphism group of $\tau$.
%%%%%%%%%%%%%%%%%%%%%%%%%%%%%%%
By Takesaki's theorem on conditional expectations \cite{T1972a,T2003a}, there exists a $\tau$-preserving conditional expectation
$$
\mathcal{P}\colon \mathcal{B}(\mathcal{K})\longrightarrow \phi(\mathscr{A})
$$
onto $\phi(\mathscr{A})$.
%%%%%%%%%%%%%%%%%%%%%%%%%%%%%%%
We then define
$$
\mathcal{E}\colon \mathcal{B}(\mathcal{K})\longrightarrow \mathscr{A},
\qquad
\mathcal{E}:=\phi^{-1}\circ\mathcal{P}.
$$
The map $\mathcal{E}$ is completely positive and unital, because $\mathcal{P}$ is a conditional expectation and $\phi^{-1}$ is a $*$-isomorphism from $\phi(\mathscr{A})$ onto $\mathscr{A}$.
%%%%%%%%%%%%%%%%%%%%%%%%%%%%%%%
Moreover,
$$
\mathcal{E}\circ\phi
=
\phi^{-1}\circ\mathcal{P}\circ\phi
=
\mathrm{id}_{\mathscr{A}},
$$
because $\mathcal{P}$ restricts to the identity on $\phi(\mathscr{A})$.
%%%%%%%%%%%%%%%%%%%%%%%%%%%%%%%
Furthermore, since $\mathcal{P}$ is $\tau$-preserving and $\phi^{*}\tau=\sigma$ by equation \eqref{eqn: rational states are retracts of trace}, we obtain, for every $\mathbf{A}\in\mathcal{B}(\mathcal{K})$,
$$
(\mathcal{E}^{*}\sigma)(\mathbf{A})
=
\sigma(\mathcal{E}(\mathbf{A}))
=
\sigma\bigl(\phi^{-1}(\mathcal{P}(\mathbf{A}))\bigr)
=
\tau\bigl(\phi(\phi^{-1}(\mathcal{P}(\mathbf{A})))\bigr)
=
\tau(\mathcal{P}(\mathbf{A}))
=
\tau(\mathbf{A}).
$$
Therefore, $\mathcal{E}^{*}\sigma=\tau$.
%%%%%%%%%%%%%%%%%%%%%%%%%%%%%%%
It follows that the CPU map $\mathcal{E}\colon\mathcal{B}(\mathcal{K})\rightarrow\mathscr{A}$ defines a morphism
$$
\widehat{\mathcal{E}}\colon(\mathscr{A},\sigma)\longrightarrow(\mathcal{B}(\mathcal{K}),\tau)
$$
in $\mathsf{fNCT}$, while the injective unital $*$-homomorphism $\phi$ defines a morphism
$$
\widehat{\phi}\colon(\mathcal{B}(\mathcal{K}),\tau)\longrightarrow(\mathscr{A},\sigma)
$$
in $\mathsf{fNCT}$.
%%%%%%%%%%%%%%%%%%%%%%%%%%%%%%%
The identity $\mathcal{E}\circ\phi=\mathrm{id}_{\mathscr{A}}$ means that $\widehat{\mathcal{E}}$ is a split monomorphism with left inverse $\widehat{\phi}$.
Consequently, the invariance property in equation \eqref{eqn: categorical invariance condition} together with proposition \ref{prop: fields of covariances at the tracial state of B(H)} implies the validity of equation \eqref{eqn: monotone covariance on faithful rational traces}.
\end{proof}

To characterize $\mathfrak{C}_{\sigma}$ for a tracial  state $\sigma$ that is not faithful rational, we now impose and exploit the continuity condition as in definition \ref{defn: continuity for fields of covariances}.
%%%%%%%%%%%%%%

\begin{proposition}\label{prop: classification of invariant covariances for any algebra and tracial states}
Let $\mathfrak{C}\colon \mathsf{fNCT}\rightsquigarrow\mathsf{Hilb}$ be a continuous field of covariances as in definitions \ref{defn: field of covariances} and \ref{defn: continuity for fields of covariances}, and let $\alpha,\beta>0$ be the constants appearing in proposition \ref{prop: field of covariances on faithful rational tracial states}.
%%%%%%%%%%%%%%
If $(\mathscr{A},\sigma)\in\mathsf{fNCT}_{0}$, with $\mathbf{p}$ the support projection of $\sigma$, then it holds
\begin{equation}\label{eqn: monotone covariance on tracial states}
 \mathfrak{C}_{\sigma}(\xi,\eta)=\beta\langle\xi\mid \eta\rangle_{\sigma} + (\alpha-\beta)\langle\xi\mid \xi_{\mathbb{I}}^{\sigma}\rangle_{\sigma}\langle\xi_{\mathbb{I}}^{\sigma}\mid \eta\rangle_{\sigma}
\end{equation}
for all $\xi,\eta\in\mathcal{H}_{\sigma}$.
%%%%%%%%%%%%%%%%%%
In particular, writing $\xi\equiv\xi_{\mathbf{a}}=\xi_{\mathbf{ap}}$ and $\eta\equiv\xi_{\mathbf{b}}=\xi_{\mathbf{bp}}$, equation \eqref{eqn: monotone covariance on tracial states} reads
\begin{equation}\label{eqn: monotone covariance on tracial states with support}
 \mathfrak{C}_{\sigma}(\xi_{\mathbf{a}},\xi_{\mathbf{b}})=\beta\,\sigma((\mathbf{ap})^{\dagger}\mathbf{bp})+(\alpha-\beta)\overline{\sigma(\mathbf{ap})}\,\sigma(\mathbf{bp}).
\end{equation}
%%%%%%%%%
\end{proposition}
 
\begin{proof}
%%%%%%%%%%%%%%%%%%
Without loss of generality, let $\mathscr{A}$ be as in equation \eqref{eqn: isomorphism of finite-dimensional algebras}.
%%%%%%%%%%%%%%%%%%%%%
By suitably arranging the order of the summands in equation \eqref{eqn: isomorphism of finite-dimensional algebras}, a tracial state on $\mathscr{A}$ can be written as
\begin{equation}\label{eqn: tracial state in finite dimensions}
\sigma=(p_{1}\tau_{1},\cdots ,p_{r}\tau_{r},\mathbf{0},\cdots,\mathbf{0}),
\end{equation}
where $r\leq N$, $p_{j}>0$ for $j\leq r$, and $\sum_{j=1}^{r}p_{j}=1$.
%%%%%%%%%%%%%%%%%%
Its support projection is
$p=(I_{1},\ldots,I_{r},0,\ldots,0)$.

We construct a sequence of faithful rational tracial states converging to
$\sigma$. Suppose first that $r<N$. For every $j<r$, choose a sequence
$\{(p_{j})_{n}\}_{n\in\mathbb{N}}$ of strictly positive rational numbers
converging from below to $p_{j}$, and choose a sequence
$\{q_{n}\}_{n\in\mathbb{N}}$ of strictly positive rational numbers converging
to zero. Set
\begin{equation}
(p_{j})_{n}
=
\frac{q_{n}}{N-r}
\end{equation}
for $j>r$, and
\begin{equation}
(p_{r})_{n}
=
1-q_{n}-\sum_{j=1}^{r-1}(p_{j})_{n}.
\end{equation}
Since $(p_{r})_{n}\to p_{r}>0$, after possibly discarding finitely many terms,
all the coefficients $(p_{j})_{n}$ are strictly positive rational numbers,
their sum is equal to one, and $(p_{j})_{n}\to p_{j}$ for every $j\leq r$,
while $(p_{j})_{n}\to0$ for every $j>r$.

Suppose now that $r=N$. For every $j<N$, choose a sequence
$\{(p_{j})_{n}\}_{n\in\mathbb{N}}$ of strictly positive rational numbers
converging from below to $p_{j}$, and set
\begin{equation}
(p_{N})_{n}
=
1-\sum_{j=1}^{N-1}(p_{j})_{n}.
\end{equation}
Then $(p_{N})_{n}$ is strictly positive and rational, and
$(p_{N})_{n}\to p_{N}$. This also covers the case $N=r=1$, in which
$(p_{1})_{n}=1$.

In either case, define
\begin{equation}
\sigma_{n}
=
\bigl(
(p_{1})_{n}\tau_{1},
\ldots,
(p_{N})_{n}\tau_{N}
\bigr).
\end{equation}
Each $\sigma_{n}$ is a faithful rational tracial state and
$\|\sigma_{n}-\sigma\|_{\mathscr{A}^{*}}\to0$. Moreover, since every
$\sigma_{n}$ is tracial, its modular flow is trivial and hence leaves the
support projection $p$ invariant. Therefore, $\{\sigma_{n}\}_{n\in\mathbb N}$
is a support-compatible sequence for $\sigma$ in the sense of definition \ref{defn: commuting sequence}.
%%%%%%%%%%%%%%%%%%%%%%%%%

Since $\mathfrak{C}$ is a continuous field of covariances as in definitions \ref{defn: field of covariances} and \ref{defn: continuity for fields of covariances} by assumption, it holds
\begin{equation}
\begin{split}
\mathfrak{C}_{\sigma}(\xi_{\mathbf{a}},\xi_{\mathbf{a}})&=\lim_{n\rightarrow\infty} \mathfrak{C}_{\sigma_{n}}(\xi_{\mathbf{ap}},\xi_{\mathbf{ap}})\stackrel{\eqref{eqn: monotone covariance on faithful rational traces}}{=}\lim_{n\rightarrow\infty} \beta\sigma_{n}(\mathbf{pa}^{\dagger}\mathbf{ap})+ (\alpha-\beta)\vert\sigma_{n}(\mathbf{ap})\vert^{2}=\\ 
& 
= \beta\sigma(\mathbf{pa}^{\dagger}\mathbf{ap})+ (\alpha-\beta)\vert\sigma(\mathbf{ap})\vert^{2} \stackrel{\mbox{\eqref{eqn: support projection of normal state}}}{=}\beta\sigma(\mathbf{a}^{\dagger}\mathbf{a})+ (\alpha-\beta)\vert\sigma(\mathbf{a})\vert^{2}.
\end{split}
\end{equation}
%%%%%%%%%%%%%%%%%%%%%%%%%%%%%%%%%%%%%%%%%%%%%%%%%%%%%%%%%%
Equation \eqref{eqn: monotone covariance on tracial states} then follows from the   complex  polarization identity.
%%%%%%%%%%%%%%%%%%%%%%%%
For the convention used here, where the inner product is linear in the second argument, this means that if $Q(\xi):=\mathfrak{C}_{\sigma}(\xi,\xi)$, then
\begin{equation}
\mathfrak{C}_{\sigma}(\xi,\eta)=\frac{1}{4}\left(Q(\xi+\eta)-Q(\xi-\eta)-\mathrm{i}Q(\xi+\mathrm{i}\eta)+\mathrm{i}Q(\xi-\mathrm{i}\eta)\right).
\end{equation}
%%%%%%%%%%%%%%%%%%
\end{proof}

\begin{remark}
The reduction to the supported vectors
$\xi_{\mathbf{a}\mathbf{p}}^{\sigma_n}$
in definition
\ref{defn: continuity for fields of covariances}
is not needed for the proof of proposition
\ref{prop: classification of invariant covariances for any algebra and tracial states}.
Indeed, suppose that continuity were instead imposed directly on
$\xi_{\mathbf{a}}^{\sigma_n}$.
For the sequence of faithful rational tracial states
$\{\sigma_n\}_{n\in\mathbb{N}}$
used in the proof, one would obtain
\begin{equation}
\begin{split}
\mathfrak{C}_{\sigma}
\left(
\xi_{\mathbf{a}}^{\sigma},
\xi_{\mathbf{a}}^{\sigma}
\right)
&=
\lim_{n\rightarrow\infty}
\mathfrak{C}_{\sigma_n}
\left(
\xi_{\mathbf{a}}^{\sigma_n},
\xi_{\mathbf{a}}^{\sigma_n}
\right)
\\
&=
\lim_{n\rightarrow\infty}
\left(
\beta\,\sigma_n(\mathbf{a}^{\dagger}\mathbf{a})
+
(\alpha-\beta)
\left|\sigma_n(\mathbf{a})\right|^{2}
\right)
\\
&=
\beta\,\sigma(\mathbf{a}^{\dagger}\mathbf{a})
+
(\alpha-\beta)
\left|\sigma(\mathbf{a})\right|^{2}.
\end{split}
\end{equation}
The conclusion then follows from the complex polarization identity.
Thus, the use of the support projection in the continuity condition is not
required for the extension from faithful rational tracial states to arbitrary
tracial states.
Its role becomes essential only in the treatment of non-faithful,
non-tracial states, as explained in remark
\ref{rem: continuity on reduced directions is necessary}.
\end{remark}

Putting together the results of this subsection, we obtain a complete characterization of continuous fields of covariances on $\mathsf{fNCT}$.
%%%%%%%%%%%%%%%%%%%%%%%%%%%%%%%%

\begin{theorem}\label{thm: full classification of fields of covariances on fNCT}		
A functor $\mathfrak{C}\colon \mathsf{fNCT}\rightsquigarrow\mathsf{Hilb}$ is a continuous field of covariances as in definitions \ref{defn: field of covariances} and \ref{defn: continuity for fields of covariances} \textbf{if and only if} $\mathfrak{C}_{0}(\mathscr{A},\sigma)\equiv \mathcal{H}_{\sigma}^{\mathfrak{C}}$ is the \textup{GNS} Hilbert space of $\sigma$ endowed with the alternative Hilbert product determined by the sesquilinear form 
\begin{equation}
 \mathfrak{C}_{\sigma}(\xi,\eta)=\beta\langle\xi\mid\eta\rangle_{\sigma} + (\alpha -\beta)\langle\xi\mid \xi_{\mathbb{I}}^{\sigma}\rangle_{\sigma}\langle\xi^{\sigma}_{\mathbb{I}}\mid\eta\rangle_{\sigma},
\end{equation}
where $\alpha,\beta\in(0,\infty)$. 
%%%%%%%%%%%%%%%%%%%%%%%%%%%%
\end{theorem}

\begin{proof}
The   \textbf{if} part amounts to a direct check, and the \textbf{only if} part follows from proposition
\ref{prop: classification of invariant covariances for any algebra and tracial states}.
%%%%%%%%%%%%%%%%%%%%%%%%%%%%%%%%%%%%%%%%%%%%%%

\end{proof}

\begin{remark}
Note that the choice $\alpha\neq\beta$ only affects the subspace generated by the identity at each \textup{GNS} Hilbert space.
%%%%%%%%%%%%%%%%%%%%%%%%%%%%%%%%%%%%%%%%%%
Moreover, when $\alpha=\beta=1$ and we focus on the subcategory $\mathsf{fCP}$, the covariance $\mathfrak{C}_{\rho}$ reduces to the complex statistical covariance in equation \eqref{eqn: statistical covariance} in the sense that
\begin{equation} 
\mathrm{Cov}_{\rho}(X,Y)=\mathfrak{C}_{\rho}(\mathbf{P}(X),\mathbf{P}(Y)),
\end{equation}
where $\mathbf{P}$ is the $\mathfrak{C}_{\rho}$-orthogonal projection on the orthogonal complement of the vector subspace generated by the identity.
%%%%%%%%%%%%%%%%%%%%%%%%%%

\end{remark}

\subsection{Faithful non-tracial states}\label{subsec: faithful states}

Let us recall that the centralizer of $\rho$ is the unital subalgebra of $\mathscr{A}$ given by
\begin{equation}\label{eqn: centralizer of a state}
\mathscr{M}_{\rho}:=\left\{\mathbf{a}\in\mathscr{A}\mid\;\rho(\mathbf{ab})=\rho(\mathbf{ba})\;\forall\mathbf{b}\in\mathscr{A}\right\}.
\end{equation}
%%%%%%%%%%%%%%%%%%%%%%%%
When $\rho$ is faithful, the centralizer can equivalently be expressed in terms of the modular flow as
\begin{equation}\label{eqn: centralizer of state modular automorphism}
\mathscr{M}_{\rho}= \left\{\mathbf{a}\in\mathscr{A}\mid\;\Phi_{t}^{\rho}(\mathbf{a})=\mathbf{a}\;\forall t\in\mathbb{R}\right\},
\end{equation}
where $\Phi_{t}^{\rho}$ is the modular flow of $\rho$ as in equation \eqref{eqn: modular flow}.
%%%%%%%%%%%%%%%%%%%%%%%%

The \textup{GNS} Hilbert space $\mathcal{H}_{\rho}$ can be decomposed as the direct sum
\begin{equation}\label{eqn: decomposition of GNS Hilbert space at faithful state on A}
\mathcal{H}_{\rho}\cong\mathscr{M}_{\rho}\oplus \mathscr{K}_{\rho},
\end{equation}
where $\mathscr{M}_{\rho}$ is the centralizer of $\rho$, and $\mathscr{K}_{\rho}$ is its orthogonal complement with respect to the \textup{GNS} Hilbert product.
%%%%%%%%%%%%%%%%%%%%%%%%%%%%%
The  following lemma shows that the covariance operator maps each summand in equation \eqref{eqn: decomposition of GNS Hilbert space at faithful state on A} into itself.
%%%%%%%%%%%%%%%%%%%

\begin{lemma}\label{lem: commutation}
Let $\mathfrak{C}$ be a covariance field on $\mathsf{fNCP}$ as in definition \ref{defn: field of covariances}.
%%%%%%%%%%%%%%%%%%%%
Let $\rho$ be a faithful state on $\mathscr{A}$. 
%%%%%%%%%%%%%%%%
Then the covariance operator $\mathbf{T}_\rho$ determined by $\mathfrak{C}_{\rho}$ commutes with the modular operator $\Delta_\rho$. 
%%%%%%%%%%%%%%%%%%%%%%%
In particular, $\mathbf{T}_\rho$  maps both subspaces $\mathscr{M}_{\rho}$ and $\mathscr{K}_{\rho}$ in decomposition in equation \eqref{eqn: decomposition of GNS Hilbert space at faithful state on A} into themselves.
\end{lemma}

\begin{proof}
Let $\mathrm{Aut}_{\rho}(\mathscr{A})$ denote the group of automorphisms of $\mathscr{A}$ preserving $\rho$.
%%%%%%%%%%%%%%%%%%%%%%%%
Because of equations \eqref{eqn: categorical invariance condition} and \eqref{eqn: covariance operator commutes with symmetries}, it holds
\begin{equation}\label{eqn: covariance operator commutes with rho-preserving automorphisms}
[\mathbf{T}_{\rho},\tilde{\Phi}]=\mathbf{0}
\end{equation}
for all unitary operators $\tilde{\Phi}$ on $\mathcal{H}_{\rho}$ induced by automorphisms $\Phi\in\mathrm{Aut}_{\rho}(\mathscr{A})$.
%%%%%%%%%%%%%%%%%%%%%%%%%%%%%%%%%%%%
In particular, the covariance operator $\mathbf{T}_{\rho}$ commutes with $\Delta_{\rho}$ because this operator generates the modular flow $\Phi_{t}^{\rho}$ of $\rho$ which is in $\mathrm{Aut}_{\rho}(\mathscr{A})$ for all $t\in\mathbb{R}$.
%%%%%%%%%%%%%%%%%%%%%%%%%%%%%%%%
Consequently, we can simultaneously diagonalize $\mathbf{T}_{\rho}$ and $\Delta_{\rho}$.
%%%%%%%%%%%%%%%%%%%%%%%%%%%%%%
Moreover, $\mathbf{T}_{\rho}$ cannot intertwine subspaces of $\mathcal{H}_{\rho}$ associated with different eigenvalues of $\Delta_{\rho}$, which means that $\mathbf{T}_{\rho}$ maps each summand  in equation \eqref{eqn: decomposition of GNS Hilbert space at faithful state on A} into itself.
%%%%%%%%%%%%%%%%%%%%%%%%%%%%%%%%%%%%%%%%%%
\end{proof}

Because of theorem  \ref{thm: full classification of fields of covariances on fNCT}, we only need to understand the cases where $\rho$ is not a tracial state.
%%%%%%%%%%%%%%%%%%%%%%%%%%%%
Because of lemma \ref{lem: commutation}, we need to characterize how the covariance operator $\mathbf{T}_{\rho}$ acts on  $\mathscr{M}_{\rho}\subset\mathcal{H}_{\rho}$  and  $\mathscr{K}_{\rho}\subset\mathcal{H}_{\rho}$ separately. 
%%%%%%%%%%%%%%%%%%%%%%%%%%%%%%%
We start with the centralizer $\mathscr{M}_{\rho}$.
%%%%%%%%%%%%%%%%%%%%%%

\begin{proposition}\label{prop: covariance operator on the centralizer}
Let $\mathfrak{C}\colon \mathsf{fNCP}\rightsquigarrow\mathsf{Hilb}$ be a continuous field of covariances as in definitions \ref{defn: field of covariances} and \ref{defn: continuity for fields of covariances}, and let  $\alpha,\beta>0$ be the constants appearing in theorem \ref{thm: full classification of fields of covariances on fNCT}.
%%%%%%%%%%%%%%%%%%%%%%%%%%
If $(\mathscr{A},\rho)\in\mathsf{fNCP}_{0}$ with $\rho$ a faithful state on $\mathscr{A}$, then the covariance operator $\mathbf{T}_{\rho}$ restricted to $\mathscr{M}_{\rho}$  reads
\begin{equation}\label{eqn: covariance operator on the centralizer}
 \mathbf{T}_{\rho}|_{\mathscr{M}_{\rho}}=\beta\mathbb{I}_{\mathscr{M}_{\rho}} + (\alpha-\beta)|\xi^{\rho}_{\mathbb{I}}\rangle_{\rho}\langle\xi^{\rho}_{\mathbb{I}}|.
\end{equation}
%%%%%%%%%%%%%%%%%%%%%%%%%%%%%%%%%%%%%%

\end{proposition}
\begin{proof}

Since $\mathscr{M}_{\rho}$ is a subalgebra of $\mathscr{A}$ that is invariant under the modular flow of $\rho$, there is a conditional expectation $\mathcal{E}\colon \mathscr{A}\rightarrow\mathscr{M}_{\rho}$ such that $\mathcal{E}\circ \mathrm{i}=\mathrm{id}_{\mathscr{M}_{\rho}}$, where $\mathrm{i}\colon \mathscr{M}_{\rho}\rightarrow\mathscr{A}$ is the subset inclusion.
%%%%%%%%%%%%%%
In particular, $\mathrm{i}^{*}(\rho)\equiv\sigma$ is a faithful tracial state on $\mathscr{M}_{\rho}$ such that $\mathcal{E}^{*}(\sigma)=\rho$, and thus the invariance condition in equation \eqref{eqn: categorical invariance condition}  together with theorem \ref{thm: full classification of fields of covariances on fNCT} imply that the restriction of $\mathbf{T}_{\rho}$ to $\mathscr{M}_{\rho}$ is as in equation \eqref{eqn: covariance operator on the centralizer} as required.
%%%%%%%%%%%%%%%%%%%%%%%%%%%%%%%%%%%
\end{proof}

To understand what happens on $\mathscr{K}_{\rho}$, we start considering the case in which the restriction of the modular operator there has simple spectrum (\textit{i.e.}, all the eigenvalues are different).
%%%%%%%%%%%%%%%%%%%%%%

\begin{proposition}\label{prop: fields of covariances on faithful states in fNCP with non-degenerate modular operator}
Let $\mathfrak{C}\colon \mathsf{fNCP}\rightsquigarrow\mathsf{Hilb}$ be a continuous field of covariances as in definitions \ref{defn: field of covariances} and \ref{defn: continuity for fields of covariances}, and let $\alpha,\beta>0$ be the constants appearing in proposition \ref{prop: covariance operator on the centralizer}.
%%%%%%%%%%%%%%%%%%%
If $(\mathscr{A},\rho)\in\mathsf{fNCP}_{0}$ with $\rho$ a faithful state on $\mathscr{A}$ such that its modular operator $\Delta_{\rho}$ restricted to the orthogonal complement $\mathscr{K}_{\rho}$ of the centralizer $\mathscr{M}_{\rho}$ in the \textup{GNS} Hilbert space $\mathcal{H}_{\rho}$ has simple spectrum, then there exists a $\rho$-dependent function $F_{\rho}$ on $\mathrm{sp}(\Delta_{\rho}|_{\mathscr{K}_{\rho}})\cup\{1\}$, with values in $(0,\infty)$ and with $F_{\rho}(1)=\beta$, such that
\begin{equation}\label{eqn: fields of covariances on faithful fNCP}
\mathfrak{C}_{\rho}(\xi,\eta)=\langle\xi\mid F_{\rho}(\Delta_{\rho})(\eta)\rangle_{\rho} + (\alpha -F_{\rho}(1))\langle\xi\mid\xi^{\rho}_{\mathbb{I}}\rangle_{\rho}\langle\xi_{\mathbb{I}}^{\rho}\mid\eta\rangle_{\rho}
\end{equation}
for all $\xi,\eta\in\mathcal{H}_{\rho}$.
%%%%%%%%%%%%%%%%%%%%%%%%%%%%%%%%%%%%%%
\end{proposition}

\begin{proof}
Proposition \ref{prop: covariance operator on the centralizer} gives the explicit form on $\mathscr{M}_{\rho}$.
%%%%%%%%%%%%%%%%%%%%%
On the other hand, since $\Delta_{\rho}|_{\mathscr{K}_{\rho}}$  has a simple spectrum by assumption, it generates a \textit{maximally Abelian subalgebra} (\textit{masa}) of $\mathcal{B}(\mathscr{K}_{\rho})$, that is, a $C^*$-subalgebra $\mathscr{C}$ that is equal to its own commutant $\mathscr{C}'$ (\textit{i.e.}, the subset of  all elements in $\mathcal{B}(\mathscr{K}_{\rho})$ commuting with elements in $\mathscr{C}$).
%%%%%%%%%%%%%
Consequently, the fact that $\mathbf{T}_{\rho}|_{ \mathscr{K}_{\rho}}$ commutes with $\Delta_{\rho}|_{\mathscr{K}_{\rho}}$ because of lemma \ref{lem: commutation} implies that $\mathbf{T}_{\rho}|_{\mathscr{K}_{\rho}}$ is an \textit{a priori} unknown function $F_{\rho}$ of $\Delta_{\rho}|_{\mathscr{K}_{\rho}}$, and we conclude that 
 \begin{equation}\label{eqn: fields of covariances when Deltarho generates a masa}
\mathfrak{C}_{\rho}(\xi,\eta)=\langle\xi\mid F_{\rho}(\Delta_{\rho})(\eta)\rangle_{\rho} + (\alpha -F_{\rho}(1))\langle\xi\mid\xi_{\mathbb{I}}^{\rho}\rangle_{\rho}\langle\xi^{\rho}_{\mathbb{I}}\mid\eta\rangle_{\rho}
\end{equation}
with $F_{\rho}(1)=\beta$.
%%%%%%%%%%%%%%%%%%%%%%%%%%
The positivity of $F_{\rho}$ follows from the positivity of $\mathfrak{C}$.
%%%%%%%%%%%%%%%%%%%%%%%%%%%%%
\end{proof} 

The algebras $\mathscr{A}=M_{2}(\mathbb{C})$ and $\mathscr{A}=M_{2}(\mathbb{C})\oplus \mathbb{C}$ are instrumental in providing a characterization of continuous covariances for faithful states on arbitrary finite-dimensional algebras.
%%%%%%%%%%%%%%%%%%%%%%%%%
Proposition \ref{prop: fields of covariances on faithful states in fNCP with non-degenerate modular operator} allows a complete characterization in the case $\mathscr{A}=M_{2}(\mathbb{C})$  however, the case of $M_2(\mathbb{C}) \oplus \mathbb{C}$ falls outside the scope of this proposition.
%%%%%%%%%%%%%%%%%%%%%%%%%
We study both cases in detail in the next proposition.

\begin{proposition}\label{prop: covariance at qubit and augmented qubit}
Let $\mathfrak{C}\colon \mathsf{fNCP}\rightsquigarrow\mathsf{Hilb}$ be a continuous field of covariances as in Definitions \ref{defn: field of covariances} and \ref{defn: continuity for fields of covariances}, and let $\alpha,\beta>0$ be the constants appearing in proposition \ref{prop: fields of covariances on faithful states in fNCP with non-degenerate modular operator}.

\begin{enumerate}

\item There exists a function $F\colon(0,\infty)\rightarrow(0,\infty)$, with $F(1)=\beta$ and depending only on $\mathfrak{C}$, such that, for every faithful state $\rho$ on $M_{2}(\mathbb{C})$, it holds
\begin{equation}\label{eqn: fields of covariances on faithful qubit}
\mathfrak{C}_{\rho}(\xi,\eta)
=
\left\langle
\xi
\mid
F(\Delta_{\rho})(\eta)
\right\rangle_{\rho}
+
(\alpha-F(1))
\left\langle
\xi
\mid
\xi_{\mathbb{I}}^{\rho}
\right\rangle_{\rho}
\left\langle
\xi_{\mathbb{I}}^{\rho}
\mid
\eta
\right\rangle_{\rho}
\end{equation}
for all $\xi,\eta\in\mathcal{H}_{\rho}$.

\item Every faithful state $\rho$ on $M_{2}(\mathbb{C})\oplus\mathbb{C}$ can be written uniquely in the form
\begin{equation}\label{eqn: faithful state on M2(C)+C}
\rho(\mathbf{a},z)
=
\lambda\widetilde{\rho}(\mathbf{a})
+
(1-\lambda)z,
\end{equation}
where $\lambda\in(0,1)$ and $\widetilde{\rho}$ is a faithful state on $M_{2}(\mathbb{C})$.
%%%%%%%%%%%%%%%%%%%%%%%%%%%%%%%%%%%%
Moreover, there exists a family of functions
\begin{equation}
\left\{
F_{\lambda}\colon(0,\infty)\rightarrow (0,\infty)
\right\}_{\lambda\in(0,1)}
\end{equation}
depending only on $\mathfrak{C}$, such that for every faithful state $\rho$ written as in equation \eqref{eqn: faithful state on M2(C)+C}, it holds
\begin{equation}\label{eqn: fields of covariances on faithful M2C+C}
\mathfrak{C}_{\rho}(\xi,\eta)
=
\left\langle
\xi
\mid
F_{\lambda}(\Delta_{\rho})(\eta)
\right\rangle_{\rho}
+
(\alpha-F_{\lambda}(1))
\left\langle
\xi
\mid
\xi_{\mathbb{I}}^{\rho}
\right\rangle_{\rho}
\left\langle
\xi_{\mathbb{I}}^{\rho}
\mid
\eta
\right\rangle_{\rho}
\end{equation}
for all $\xi,\eta\in\mathcal{H}_{\rho}$.
%%%%%%%%%%%%%%%%%%%%%%
\end{enumerate}
\end{proposition}

\begin{proof}
We first consider the case of $M_{2}(\mathbb{C})$. 
%%%%%%%%%%%%%%%%%%%%%%%
Writing
\begin{equation}
\mathbf{e}_{jk}
=
|j\rangle\langle k|,
\qquad
j,k\in\{1,2\},
\end{equation}
for every $t\in(0,\infty)$, let $\rho_{t}$ be the faithful state on $M_{2}(\mathbb{C})$ whose density operator is
\begin{equation}\label{eqn: faithful state qubit}
\varrho_{t}
=
\frac{t}{1+t}\mathbf{e}_{11}
+
\frac{1}{1+t}\mathbf{e}_{22}.
\end{equation}
%%%%%%%%%%%%%%%%%%%%%%%%%
If $t\neq1$, the orthogonal complement $\mathscr{K}_{\rho_{t}}$ of the centralizer is spanned by $\xi_{\mathbf{e}_{12}}^{\rho_{t}}$ and $\xi_{\mathbf{e}_{21}}^{\rho_{t}}$, and the modular operator satisfies
\begin{equation}
\Delta_{\rho_{t}}
\left(
\xi_{\mathbf{e}_{12}}^{\rho_{t}}
\right)
=
t
\xi_{\mathbf{e}_{12}}^{\rho_{t}}
\end{equation}
and
\begin{equation}
\Delta_{\rho_{t}}
\left(
\xi_{\mathbf{e}_{21}}^{\rho_{t}}
\right)
=
t^{-1}
\xi_{\mathbf{e}_{21}}^{\rho_{t}}.
\end{equation}
%%%%%%%%%%%%%%%%%%
Consequently, the restriction of $\Delta_{\rho_{t}}$ to $\mathscr{K}_{\rho_{t}}$ has non-degenerate spectrum. 
%%%%%%%%%%%%%%%%%%%
Proposition \ref{prop: fields of covariances on faithful states in fNCP with non-degenerate modular operator} therefore implies that
\begin{equation}
\mathbf{T}_{\rho_{t}}
\left(
\xi_{\mathbf{e}_{12}}^{\rho_{t}}
\right)
=
F_{\rho_{t}}(t)
\xi_{\mathbf{e}_{12}}^{\rho_{t}}.
\end{equation}
%%%%%%%%%%%%%%%%%%%%%%%%%%%%%
We define
\begin{equation}\label{eqn: definition universal qubit F}
F(t)
=
F_{\rho_{t}}(t),
\qquad
t\neq1,
\end{equation}
and set $F(1)=\beta$.
%%%%%%%%%%%%%%%%%%%%%
The positivity of $F$ follows from the positivity of $\mathfrak{C}$ and $\mathbf{T}_{\rho}$.
%%%%%%%%%%%%%%%%%%%%%%%%%%%%
Let now $\rho$ be an arbitrary faithful nontracial state on $M_{2}(\mathbb{C})$, and let
\begin{equation}
\varrho
=
p_{1}|v_{1}\rangle\langle v_{1}|
+
p_{2}|v_{2}\rangle\langle v_{2}|
\end{equation}
be the spectral decomposition of its density operator. Set
\begin{equation}
t
=
\frac{p_{1}}{p_{2}}.
\end{equation}
Since $p_{1}+p_{2}=1$, one has
\begin{equation}
p_{1}
=
\frac{t}{1+t},
\qquad
p_{2}
=
\frac{1}{1+t}.
\end{equation}
%%%%%%%%%%%%%%%%%%%%%%%%%%%%%%%%
Let $\mathbf{u}$ be a unitary operator such that
\begin{equation}
\mathbf{u}|j\rangle
=
|v_{j}\rangle,
\qquad
j=1,2,
\end{equation}
and consider the automorphism
\begin{equation}
\Phi(\mathbf{a})
=
\mathbf{u}\mathbf{a}\mathbf{u}^{\dagger}
\end{equation}
of $M_{2}(\mathbb{C})$. A direct computation gives
\begin{equation}
\Phi^{*}\rho
=
\rho_{t}.
\end{equation}
The equivariance of the covariance operators therefore implies that the eigenvalue of $\mathbf{T}_{\rho}$ corresponding to $\xi_{|v_{1}\rangle\langle v_{2}|}^{\rho}$ is equal to $F(t)$.
%%%%%%%%%%%%%%%%%%%%%%%%%%%%%
Applying the same argument after interchanging the two eigenvectors shows that the eigenvalue of $\mathbf{T}_{\rho}$ corresponding to $\xi_{|v_{2}\rangle\langle v_{1}|}^{\rho}$ is equal to $F(t^{-1})$.
%%%%%%%%%%%%%%%%%%%%%%%%%%%%%%%%
Together with Proposition \ref{prop: covariance operator on the centralizer}, this proves equation \eqref{eqn: fields of covariances on faithful qubit} for every faithful nontracial state on $M_{2}(\mathbb{C})$.
%%%%%%%%%%%%%%%%%%%%%%%%%%%%%%
If $\rho$ is tracial, then $\Delta_{\rho}=\mathbb{I}$, and equation \eqref{eqn: fields of covariances on faithful qubit} follows directly from theorem \ref{thm: full classification of fields of covariances on fNCT}, because $F(1)=\beta$.
%%%%%%%%%%%%%%%%%%%%%%%%%%

\vspace{0.3cm}

We next consider the case of $M_{2}(\mathbb{C})\oplus\mathbb{C}$.
%%%%%%%%%%%%%%%%%%%%%%%%%%%%%
Let $\rho$ be a faithful state on $M_{2}(\mathbb{C})\oplus\mathbb{C}$, and set $\lambda=\rho(\mathbb{I},0)$.
%%%%%%%%%%%%%%%%%%%%%%%%%%%
Since $(\mathbb{I},0)$ and $(0,1)$ are nonzero positive elements and $(\mathbb{I},0)+(0,1)=(\mathbb{I},1)$, the faithfulness and normalization of $\rho$ imply $0<\lambda<1$ and $\rho(0,1)=1-\lambda$.
%%%%%%%%%%%%%%%%%%%%
The map $\varphi(\mathbf{a})=\rho(\mathbf{a},0)$ is a positive linear functional on $M_{2}(\mathbb{C})$ satisfying $\varphi(\mathbb{I})=\lambda$.
%%%%%%%%%%%%%%%%%%%%%%%%%%%%%%%%%%%%%%
Consequently, the functional
\begin{equation}\label{eqn: reduced qubit from augmented qubit state}
\widetilde{\rho}(\mathbf{a})
=
\frac{1}{\lambda}\rho(\mathbf{a},0)
\end{equation}
is a state on $M_{2}(\mathbb{C})$, and
\begin{equation}
\rho(\mathbf{a},z)
=
\lambda\widetilde{\rho}(\mathbf{a})
+
(1-\lambda)z.
\end{equation}
%%%%%%%%%%%%%%%%%%%%%%%%%%%
If $\mathbf{a}\in M_{2}(\mathbb{C})$ is positive and $\widetilde{\rho}(\mathbf{a})=0$, then
\begin{equation}
\rho(\mathbf{a},0)
=
\lambda\widetilde{\rho}(\mathbf{a})
=
0.
\end{equation}
The faithfulness of $\rho$ implies $\mathbf{a}=0$, and hence $\widetilde{\rho}$ is faithful.
%%%%%%%%%%%%%%%%%%%%%%%%%%%%%%%%%%%%%%%
Conversely, if $\lambda\in(0,1)$ and $\widetilde{\rho}$ is a faithful state on $M_{2}(\mathbb{C})$, then $\rho(\mathbf{a},z)=\lambda\widetilde{\rho}(\mathbf{a})+(1-\lambda)z$ defines a faithful state on $M_{2}(\mathbb{C})\oplus\mathbb{C}$. 
%%%%%%%%%%%%%%%%%
Indeed, if $(\mathbf{a},z)\geq0$ and $\rho(\mathbf{a},z)=0$, then $\lambda\widetilde{\rho}(\mathbf{a})=0$ and $(1-\lambda)z=0$.
%%%%%%%%%%%%%
It follows that $\mathbf{a}=0$ and $z=0$.
%%%%%%%%%%%%%%%%%%%%%%%%%%%
The uniqueness of $\lambda$ and $\widetilde{\rho}$ follows from $\lambda=\rho(\mathbb{I},0)$ and from equation \eqref{eqn: reduced qubit from augmented qubit state}.
%%%%%%%%%%%%%%%%%%%%%%%%%%%%%%

Fix now $\lambda\in(0,1)$. 
%%%%%%%%%%%%%%%%%%%%%%%%%%%%%%%%
For every $t\in(0,\infty)$, define the faithful state $\rho_{\lambda,t}$ on $M_{2}(\mathbb{C})\oplus\mathbb{C}$ by
\begin{equation}
\rho_{\lambda,t}(\mathbf{a},z)
=
\lambda\rho_{t}(\mathbf{a})
+
(1-\lambda)z.
\end{equation}
%%%%%%%%%%%%%%%%%%%%%%
Suppose first that $t\neq1$. 
%%%%%%%%%%%%%%%%%
The orthogonal complement $\mathscr{K}_{\rho_{\lambda,t}}$ of the centralizer is spanned by $\xi_{(\mathbf{e}_{12},0)}^{\rho_{\lambda,t}}$ and $\xi_{(\mathbf{e}_{21},0)}^{\rho_{\lambda,t}}$, and the modular operator satisfies
\begin{equation}
\Delta_{\rho_{\lambda,t}}
\left(
\xi_{(\mathbf{e}_{12},0)}^{\rho_{\lambda,t}}
\right)
=
t
\xi_{(\mathbf{e}_{12},0)}^{\rho_{\lambda,t}}
\end{equation}
and
\begin{equation}
\Delta_{\rho_{\lambda,t}}
\left(
\xi_{(\mathbf{e}_{21},0)}^{\rho_{\lambda,t}}
\right)
=
t^{-1}
\xi_{(\mathbf{e}_{21},0)}^{\rho_{\lambda,t}}.
\end{equation}
%%%%%%%%%%%%%%%%%%%%%%%%%%%%%%%%%
Thus the restriction of $\Delta_{\rho_{\lambda,t}}$ to $\mathscr{K}_{\rho_{\lambda,t}}$ has non-degenerate spectrum. 
%%%%%%%%%%%%%%%%%%%%%%
Proposition \ref{prop: fields of covariances on faithful states in fNCP with non-degenerate modular operator} therefore implies that
\begin{equation}
\mathbf{T}_{\rho_{\lambda,t}}
\left(
\xi_{(\mathbf{e}_{12},0)}^{\rho_{\lambda,t}}
\right)
=
F_{\rho_{\lambda,t}}(t)
\xi_{(\mathbf{e}_{12},0)}^{\rho_{\lambda,t}},
\end{equation}
and we define
\begin{equation}\label{eqn: definition of F lambda}
F_{\lambda}(t)
=
F_{\rho_{\lambda,t}}(t),
\qquad
t\neq1,
\end{equation}
and set $F_{\lambda}(1)=\beta$.
%%%%%%%%%%%%%%
The positivity of $F_{\lambda}$ follows from the positivity of $\mathfrak{C}$ and $\mathbf{T}_{\rho}$.
%%%%%%%%%%%%%%%%%%%%%%%%%%%%
Let now $\rho$ be an arbitrary faithful state on $M_{2}(\mathbb{C})\oplus\mathbb{C}$ with fixed weight $\lambda$. 
%%%%%%%%%%%%%%%%%%%%
By the first part of the proof, it is uniquely of the form in equation \eqref{eqn: faithful state on M2(C)+C} for a faithful state $\widetilde{\rho}$ on $M_{2}(\mathbb{C})$.
%%%%%%%%%%%%%%%%%%%%%%%%%%%%%%%%
Assume that $\widetilde{\rho}$ is nontracial, and let
\begin{equation}
\widetilde{\varrho}
=
p_{1}|v_{1}\rangle\langle v_{1}|
+
p_{2}|v_{2}\rangle\langle v_{2}|
\end{equation}
be the spectral decomposition of its density operator. %%%%%%%%%%%%%%%%
Set $t=\frac{p_{1}}{p_{2}}$.
%%%%%%%%%%%%%%%%%%%%%%%%%%%
Since $p_{1}+p_{2}=1$, one has
\begin{equation}
p_{1}
=
\frac{t}{1+t},
\qquad
p_{2}
=
\frac{1}{1+t}.
\end{equation}
%%%%%%%%%%%%%%%%%%%%%%%%%%%%%%
Let $\mathbf{u}$ be a unitary operator satisfying
\begin{equation}
\mathbf{u}|j\rangle
=
|v_{j}\rangle,
\qquad
j=1,2,
\end{equation}
and consider the automorphism
\begin{equation}
\Phi(\mathbf{a},z)
=
(\mathbf{u}\mathbf{a}\mathbf{u}^{\dagger},z)
\end{equation}
of $M_{2}(\mathbb{C})\oplus\mathbb{C}$. 
%%%%%%%%%%%%%%%%%%%%%%%%%%%%%
A direct computation gives
\begin{equation}
\Phi^{*}\rho
=
\rho_{\lambda,t}.
\end{equation}
%%%%%%%%%%%%%%%%%%%%%%%%
The equivariance of the covariance operators therefore implies that the eigenvalue of $\mathbf{T}_{\rho}$ corresponding to $\xi_{(|v_{1}\rangle\langle v_{2}|,0)}^{\rho}$ is equal to $F_{\lambda}(t)$.
%%%%%%%%%%%%%%%%%%%%%%%%%%%%
Applying the same argument after interchanging the two eigenvectors shows that the eigenvalue of $\mathbf{T}_{\rho}$ corresponding to $\xi_{(|v_{2}\rangle\langle v_{1}|,0)}^{\rho}$ is equal to $F_{\lambda}(t^{-1})$.
%%%%%%%%%%%%%%%%%%%%%%%%%%%%%%%%%%%
Together with Proposition \ref{prop: covariance operator on the centralizer}, this proves equation \eqref{eqn: fields of covariances on faithful M2C+C} whenever $\widetilde{\rho}$ is nontracial.
%%%%%%%%%%%%%%%%%%%%%%%%%%%%%%%%
If $\widetilde{\rho}$ is tracial, then $\rho$ is tracial and $\Delta_{\rho}=\mathbb{I}$.
%%%%%%%%%%%%%%%%%%
In this case, equation \eqref{eqn: fields of covariances on faithful M2C+C} follows from theorem \ref{thm: full classification of fields of covariances on fNCT}, because $F_{\lambda}(1)=\beta$.
\end{proof}

We are now ready to give a full characterization of continuous field of covariances at all faithful states on arbitrary finite-dimensional algebras.
%%%%%%%%%%%%%%%%%%%%%%%%%%%%%%%%%%%%%
 
\begin{proposition}\label{prop: fields of covariances on faithful states in fNCP}
Let $\mathfrak{C}\colon \mathsf{fNCP}\rightsquigarrow\mathsf{Hilb}$ be a continuous field of covariances as in definitions \ref{defn: field of covariances} and \ref{defn: continuity for fields of covariances}, let $\alpha,\beta>0$ be the constants appearing in  proposition \ref{prop: covariance operator on the centralizer}, and let $F$ and the family $\{F_{\lambda}\}_{\lambda\in(0,1)}$ be as in
proposition \ref{prop: covariance at qubit and augmented qubit}.
%%%%%%%%%%%%%%%%%%
If $(\mathscr{A},\rho)\in\mathsf{fNCP}_{0}$ with $\rho$ a faithful state on $\mathscr{A}$, then it holds
\begin{equation}\label{eqn: fields of covariances on faithful fNCP 2}
\mathfrak{C}_{\rho}(\xi,\eta)=\langle\xi\mid F(\Delta_{\rho})(\eta)\rangle_{\rho} + (\alpha -F(1))\langle\xi\mid\xi_{\mathbb{I}}^{\rho}\rangle_{\rho}\langle\xi_{\mathbb{I}}^{\rho}\mid\eta\rangle_{\rho},
\end{equation}
for all $\xi,\eta\in\mathcal{H}_{\rho}$.
%%%%%%%%%%%%%%%%%%%%%%%%%%%%%%%%%%%%%
\end{proposition}

\begin{proof}
Proposition \ref{prop: covariance operator on the centralizer} determines the restriction of $\mathbf{T}_{\rho}$ to $\mathscr{M}_{\rho}$.
%%%%%%%%%%%%%%%%%%%%%%%%%%%%%%%%
Since $F(1)=\beta$ by proposition \ref{prop: covariance at qubit and augmented qubit}, it remains to prove that
\begin{equation}
\mathbf{T}_{\rho}|_{\mathscr{K}_{\rho}}
=
F\left(
\Delta_{\rho}|_{\mathscr{K}_{\rho}}
\right).
\end{equation}
%%%%%%%%%%%%%%%%%%%%%%%%%%%%

We first consider the case $\mathscr{A}=M_{N}(\mathbb{C})$ with $N>2$. 
%%%%%%%%%%%%%%%%%%%%%%%
Let $\varrho$ be the
density operator associated with $\rho$, and choose an orthonormal
basis $\{|j\rangle\}_{j=1}^{N}$ in which
\begin{equation}
\varrho
=
\sum_{j=1}^{N}
p_{j}\mathbf{e}_{jj},
\qquad
p_{j}>0,
\qquad
\sum_{j=1}^{N}p_{j}=1,
\end{equation}
where $\mathbf{e}_{jk}:=|j\rangle\langle k|$. 
%%%%%%%%%%%%%%%%
We also write $\xi_{jk}^{\rho}:=\xi_{\mathbf{e}_{jk}}^{\rho}$.
%%%%%%%%%%%%%%%%%%%%%%%%%%%
The modular operator acts as
\begin{equation}\label{eqn: modular operator on matrix units faithful}
\Delta_{\rho}
\left(
\xi_{jk}^{\rho}
\right)
=
\frac{p_{j}}{p_{k}}
\xi_{jk}^{\rho}.
\end{equation}
%%%%%%%%%%%%%%%%%%%%%%%%%%%%%%
Consequently, $\mathscr{K}_{\rho}$ is spanned by the vectors
$\xi_{jk}^{\rho}$ for which $p_{j}\neq p_{k}$.
%%%%%%%%%%%%%%%%%
We show that each of these vectors is an eigenvector of
$\mathbf{T}_{\rho}$. 
%%%%%%%%%%%%%%%%%%%%%%%%%%%%%
For $\mathbf{z}=(z_{1},\ldots,z_{N})\in\mathbb{T}^{N}$, define the diagonal unitary and the corresponding automorphism by
\begin{equation}
\mathbf{u}_{\mathbf{z}}
:=
\sum_{j=1}^{N}
z_{j}\mathbf{e}_{jj},
\qquad
\Phi_{\mathbf{z}}(\mathbf{a})
:=
\mathbf{u}_{\mathbf{z}}^{\dagger}
\mathbf{a}
\mathbf{u}_{\mathbf{z}}.
\end{equation}
Since $\mathbf{u}_{\mathbf{z}}$ commutes with $\varrho$,
$\Phi_{\mathbf{z}}$ preserves $\rho$. Its induced unitary on
$\mathcal{H}_{\rho}$ satisfies
\begin{equation}
\widetilde{\Phi}_{\mathbf{z}}
\left(
\xi_{jk}^{\rho}
\right)
=
\overline{z}_{j}z_{k}
\xi_{jk}^{\rho}.
\end{equation}
For $j\neq k$, the characters
$\mathbf{z}\mapsto\overline{z}_{j}z_{k}$ are distinct for distinct
ordered pairs $(j,k)$. Equation
\eqref{eqn: covariance operator commutes with symmetries} therefore
implies that every one-dimensional subspace
$\mathbb{C}\xi_{jk}^{\rho}$ with $j\neq k$ is invariant under
$\mathbf{T}_{\rho}$. In particular, by positivity of $\mathbf{T}_\rho$ there exist positive numbers
$c_{jk}^{\rho}$ such that
\begin{equation}\label{eqn: covariance operator diagonal on matrix units}
\mathbf{T}_{\rho}
\left(
\xi_{jk}^{\rho}
\right)
=
c_{jk}^{\rho}
\xi_{jk}^{\rho}
\end{equation}
whenever $p_{j}\neq p_{k}$.
%%%%%%%%%%%%%%%%%%%%%%%%%%%%%%

We now determine $c_{jk}^{\rho}$ by reducing each ordered pair of eigenvectors to the algebra
$M_{2}(\mathbb{C})\oplus\mathbb{C}$. 
%%%%%%%%%%%%%%%%%%%%%%%%%%%%
Fix $r\neq s$ such that
$p_{r}\neq p_{s}$, and define the unital $*$-homomorphism $\mathrm{i}_{rs}\colon
M_{2}(\mathbb{C})\oplus\mathbb{C}\to
M_{N}(\mathbb{C})$ as
\begin{equation}
\mathrm{i}_{rs}(A,z)
:=
A_{11}\mathbf{e}_{rr}
+
A_{12}\mathbf{e}_{rs}
+
A_{21}\mathbf{e}_{sr}
+
A_{22}\mathbf{e}_{ss}
+
z\sum_{j\neq r,s}\mathbf{e}_{jj}.
\end{equation}
%%%%%%%%%%%%%%%%%%%%%%%%%%%%%
The pullback of $\rho$ through $\mathrm{i}_{rs}$ is the state $\sigma_{rs}:=\mathrm{i}_{rs}^{*}\rho$ given by
\begin{equation}
\sigma_{rs}(A,z)
=
p_{r}A_{11}
+
p_{s}A_{22}
+
(1-p_{r}-p_{s})z.
\end{equation}
%%%%%%%%%%%%%%%
Setting $\lambda_{rs}:=p_{r}+p_{s}$ and
\begin{equation}
\widetilde{\sigma}_{rs}(A)
:=
\frac{p_{r}}{p_{r}+p_{s}}A_{11}
+
\frac{p_{s}}{p_{r}+p_{s}}A_{22},
\end{equation}
then
\begin{equation}\label{eqn: pair state as augmented qubit}
\sigma_{rs}(A,z)
=
\lambda_{rs}
\widetilde{\sigma}_{rs}(A)
+
(1-\lambda_{rs})z.
\end{equation}
Since $N>2$ and $\rho$ is faithful, one has
$0<\lambda_{rs}<1$, and $\widetilde{\sigma}_{rs}$ is faithful (see proposition \ref{prop: covariance at qubit and augmented qubit}).
%%%%%%%%%%%%%%%%%%%%%%%%%%%%%%%%%%%
The inclusion $\mathrm{i}_{rs}$ admits a state-preserving CPU left
inverse. 
%%%%%%%%%%%%%%%%%%%%%%%
Namely, define 
\begin{equation}\label{eqn: conditional expectation pair augmented qubit}
\mathcal{E}_{rs}^{\rho}(\mathbf{a})
:=
\left(
\mathrm{P}_{rs}\mathbf{a}\mathrm{P}_{rs},
\frac{
\rho(\mathrm{Q}_{rs}\mathbf{a}\mathrm{Q}_{rs})
}{
\rho(\mathrm{Q}_{rs})
}
\right),
\end{equation}
where $\mathrm{P}_{rs} = \mathbf{e}_{rr}+\mathbf{e}_{ss}$ and $\mathrm{Q}_{rs}=\mathbf{1}-\mathrm{P}_{rs}$. The first component is a compression and hence is completely positive, while the second component is completely positive because it is a state on
$M_{N}(\mathbb{C})$. 
%%%%%%%%%%%%%
Thus $\mathcal{E}_{rs}^{\rho}$ is CPU, and a direct
calculation gives
\begin{equation}
\mathcal{E}_{rs}^{\rho}\circ\mathrm{i}_{rs}
=
\operatorname{id}_{M_{2}(\mathbb{C})\oplus\mathbb{C}},
\qquad
\left(
\mathcal{E}_{rs}^{\rho}
\right)^{*}
\sigma_{rs}
=
\rho.
\end{equation}
Therefore,
$(M_{2}(\mathbb{C})\oplus\mathbb{C},\sigma_{rs})$ is a retract of
$(M_{N}(\mathbb{C}),\rho)$  in the sense of definition
\ref{defn: split monomorphisms}, and equation
\eqref{eqn: categorical invariance condition} gives
\begin{equation}\label{eqn: covariance pair equals augmented qubit}
\mathfrak{C}_{\rho}
\left(
\xi_{rs}^{\rho},
\xi_{rs}^{\rho}
\right)
=
\mathfrak{C}_{\sigma_{rs}}
\left(
\xi_{(\mathbf{e}_{12},0)}^{\sigma_{rs}},
\xi_{(\mathbf{e}_{12},0)}^{\sigma_{rs}}
\right),
\end{equation}
because
\begin{equation}
\widetilde{\mathrm{i}}_{rs}
\left(
\xi_{(\mathbf{e}_{12},0)}^{\sigma_{rs}}
\right)
=
\xi_{rs}^{\rho}.
\end{equation}
%%%%%%%%%%%%%%%%%%%%%%%%
The  eigenvalues of $\widetilde{\sigma}_{rs}$ corresponding to
the first and second basis vectors are
$p_{r}/\lambda_{rs}$ and $p_{s}/\lambda_{rs}$, respectively. 
%%%%%%%%%%%%%%%%%%%%%%%%
Hence
the modular eigenvalue associated with
$\xi_{(\mathbf{e}_{12},0)}^{\sigma_{rs}}$ is $p_{r}/p_{s}$.
Equation \eqref{eqn: fields of covariances on faithful M2C+C} therefore gives
\begin{equation}\label{eqn: intermediate covariance coefficient on matrix algebra}
\mathfrak{C}_{\rho}
\left(
\xi_{rs}^{\rho},
\xi_{rs}^{\rho}
\right)=\mathfrak{C}_{\sigma_{rs}}
\left(
\xi_{(\mathbf{e}_{12},0)}^{\sigma_{rs}},
\xi_{(\mathbf{e}_{12},0)}^{\sigma_{rs}}
\right)
=
F_{\lambda_{rs}}
\left(
\frac{p_{r}}{p_{s}}
\right)
\left\|
\xi_{(\mathbf{e}_{12},0)}^{\sigma_{rs}}
\right\|_{\sigma_{rs}}^{2}
=
p_{s}
F_{p_{r}+p_{s}}
\left(
\frac{p_{r}}{p_{s}}
\right)
\end{equation}
%%%%%%%%%%%%%%%%%%%%%%%%
Since
$\|\xi_{rs}^{\rho}\|_{\rho}^{2}=p_{s}$, equations
\eqref{eqn: covariance operator diagonal on matrix units} and
\eqref{eqn: intermediate covariance coefficient on matrix algebra}
yield
\begin{equation}\label{eqn: intermediate covariance operator on matrix units}
\mathbf{T}_{\rho}
\left(
\xi_{rs}^{\rho}
\right)
=
F_{p_{r}+p_{s}}
\left(
\frac{p_{r}}{p_{s}}
\right)
\xi_{rs}^{\rho}.
\end{equation}
%%%%%%%%%%%%%%%%%%%%%%%

Therefore we know that for the object $(M_N(\mathbb{C}),\rho)$ $\mathbf{T}_{\rho}$ is diagonal on $\mathcal{K}_{\rho}$ with eigenvectors $\xi^{\rho}_{rs}$ and corresponding eigenvalues $F_{\lambda_{rs}}(\frac{p_r}{p_s})$, being $\lambda_{rs}=p_r+p_s$.
%%%%%%%%%%%%%%%%%%%%%%%%%%%%%%%%
Now we prove that the family $\{F_{\lambda}\}_{\lambda\in(0,1)}$ is constant.
Fix $t\in(0,\infty)\setminus\{1\}$ throughout; the case $t=1$ is handled at the end.

For every $k\geq 2$ and every vector
$\boldsymbol{\mu}=(\mu_{1},\ldots,\mu_{k})$ with $\mu_{j}>0$ and
$\sum_{j=1}^{k}\mu_{j}=1$, let $\sigma_{\boldsymbol{\mu}}$ be the faithful
state on $M_{k}(\mathbb{C})$ whose density matrix is
$\varsigma_{\boldsymbol{\mu}}=\sum_{j=1}^{k}\mu_{j}\mathbf{e}_{jj}$
with respect to a fixed system of matrix units
$\{\mathbf{e}_{jl}\}$ of $M_{k}(\mathbb{C})$.
Define
\begin{equation}\label{eqn: tensor amplification maps}
\mathrm{i}_{k}\colon M_{2}(\mathbb{C})\rightarrow
M_{2}(\mathbb{C})\otimes M_{k}(\mathbb{C}),
\qquad
\mathrm{i}_{k}(\mathbf{a})=\mathbf{a}\otimes\mathbf{1}_{k},
\end{equation}
and
\begin{equation}\label{eqn: tensor amplification expectation}
\mathcal{E}_{\sigma_{\boldsymbol{\mu}}}
:=
\mathrm{id}_{M_{2}(\mathbb{C})}\otimes\sigma_{\boldsymbol{\mu}}
\colon
M_{2}(\mathbb{C})\otimes M_{k}(\mathbb{C})\rightarrow M_{2}(\mathbb{C}),
\qquad
\mathcal{E}_{\sigma_{\boldsymbol{\mu}}}(\mathbf{a}\otimes\mathbf{b})
=
\sigma_{\boldsymbol{\mu}}(\mathbf{b})\,\mathbf{a},
\end{equation}
extended by linearity.
The map $\mathrm{i}_{k}$ is a unital $*$-homomorphism, hence CPU, and
\begin{equation}
\left(\rho_{t}\otimes\sigma_{\boldsymbol{\mu}}\right)
\left(\mathrm{i}_{k}(\mathbf{a})\right)
=
\rho_{t}(\mathbf{a})\,\sigma_{\boldsymbol{\mu}}(\mathbf{1}_{k})
=
\rho_{t}(\mathbf{a}),
\end{equation}
that is, $\mathrm{i}_{k}^{*}(\rho_{t}\otimes\sigma_{\boldsymbol{\mu}})=\rho_{t}$.
The map $\mathcal{E}_{\sigma_{\boldsymbol{\mu}}}$ is the tensor product of
the CPU maps $\mathrm{id}_{M_{2}(\mathbb{C})}$ and
$\sigma_{\boldsymbol{\mu}}\colon M_{k}(\mathbb{C})\rightarrow\mathbb{C}$,
hence CPU; it is unital because
$\sigma_{\boldsymbol{\mu}}(\mathbf{1}_{k})=1$; and on elementary tensors
\begin{equation}
\rho_{t}\left(
\mathcal{E}_{\sigma_{\boldsymbol{\mu}}}(\mathbf{a}\otimes\mathbf{b})
\right)
=
\rho_{t}(\mathbf{a})\,\sigma_{\boldsymbol{\mu}}(\mathbf{b})
=
\left(\rho_{t}\otimes\sigma_{\boldsymbol{\mu}}\right)
(\mathbf{a}\otimes\mathbf{b}),
\end{equation}
so that
$\mathcal{E}_{\sigma_{\boldsymbol{\mu}}}^{*}(\rho_{t})
=\rho_{t}\otimes\sigma_{\boldsymbol{\mu}}$.
Finally,
$\mathcal{E}_{\sigma_{\boldsymbol{\mu}}}\left(\mathrm{i}_{k}(\mathbf{a})\right)
=\sigma_{\boldsymbol{\mu}}(\mathbf{1}_{k})\,\mathbf{a}=\mathbf{a}$.
Therefore $(M_{2}(\mathbb{C}),\rho_{t})$ is a retract of
$(M_{2}(\mathbb{C})\otimes M_{k}(\mathbb{C}),
\rho_{t}\otimes\sigma_{\boldsymbol{\mu}})$ in the sense of definition
\ref{defn: split monomorphisms}, and equation
\eqref{eqn: categorical invariance condition} gives
\begin{equation}\label{eqn: tensor sandwich equality}
\mathfrak{C}_{\rho_{t}}
\left(
\xi_{12}^{\rho_{t}},
\xi_{12}^{\rho_{t}}
\right)
=
\mathfrak{C}_{\rho_{t}\otimes\sigma_{\boldsymbol{\mu}}}
\left(
\xi_{\mathbf{e}_{12}\otimes\mathbf{1}_{k}}
^{\rho_{t}\otimes\sigma_{\boldsymbol{\mu}}},
\xi_{\mathbf{e}_{12}\otimes\mathbf{1}_{k}}
^{\rho_{t}\otimes\sigma_{\boldsymbol{\mu}}}
\right),
\end{equation}
because
$\widetilde{\mathrm{i}}_{k}
\left(\xi_{12}^{\rho_{t}}\right)
=\xi_{\mathbf{e}_{12}\otimes\mathbf{1}_{k}}
^{\rho_{t}\otimes\sigma_{\boldsymbol{\mu}}}$.

We now evaluate the right-hand side of equation
\eqref{eqn: tensor sandwich equality}.
Identify $M_{2}(\mathbb{C})\otimes M_{k}(\mathbb{C})$ with
$M_{2k}(\mathbb{C})$ through the product basis
$\{|i\rangle\otimes|j\rangle\}$; the density matrix of
$\rho_{t}\otimes\sigma_{\boldsymbol{\mu}}$ is diagonal in this basis,
with weight $p_{i}\mu_{j}$ attached to the index $(i,j)$, where
$p_{1}=\tfrac{t}{1+t}$ and $p_{2}=\tfrac{1}{1+t}$.
Decompose
\begin{equation}\label{eqn: decomposition of amplified vector}
\xi_{\mathbf{e}_{12}\otimes\mathbf{1}_{k}}
^{\rho_{t}\otimes\sigma_{\boldsymbol{\mu}}}
=
\sum_{j=1}^{k}
\xi_{\mathbf{e}_{12}\otimes\mathbf{e}_{jj}}
^{\rho_{t}\otimes\sigma_{\boldsymbol{\mu}}}.
\end{equation}
Each element $\mathbf{e}_{12}\otimes\mathbf{e}_{jj}$ is, under the above
identification, an off-diagonal matrix unit of $M_{2k}(\mathbb{C})$
connecting the indices $(1,j)$ and $(2,j)$, whose weights are
$\mu_{j}p_{1}$ and $\mu_{j}p_{2}$.
Since $t\neq1$, these two weights are different, so equation
\eqref{eqn: covariance operator diagonal on matrix units} applies to the
object
$(M_{2k}(\mathbb{C}),\rho_{t}\otimes\sigma_{\boldsymbol{\mu}})$
and shows that each summand in equation
\eqref{eqn: decomposition of amplified vector} is an eigenvector of
$\mathbf{T}_{\rho_{t}\otimes\sigma_{\boldsymbol{\mu}}}$; moreover, since
the weight sum is $\mu_{j}p_{1}+\mu_{j}p_{2}=\mu_{j}\in(0,1)$ (here
$k\geq2$ is used), equation
\eqref{eqn: intermediate covariance operator on matrix units} identifies
the corresponding eigenvalue as $F_{\mu_{j}}(t)$.
In particular, for two distinct summands $\xi_{u}$ and $\xi_{v}$ of the
decomposition,
\begin{equation}
\mathfrak{C}_{\rho_{t}\otimes\sigma_{\boldsymbol{\mu}}}
(\xi_{u},\xi_{v})
=
\langle
\xi_{u}\mid
\mathbf{T}_{\rho_{t}\otimes\sigma_{\boldsymbol{\mu}}}(\xi_{v})
\rangle_{\rho_t\otimes\sigma_{\boldsymbol{\mu}}}
=
F_{\mu_{v}}(t)\,
\langle\xi_{u}\mid\xi_{v}\rangle_{\rho_t\otimes\sigma_{\boldsymbol{\mu}}}
=
0,
\end{equation}
because distinct matrix units are orthogonal in the \textup{GNS} Hilbert
product. Hence the cross terms in the expansion of the quadratic form
vanish, and, using
$\|\xi_{\mathbf{e}_{12}\otimes\mathbf{e}_{jj}}\|^{2}
=(\rho_{t}\otimes\sigma_{\boldsymbol{\mu}})
\left((\mathbf{e}_{22}\otimes\mathbf{e}_{jj})\right)
=\tfrac{\mu_{j}}{1+t}$,
\begin{equation}\label{eqn: amplified covariance as a sum}
\mathfrak{C}_{\rho_{t}\otimes\sigma_{\boldsymbol{\mu}}}
\left(
\xi_{\mathbf{e}_{12}\otimes\mathbf{1}_{k}},
\xi_{\mathbf{e}_{12}\otimes\mathbf{1}_{k}}
\right)
=
\sum_{j=1}^{k}
F_{\mu_{j}}(t)\,
\frac{\mu_{j}}{1+t}.
\end{equation}
On the other hand, equation
\eqref{eqn: fields of covariances on faithful qubit} gives
$\mathfrak{C}_{\rho_{t}}(\xi_{12}^{\rho_{t}},\xi_{12}^{\rho_{t}})
=\tfrac{1}{1+t}F(t)$,
since
$\langle\xi_{\mathbb{I}}^{\rho_{t}}\mid\xi_{12}^{\rho_{t}}\rangle_{\rho_t}
=\rho_{t}(\mathbf{e}_{12})=0$
and
$\|\xi_{12}^{\rho_{t}}\|^{2}=\tfrac{1}{1+t}$.
Substituting into equation \eqref{eqn: tensor sandwich equality} and
multiplying by $1+t$, we arrive at the partition identity
\begin{equation}\label{eqn: partition identity}
\sum_{j=1}^{k}
\mu_{j}\,F_{\mu_{j}}(t)
=
F(t)
\qquad
\text{for every }k\geq2
\text{ and every }\boldsymbol{\mu}
\text{ as above.}
\end{equation}

Define
\begin{equation}
h\colon(0,1)\rightarrow(0,\infty),
\qquad
h(\mu):=\mu\,F_{\mu}(t),
\qquad
c:=F(t)>0.
\end{equation}
Equation \eqref{eqn: partition identity} states that
$\sum_{j=1}^{k}h(\mu_{j})=c$ for every partition of unity of length
$k\geq2$; only the lengths $k=2$ and $k=3$ will be used.
Comparing the length-$3$ partition
$(\mu_{1},\mu_{2},1-\mu_{1}-\mu_{2})$ with the length-$2$ partition
$(\mu_{1}+\mu_{2},1-\mu_{1}-\mu_{2})$ and subtracting yields
\begin{equation}\label{eqn: restricted additivity}
h(\mu_{1})+h(\mu_{2})=h(\mu_{1}+\mu_{2})
\qquad
(\mu_{1},\mu_{2}>0,\ \mu_{1}+\mu_{2}<1).
\end{equation}
For $0<\mu<\mu'<1$, equation \eqref{eqn: restricted additivity} gives
$h(\mu')=h(\mu)+h(\mu'-\mu)\geq h(\mu)$ because
$h(\mu'-\mu)\geq0$; thus $h$ is nondecreasing on $(0,1)$.
Fix $n\geq2$. By induction using equation
\eqref{eqn: restricted additivity} (all partial sums $\tfrac{m}{n}$ with
$m<n$ stay in $(0,1)$), one has
$h\!\left(\tfrac{m}{n}\right)=m\,h\!\left(\tfrac{1}{n}\right)$
for $1\leq m\leq n-1$.
The length-$2$ instance of equation \eqref{eqn: partition identity} with
$\boldsymbol{\mu}=\left(\tfrac{1}{n},\tfrac{n-1}{n}\right)$
then gives
$h\!\left(\tfrac{1}{n}\right)+(n-1)h\!\left(\tfrac{1}{n}\right)=c$,
that is, $h\!\left(\tfrac{1}{n}\right)=\tfrac{c}{n}$, whence
\begin{equation}
h(q)=c\,q
\qquad
\text{for every rational }q\in(0,1).
\end{equation}
Let $\mu\in(0,1)$ be irrational and choose rationals $q<\mu<q'$ in
$(0,1)$. By monotonicity,
$c\,q=h(q)\leq h(\mu)\leq h(q')=c\,q'$, and letting
$q\uparrow\mu$, $q'\downarrow\mu$ gives $h(\mu)=c\,\mu$.
Therefore
\begin{equation}\label{eqn: F lambda equals F}
F_{\lambda}(t)=F(t)
\end{equation}
for every $\lambda\in(0,1)$ and every
$t\in(0,\infty)\setminus\{1\}$. The same identity holds at $t=1$ because
$F_{\lambda}(1)=\beta=F(1)$ by definition. Hence
$F_{\lambda}=F$ for every $\lambda\in(0,1)$.
%%%%%%%%%%%%%%%%%%%%%%%%%%%%%

Consequently, we obtain
\begin{equation}
\mathbf{T}_{\rho}
\left(
\xi_{rs}^{\rho}
\right)
=
F\left(
\frac{p_{r}}{p_{s}}
\right)
\xi_{rs}^{\rho}
\end{equation}
whenever $p_{r}\neq p_{s}$, which means that 
\begin{equation}
\mathbf{T}_{\rho}|_{\mathscr{K}_{\rho}}
=
F\left(
\Delta_{\rho}|_{\mathscr{K}_{\rho}}
\right).
\end{equation}
On the centralizer, Proposition
\ref{prop: covariance operator on the centralizer}, together with
$F(1)=\beta$, gives
\begin{equation}
\mathbf{T}_{\rho}|_{\mathscr{M}_{\rho}}
=
F(1)\mathbb{I}_{\mathscr{M}_{\rho}}
+
(\alpha-F(1))
\left|
\xi_{\mathbb{I}}^{\rho}
\right\rangle
\left\langle
\xi_{\mathbb{I}}^{\rho}
\right|.
\end{equation}
Since $\Delta_{\rho}$ is the identity on $\mathscr{M}_{\rho}$, the two
restrictions combine to give
\begin{equation}\label{eqn: fields of covariances on faithful states of matrix algebras}
\begin{split}
\mathfrak{C}_{\rho}(\xi,\eta)
&=
\left\langle
\xi
\mid
F(\Delta_{\rho})(\eta)
\right\rangle_{\rho}
\\
&\quad+
(\alpha-F(1))
\left\langle
\xi
\mid
\xi_{\mathbb{I}}^{\rho}
\right\rangle_{\rho}
\left\langle
\xi_{\mathbb{I}}^{\rho}
\mid
\eta
\right\rangle_{\rho}
\end{split}
\end{equation}
for every faithful state on $M_{N}(\mathbb{C})$ with $N>2$ and all
$\xi,\eta\in\mathcal{H}_{\rho}$. 
%%%%%%%%%%%%%%%%%%%%%%%%%%%%%
For $N=2$, equation
\eqref{eqn: fields of covariances on faithful states of matrix algebras}
is precisely equation
\eqref{eqn: fields of covariances on faithful qubit} in Proposition
\ref{prop: covariance at qubit and augmented qubit}. For $N=1$, every
state is tracial, and the conclusion follows from Proposition
\ref{prop: covariance operator on the centralizer}. Thus
\eqref{eqn: fields of covariances on faithful states of matrix algebras}
holds for faithful states on every full matrix algebra $M_{N}(\mathbb{C})$, and thus on every algebra of the form $\mathcal{B}(\mathcal{H})$ with $\mathrm{dim}(\mathcal{H})<\infty$ because of the invariance of $\mathfrak{C}_{\rho}$ under automorphisms.
%%%%%%%%%%%%%%%%%%%%%%%%%%%%

We finally pass to an arbitrary finite-dimensional $C^{*}$-algebra.
%%%%%%%%%%%%%%%%%%%%%%%%%%%%%%%%%%%%%%
Without loss of generality, we may take
\begin{equation}
\mathscr{A}
=
\bigoplus_{\ell=1}^{L}
\mathcal{B}(\mathcal{H}_{\ell}),
\end{equation}
and let $\rho$ be faithful. 
%%%%%%%%%%%%%%%%%%%%%%
There exist strictly positive operators
$D_{\ell}\in\mathcal{B}(\mathcal{H}_{\ell})$ such that
\begin{equation}
\rho(\mathbf{a}_{1},\ldots,\mathbf{a}_{L})
=
\sum_{\ell=1}^{L}
\operatorname{Tr}
\left(
D_{\ell}\mathbf{a}_{\ell}
\right),
\qquad
\sum_{\ell=1}^{L}
\operatorname{Tr}(D_{\ell})
=
1.
\end{equation}
Set
\begin{equation}
\mathcal{K}
:=
\bigoplus_{\ell=1}^{L}
\mathcal{H}_{\ell},
\qquad
D
:=
\bigoplus_{\ell=1}^{L}
D_{\ell},
\end{equation}
and let
$\phi\colon\mathscr{A}\rightarrow\mathcal{B}(\mathcal{K})$ be the
canonical block-diagonal embedding. Define the faithful state
$\omega$ on $\mathcal{B}(\mathcal{K})$ by
\begin{equation}
\omega(\mathbf{x})
:=
\operatorname{Tr}(D\mathbf{x}).
\end{equation}
By construction,
\begin{equation}
\phi^{*}\omega
=
\rho.
\end{equation}

Let $P_{\ell}$ denote the orthogonal projection of $\mathcal{K}$ onto
$\mathcal{H}_{\ell}$, and define the block-diagonal conditional
expectation
\begin{equation}
\mathcal{E}(\mathbf{x})
:=
\bigoplus_{\ell=1}^{L}
P_{\ell}\mathbf{x}P_{\ell}.
\end{equation}
The map $\mathcal{E}$ is CPU and satisfies
\begin{equation}
\mathcal{E}\circ\phi
=
\operatorname{id}_{\mathscr{A}}.
\end{equation}
Since $D$ is block diagonal, one also has $\mathcal{E}^{*}\rho=\omega$, which means that $(\mathscr{A},\rho)$ is a retract of $(\mathcal{B}(\mathcal{K}),\omega)$   in the sense of definition
\ref{defn: split monomorphisms}.
%%%%%%%%%%%%%%%%%%%%%%%%%%%%%%%%%%%%%%%%%%%%%%%%%%
The induced GNS map is the isometry
\begin{equation}
\widetilde{\phi}
\left(
\xi_{\mathbf{a}}^{\rho}
\right)
=
\xi_{\phi(\mathbf{a})}^{\omega},
\end{equation}
and equation \eqref{eqn: categorical invariance condition}, followed by
polarization, therefore gives
\begin{equation}\label{eqn: covariance arbitrary algebra through matrix embedding}
\mathfrak{C}_{\rho}(\xi,\eta)
=
\mathfrak{C}_{\omega}
\left(
\widetilde{\phi}(\xi),
\widetilde{\phi}(\eta)
\right)
\end{equation}
for all $\xi,\eta\in\mathcal{H}_{\rho}$.
%%%%%%%%%%%%%%%%%%%%%%%%%%%%%%%
The isometry $\widetilde{\phi}$ intertwines the modular operators in the sense that
\begin{equation}
\begin{split}
\Delta_{\omega}
\widetilde{\phi}
\left(
\xi_{\mathbf{a}}^{\rho}
\right)
&=
\xi_{
D\phi(\mathbf{a})D^{-1}
}^{\omega}
\\
&=
\xi_{
\phi
\left(
D_{1}\mathbf{a}_{1}D_{1}^{-1},
\ldots,
D_{L}\mathbf{a}_{L}D_{L}^{-1}
\right)
}^{\omega}
\\
&=
\widetilde{\phi}
\Delta_{\rho}
\left(
\xi_{\mathbf{a}}^{\rho}
\right),
\end{split}
\end{equation}
with $\mathbf{a}=(\mathbf{a}_{1},\ldots,\mathbf{a}_{L})\in\mathscr{A}$, which means
\begin{equation}
\Delta_{\omega}\widetilde{\phi}
=
\widetilde{\phi}\Delta_{\rho},
\end{equation}
%%%%%%%%%%%%%%%%%%%%%%
Since we are considering finite-dimensional Hilbert spaces the operators $F(\Delta_{\rho})$ and $F(\Delta_{\omega})$ are defined through their expression on a basis of eigenvectors. Therefore, the equation $\Delta_{\omega}\widetilde{\phi} = \widetilde{\phi} \Delta_{\rho}$ can be extended to any function $F$ and we get
\begin{equation}\label{eqn: functional calculus intertwining matrix embedding}
F(\Delta_{\omega})
\widetilde{\phi}
=
\widetilde{\phi}
F(\Delta_{\rho}).
\end{equation}
Finally, being 
\begin{equation}
\widetilde{\phi}
\left(
\xi_{\mathbb{I}_{\mathscr{A}}}^{\rho}
\right)
=
\xi_{\mathbb{I}_{\mathcal{B}(\mathcal{K})}}^{\omega},
\end{equation}
applying equation \eqref{eqn: fields of covariances on faithful states of matrix algebras}
to the faithful state $\omega$, and using equations
\eqref{eqn: covariance arbitrary algebra through matrix embedding} and
\eqref{eqn: functional calculus intertwining matrix embedding} together with the fact that $\widetilde{\phi}$ is an isometry, gives equation \eqref{eqn: fields of covariances on faithful fNCP 2} for a generic finite-dimensional algebra.
%%%%%%%%%%%%%%%%%%%%%%%%%%%%%%%%%%%
\end{proof}

We now prove that the monotonicity condition \eqref{eqn: categorical monotonicity 3} forces the function $F$ in proposition \ref{prop: fields of covariances on faithful states in fNCP} to be operator monotone (see, \textit{e.g.}, \cite{B2007a}).
%%%%%%%%%%%%%%%%%%%%%

\begin{proposition}\label{prop: fields of covariances on faithful states in fNCP have operator monotone functions}
Let $\mathfrak{C}\colon \mathsf{fNCP}\rightsquigarrow\mathsf{Hilb}$ be a continuous field of covariances as in definitions \ref{defn: field of covariances} and \ref{defn: continuity for fields of covariances}.
%%%%%%%%%%%%%%%%
The  function $F$ in proposition \ref{prop: fields of covariances on faithful states in fNCP} is operator monotone on $(0,\infty)$.
%%%%%%%%%%%%%%%%%%%%%%%%%%%%%%%%%%%%%%

\end{proposition}

\begin{proof}

For every faithful state $\rho$, define the Hermitian form
\begin{equation}
\mathfrak{G}_{\rho}(\xi,\eta)
:=
\left\langle
\xi,
F(\Delta_{\rho})(\eta)
\right\rangle_{\rho}.
\end{equation}
According to equation
\eqref{eqn: fields of covariances on faithful fNCP 2}, we have
\begin{equation}\label{eqn: covariance modular and scalar parts}
\mathfrak{C}_{\rho}(\xi,\eta)
=
\mathfrak{G}_{\rho}(\xi,\eta)
+
\bigl(\alpha-F(1)\bigr)
\left\langle
\xi,
\xi_{\mathbb{I}}^{\rho}
\right\rangle_{\rho}
\left\langle
\xi_{\mathbb{I}}^{\rho},
\eta
\right\rangle_{\rho}.
\end{equation}

Let $\rho$ and $\sigma$ be faithful states on a finite-dimensional
$C^*$-algebra $\mathscr{A}$, let $\lambda\in(0,1)$, and set
\begin{equation}
\mu_{\lambda}
:=
\lambda\rho+(1-\lambda)\sigma.
\end{equation}
Consider the faithful state
\begin{equation}
\omega_{\lambda}
:=
\lambda\rho\oplus(1-\lambda)\sigma
\end{equation}
on $\mathscr{A}\oplus\mathscr{A}$ and the diagonal $*$-homomorphism $\Phi\colon\mathscr{A}\to\mathscr{A}\oplus\mathscr{A}$ given by
\begin{equation}\label{eqn: diagonal immersion}
\Phi(\mathbf{a})
=
\mathbf{a}\oplus\mathbf{a}.
\end{equation}
Since
\begin{equation}
\Phi^*(\omega_{\lambda})
=
\mu_{\lambda},
\end{equation}
the map $\Phi$ determines a morphism from  $(\mathscr{A}\oplus\mathscr{A},\omega_{\lambda})$ to $(\mathscr{A},\mu_{\lambda})$ in $\mathsf{fNCP}$. Therefore, the monotonicity condition
\eqref{eqn: categorical monotonicity 3} gives
\begin{equation}\label{eqn: diagonal monotonicity covariance}
\mathfrak{C}_{\omega_{\lambda}}
\left(
\widetilde{\Phi}(\xi_{\mathbf{a}}),
\widetilde{\Phi}(\xi_{\mathbf{a}})
\right)
\leq
\mathfrak{C}_{\mu_{\lambda}}
\left(
\xi_{\mathbf{a}},
\xi_{\mathbf{a}}
\right)
\end{equation}
for every $\mathbf{a}\in\mathscr{A}$.

The GNS Hilbert space of $\omega_{\lambda}$ is naturally identified
with the weighted direct sum
\begin{equation}
\mathcal{H}_{\omega_{\lambda}}
\cong
\mathcal{H}_{\rho}\oplus\mathcal{H}_{\sigma},
\end{equation}
and, under this identification,
\begin{equation}
\Delta_{\omega_{\lambda}}
=
\Delta_{\rho}\oplus\Delta_{\sigma}.
\end{equation}
Moreover,
\begin{equation}
\widetilde{\Phi}(\xi_{\mathbf{a}})
=
\xi_{\mathbf{a}\oplus\mathbf{a}}^{\omega_{\lambda}}.
\end{equation}
It follows that
\begin{equation}\label{eqn: direct sum modular form}
\begin{split}
&
\mathfrak{G}_{\omega_{\lambda}}
\left(
\widetilde{\Phi}(\xi_{\mathbf{a}}),
\widetilde{\Phi}(\xi_{\mathbf{a}})
\right)
\\
&\qquad=
\lambda
\mathfrak{G}_{\rho}
\left(
\xi_{\mathbf{a}},
\xi_{\mathbf{a}}
\right)
+
(1-\lambda)
\mathfrak{G}_{\sigma}
\left(
\xi_{\mathbf{a}},
\xi_{\mathbf{a}}
\right).
\end{split}
\end{equation}

On the other hand, the rank-one terms in
\eqref{eqn: covariance modular and scalar parts} agree exactly on the
two sides of \eqref{eqn: diagonal monotonicity covariance}. Indeed,
\begin{equation}
\begin{split}
\left\langle
\widetilde{\Phi}(\xi_{\mathbf{a}}),
\xi_{\mathbb{I}\oplus\mathbb{I}}^{\omega_{\lambda}}
\right\rangle_{\omega_{\lambda}}
&=
\lambda\rho(\mathbf{a}^{\dagger})
+
(1-\lambda)\sigma(\mathbf{a}^{\dagger})
\\
&=
\mu_{\lambda}(\mathbf{a}^{\dagger})
\\
&=
\left\langle
\xi_{\mathbf{a}},
\xi_{\mathbb{I}}^{\mu_{\lambda}}
\right\rangle_{\mu_{\lambda}}.
\end{split}
\end{equation}
Consequently, the rank-one terms cancel in
\eqref{eqn: diagonal monotonicity covariance}, and we obtain
\begin{equation}\label{eqn: concavity modular part}
\begin{split}
&
\lambda
\mathfrak{G}_{\rho}
\left(
\xi_{\mathbf{a}},
\xi_{\mathbf{a}}
\right)
+
(1-\lambda)
\mathfrak{G}_{\sigma}
\left(
\xi_{\mathbf{a}},
\xi_{\mathbf{a}}
\right)
\\
&\qquad\leq
\mathfrak{G}_{\mu_{\lambda}}
\left(
\xi_{\mathbf{a}},
\xi_{\mathbf{a}}
\right).
\end{split}
\end{equation}

We now use equation \eqref{eqn: concavity modular part} to prove that
$F$ is operator concave.

Let $\tau_{\mathscr{A}}$ be a faithful tracial state on $\mathscr{A}$,
and let $\varrho,\varsigma\in\mathscr{A}$ be the invertible density
operators associated with $\rho$ and $\sigma$, respectively:
\begin{equation}
\rho(\mathbf{a})
=
\tau_{\mathscr{A}}(\varrho\mathbf{a}),
\qquad
\sigma(\mathbf{a})
=
\tau_{\mathscr{A}}(\varsigma\mathbf{a}).
\end{equation}
Consider the algebra
\begin{equation}
\widetilde{\mathscr{A}}
:=
\mathscr{A}\otimes\mathbb{M}_{2}(\mathbb{C})
\end{equation}
endowed with the faithful tracial state
\begin{equation}
\widetilde{\tau}
:=
\tau_{\mathscr{A}}\otimes\tau_{2},
\qquad
\tau_{2}
:=
\frac{1}{2}\operatorname{Tr}.
\end{equation}
Define faithful states $\widetilde{\rho}$ and
$\widetilde{\sigma}$ on $\widetilde{\mathscr{A}}$ whose density
operators with respect to $\widetilde{\tau}$ are
\begin{equation}
\widetilde{\varrho}
=
\varrho\otimes\mathbf{e}_{11}
+
\mathbb{I}\otimes\mathbf{e}_{22},
\qquad
\widetilde{\varsigma}
=
\varsigma\otimes\mathbf{e}_{11}
+
\mathbb{I}\otimes\mathbf{e}_{22}.
\end{equation}
For $\mathbf{a}\in\mathscr{A}$, set
\begin{equation}
\widetilde{\mathbf{a}}
:=
\mathbf{a}\otimes\mathbf{e}_{12}.
\end{equation}
The restriction of the modular operator
$\Delta_{\widetilde{\rho}}$ to
$\mathscr{A}\otimes\mathbb{C}\mathbf{e}_{12}$ satisfies
\begin{equation}
\begin{split}
\Delta_{\widetilde{\rho}}
\left(
\mathbf{a}\otimes\mathbf{e}_{12}
\right)
&=
\widetilde{\varrho}
\left(
\mathbf{a}\otimes\mathbf{e}_{12}
\right)
\widetilde{\varrho}^{-1}
\\
&=
\varrho\mathbf{a}\otimes\mathbf{e}_{12}.
\end{split}
\end{equation}
Therefore,
\begin{equation}
F(\Delta_{\widetilde{\rho}})
\left(
\mathbf{a}\otimes\mathbf{e}_{12}
\right)
=
F(L_{\varrho})(\mathbf{a})
\otimes\mathbf{e}_{12},
\end{equation}
where $L_{\varrho}$ denotes left multiplication by $\varrho$.

A direct computation gives
\begin{equation}\label{eqn: amplified modular calculation}
\begin{split}
\mathfrak{G}_{\widetilde{\rho}}
\left(
\xi_{\widetilde{\mathbf{a}}},
\xi_{\widetilde{\mathbf{a}}}
\right)
&=
\widetilde{\tau}
\left(
\widetilde{\varrho}
\widetilde{\mathbf{a}}^{\dagger}
F(\Delta_{\widetilde{\rho}})
(\widetilde{\mathbf{a}})
\right)
\\
&=
\frac{1}{2}
\tau_{\mathscr{A}}
\left(
\mathbf{a}^{\dagger}
F(L_{\varrho})(\mathbf{a})
\right).
\end{split}
\end{equation}
The analogous equality holds for $\widetilde{\sigma}$. Moreover, the
density operator of
\begin{equation}
\lambda\widetilde{\rho}
+
(1-\lambda)\widetilde{\sigma}
\end{equation}
is
\begin{equation}
\bigl(
\lambda\varrho+(1-\lambda)\varsigma
\bigr)\otimes\mathbf{e}_{11}
+
\mathbb{I}\otimes\mathbf{e}_{22}.
\end{equation}
Applying equation \eqref{eqn: concavity modular part} to
$\widetilde{\rho}$, $\widetilde{\sigma}$ and
$\widetilde{\mathbf{a}}$, and using
\eqref{eqn: amplified modular calculation}, we obtain
\begin{equation}\label{eqn: concavity left multiplication}
\begin{split}
&
\tau_{\mathscr{A}}
\left(
\mathbf{a}^{\dagger}
\left[
\lambda F(L_{\varrho})
+
(1-\lambda)F(L_{\varsigma})
\right]
(\mathbf{a})
\right)
\\
&\qquad\leq
\tau_{\mathscr{A}}
\left(
\mathbf{a}^{\dagger}
F\left(
L_{\lambda\varrho+(1-\lambda)\varsigma}
\right)
(\mathbf{a})
\right).
\end{split}
\end{equation}

We first apply this inequality to density operators in a matrix
algebra. Let
\begin{equation}
\mathscr{A}
=
\mathbb{M}_{N}(\mathbb{C}),
\qquad
\tau_{\mathscr{A}}
=
\tau_{N}
:=
\frac{1}{N}\operatorname{Tr}.
\end{equation}
Since left multiplication is a $*$-representation, finite spectral calculus gives
\begin{equation}
F(L_{A})
=
L_{F(A)}
\end{equation}
for every positive definite $A\in\mathbb{M}_{N}(\mathbb{C})$.

Suppose that $A,B\in\mathbb{M}_{N}(\mathbb{C})$ are positive
definite and satisfy
\begin{equation}
\operatorname{Tr}(A)
=
\operatorname{Tr}(B)
=
N.
\end{equation}
They are density operators with respect to $\tau_N$. Equation
\eqref{eqn: concavity left multiplication} implies
\begin{equation}
\tau_N
\left(
X^{\dagger}D_{\lambda}X
\right)
\geq 0
\end{equation}
for every $X\in\mathbb{M}_{N}(\mathbb{C})$, where
\begin{equation}
D_{\lambda}
:=
F\left(
\lambda A+(1-\lambda)B
\right)
-
\lambda F(A)
-
(1-\lambda)F(B).
\end{equation}
It follows that
\begin{equation}
D_{\lambda}
\geq 0.
\end{equation}
Consequently,
\begin{equation}\label{eqn: concavity trace normalized}
F\left(
\lambda A+(1-\lambda)B
\right)
\geq
\lambda F(A)+(1-\lambda)F(B)
\end{equation}
whenever $A$ and $B$ are positive definite and have trace $N$.

We now remove the trace-normalization condition. Let
$A,B\in\mathbb{M}_{n}(\mathbb{C})$ be arbitrary positive definite
matrices. Choose $m\in\mathbb{N}$ sufficiently large that, setting
$N=n+m$, one has
\begin{equation}
N
>
\max
\left\{
\operatorname{Tr}(A),
\operatorname{Tr}(B)
\right\}.
\end{equation}
Define
\begin{equation}
a
:=
\frac{N-\operatorname{Tr}(A)}{m},
\qquad
b
:=
\frac{N-\operatorname{Tr}(B)}{m},
\end{equation}
and
\begin{equation}
\widehat{A}
:=
A\oplus a\mathbb{I}_{m},
\qquad
\widehat{B}
:=
B\oplus b\mathbb{I}_{m}.
\end{equation}
Then $\widehat{A}$ and $\widehat{B}$ are positive definite elements
of $\mathbb{M}_{N}(\mathbb{C})$ satisfying
\begin{equation}
\operatorname{Tr}(\widehat{A})
=
\operatorname{Tr}(\widehat{B})
=
N.
\end{equation}
Applying equation \eqref{eqn: concavity trace normalized} to
$\widehat{A}$ and $\widehat{B}$ gives
\begin{equation}
F\left(
\lambda\widehat{A}
+
(1-\lambda)\widehat{B}
\right)
\geq
\lambda F(\widehat{A})
+
(1-\lambda)F(\widehat{B}).
\end{equation}
Since finite spectral calculus preserves direct sums,
compression onto the first $n$ coordinates yields
\begin{equation}\label{eqn: operator concavity F}
F\left(
\lambda A+(1-\lambda)B
\right)
\geq
\lambda F(A)+(1-\lambda)F(B).
\end{equation}
Since $n$ is arbitrary, $F$ is operator concave on $(0,+\infty)$.

Finally, the positivity of $F$ implies that its operator concavity
entails its operator monotonicity. Indeed, let
\begin{equation}
0<A\leq B
\end{equation}
in $\mathbb{M}_{n}(\mathbb{C})$, and let $t\in(0,1)$. The matrix
\begin{equation}
C_t
:=
\frac{B-tA}{1-t}
=
A+\frac{B-A}{1-t}
\end{equation}
is positive definite, and
\begin{equation}
B
=
tA+(1-t)C_t.
\end{equation}
By operator concavity and positivity,
\begin{equation}
F(B)
\geq
tF(A)+(1-t)F(C_t)
\geq
tF(A).
\end{equation}
Letting $t\to 1$ gives
\begin{equation}
F(A)
\leq
F(B).
\end{equation}
Thus $F$ is operator monotone on $(0,+\infty)$.

\end{proof}

\subsection{Extension to non-faithful states}

We now investigate the general case admitting non-faithful states.
%%%%%%%%%%%%%%%%%%%%%%

\begin{proposition}\label{prop: continuous fields of covariances on fNCP}
Let $\mathfrak{C}\colon \mathsf{fNCP}\rightsquigarrow\mathsf{Hilb}$ be a continuous field of covariances in the sense of definition \ref{defn: field of covariances} and \ref{defn: continuity for fields of covariances}, let $\alpha,\beta>0$ and $F$ be as in proposition \ref{prop: fields of covariances on faithful states in fNCP}.
%%%%%%%%%%%%%%%%%%%%%
If $(\mathscr{A},\rho)\in\mathsf{fNCP}_{0}$, then there exists a continuous extension of $F$ to $[0,\infty)$ such that
\begin{equation}\label{eqn: field of covariances on non-faithful fNCP}
\mathfrak{C}_{\rho}(\xi,\eta) = \langle \xi \mid F(\Delta_{ \rho })(\eta)\rangle_{ \rho } + (\alpha -F(1))\langle\xi\mid\xi^{\rho}_{\mathbb{I}}\rangle_{\rho}\langle\xi^{\rho}_{\mathbb{I}}\mid\eta\rangle_{\rho},
\end{equation}
for all $\xi,\eta\in\mathcal{H}_{\rho}$.
%%%%%%%%%%%%%%%%%%%%%%%%%%%%
\end{proposition}

\begin{proof}
By proposition \ref{prop: fields of covariances on faithful states in fNCP}, it remains to determine the covariance at non-faithful states.
%%%%%%%%%%%%%%%%%%%%%%%%%%%%%
We first consider the case $\mathscr{A}=\mathcal{B}(\mathcal{H})$ with $N:=\dim(\mathcal{H})<+\infty$.
%%%%%%%%%%%%
Let $\rho$ be a non-faithful state on $\mathscr{A}$, let $\varrho$ be its density operator, let $\mathbf{p}$ be the support projection of
$\rho$, and set $\mathbf{q}:=\mathbb{I}-\mathbf{p}$, so that
\begin{equation}
\mathcal{H}_{\rho}
\cong
\mathscr{A}\mathbf{p}
=
\mathscr{A}_{pp}\oplus\mathscr{A}_{qp}
\equiv
\mathcal{H}_{\rho}^{pp}\oplus\mathcal{H}_{\rho}^{qp}.
\end{equation}
%%%%%%%%%%%%%%%%%%%%

We first determine the restriction of $\mathfrak{C}_{\rho}$ to
$\mathcal{H}_{\rho}^{pp}$. 
%%%%%%%%%%%%
Let $\widetilde{\rho}$ be the pullback of $\rho$ to $\mathscr{A}_{pp}$ through the natural immersion map in $\mathscr{A}$.
%%%%%%%%%%%%%%%%%%%
The state $\widetilde{\rho}$ is faithful on the unital $C^*$-algebra
$\mathscr{A}_{pp}$, whose identity is $\mathbf{p}$. 
%%%%%%%%%%%%
Consider the CPU map $\mathcal{E}:\mathscr{A}\to \mathscr{A}_{pp}$ given by
\begin{equation}
\mathcal{E}(\mathbf{a})
=
\mathbf{p}\mathbf{a}\mathbf{p}.
\end{equation}
%%%%%%%%%%%%%
It satisfies $\mathcal{E}^{*}(\widetilde{\rho})=\rho$ because of equation \eqref{eqn: support projection of normal state}.
%%%%%%%%%%%%%%%%%%
Moreover, the map $\mathrm{i}:\mathscr{A}_{pp}\to\mathscr{A}$ given by
\begin{equation}
\mathrm{i}(\mathbf{x})
=
\mathbf{x} + \widetilde{\rho}(\mathbf{x})\mathbf{q}
\end{equation}
is CPU and satisfies
\begin{equation}
\mathrm{i}^{*}(\rho)=\widetilde{\rho},
\qquad
\mathcal{E}\circ\mathrm{i}
=
\operatorname{id}_{\mathscr{A}_{pp}}.
\end{equation}
%%%%%%%%%%%%%%%%%%%%%%
Therefore, $(\mathscr{A}_{pp},\widetilde{\rho})$ is a retract of $(\mathscr{A},\rho)$   according to definition \ref{defn: split monomorphisms} and thus the invariance
condition in equation \eqref{eqn: categorical invariance condition} gives
\begin{equation}\label{eqn: covariance at non-faithful Hpp 1}
\mathfrak{C}_{\rho}
\left(\tilde{\mathrm{i}}(\xi),\tilde{\mathrm{i}}(\xi)\right)
=
\mathfrak{C}_{\widetilde{\rho}}
\left( \xi , \xi \right)
\end{equation}
for every $\xi\in\mathcal{H}_{\rho}^{pp}=\mathscr{A}_{pp}$.
%%%%%%%%%%%%%%%%%%%%
Since $\widetilde{\rho}$ is faithful, equation
\eqref{eqn: fields of covariances on faithful fNCP 2} applies to the right-hand side of equation \eqref{eqn: covariance at non-faithful Hpp 1}. 
%%%%%%%%%%%%%%%%%%
Moreover, the restriction of $\Delta_{\rho}$ to $\mathcal{H}_{\rho}^{pp}$ agrees
with the modular operator $\Delta_{\widetilde{\rho}}$, and $\xi_{\mathbb{I}}^{\rho}=\xi_{\mathbf{p}}^{\rho}$.
%%%%%%%%%%%%%%%%
Consequently, upon using the polarization identity, it holds
\begin{equation}\label{eqn: non-faithful covariance on pp}
\begin{split}
\mathfrak{C}_{\rho}
\left(
\xi,\eta
\right)
&=
\left\langle
\xi
\mid
F(\Delta_{\rho})
\left(
\eta
\right)
\right\rangle_{\rho}
\\
&\quad+
\bigl(\alpha-F(1)\bigr)
\left\langle
\xi
\mid
\xi_{\mathbb{I}}^{\rho}
\right\rangle_{\rho}
\left\langle
\xi_{\mathbb{I}}^{\rho}
\mid
\eta
\right\rangle_{\rho}
\end{split}
\end{equation}
for all $\xi,\eta\in\mathcal{H}_{\rho}^{pp}=\mathscr{A}_{pp}$.
%%%%%%%%%%%%%%%%%%%

We now determine the restriction of $\mathfrak{C}_{\rho}$ to
$\mathcal{H}_{\rho}^{qp}$. 
%%%%%%%%%%%%%
Choose an orthonormal basis
$\{|j\rangle\}_{j=1}^{N}$ diagonalizing $\varrho$, ordered so that
\begin{equation}
\varrho
=
\sum_{j=1}^{K}p_{j}\mathbf{e}_{j},
\qquad
p_{j}>0,
\qquad
K<N,
\end{equation}
where
\begin{equation}
\mathbf{e}_{j}:=|j\rangle\langle j|,
\qquad
\mathbf{p}
=
\sum_{j=1}^{K}\mathbf{e}_{j}.
\end{equation}
%%%%%%%%%%%%%%%%%%%%
We have the orthogonal decomposition
\begin{equation}
\mathcal{H}_{\rho}^{qp}
=
\bigoplus_{j=1}^{K}\mathcal{H}_{\rho}^{qj},
\end{equation}
where
\begin{equation}
\mathcal{H}_{\rho}^{qj}
:=
\left\{
\xi_{\mathbf{q}\mathbf{a}\mathbf{e}_{j}}^{\rho}
\;\middle|\;
\mathbf{a}\in\mathscr{A}
\right\}.
\end{equation}
%%%%%%%%%%%%%%%%%%%%%%%
The automorphisms implemented by unitaries of the form
\begin{equation}
\mathbf{p}+\mathbf{u}_{q},
\qquad
\mathbf{u}_{q}\in\mathcal{U}(\mathbf{q}\mathcal{H}),
\end{equation}
preserve $\rho$. 
%%%%%%%%%%%%%%%%%%
Their induced action is trivial on $\mathcal{H}_{\rho}^{pp}$ and is the standard unitary action on each
$\mathcal{H}_{\rho}^{qj}$. 
%%%%%%%%%%%%%%%
Since $\mathbf{T}_{\rho}$ commutes with these automorphisms (see equation \eqref{eqn: covariance operator commutes with symmetries}), it does not mix
$\mathcal{H}_{\rho}^{pp}$ and $\mathcal{H}_{\rho}^{qp}$.
%%%%%%%%%%%%%%%%
Similarly, the automorphisms implemented by diagonal unitaries on
$\mathbf{p}\mathcal{H}$ distinguish the subspaces
$\mathcal{H}_{\rho}^{qj}$. Hence $\mathbf{T}_{\rho}$ does not mix $\mathcal{H}_{\rho}^{qj}$ and $\mathcal{H}_{\rho}^{qk}$ for $j\neq k$. 
%%%%%%%%%%%%
Finally, the action of
$\mathcal{U}(\mathbf{q}\mathcal{H})$ on each
$\mathcal{H}_{\rho}^{qj}$ is irreducible, and  Schur's lemma therefore gives strictly positive numbers $\gamma_{j}^{\rho}$ such that
\begin{equation}\label{eqn: covariance operator on qj}
\mathbf{T}_{\rho}
\big|_{\mathcal{H}_{\rho}^{qj}}
=
\gamma_{j}^{\rho}
\mathbb{I}_{\mathcal{H}_{\rho}^{qj}}.
\end{equation}
%%%%%%%%%%%%%%%%%%%%
We determine these coefficients using the continuity of $\mathfrak{C}$ according to definition \ref{defn: continuity for fields of covariances}. 
%%%%%%%%%%%%
Let $\tau$ be the
tracial state on $\mathcal{B}(\mathcal{H})$, let
$\{\varepsilon_{n}\}_{n\in\mathbb{N}}\subset(0,1)$ satisfy $\varepsilon_{n}\to0$, and define
\begin{equation}
\rho_{n}
:=
(1-\varepsilon_{n})\rho
+
\varepsilon_{n}\tau.
\end{equation}
%%%%%%%%%%%%%%%%%%
Every $\rho_{n}$ is faithful. 
%%%%%%%%%%%%%%%%%%%%
Moreover, the density operator of
$\rho_{n}$ is a function of $\varrho$, so the states $\rho_{n}$ commute  with $\rho$, providing a support-compatible sequence for $\rho$ according to definition \ref{defn: commuting sequence}.
%%%%%%%%%%%%%%%%%%%%%%%%%%%%%%%%%
Fix $1\leq j\leq K$ and $K<r\leq N$. 
%%%%%%%%%%%%
Since $\mathbf{e}_{rj}\mathbf{p}=\mathbf{e}_{rj}$, equation \eqref{eqn: definition of continuity} gives
\begin{equation}\label{eqn: continuity coefficient qj}
p_{j}\gamma_{j}^{\rho}\stackrel{\mbox{\eqref{eqn: covariance operator on qj}}}{=}\mathfrak{C}_{\rho}
\left(
\xi_{\mathbf{e}_{rj}}^{\rho},
\xi_{\mathbf{e}_{rj}}^{\rho}
\right)
=
\lim_{n\rightarrow\infty}
\mathfrak{C}_{\rho_{n}}
\left(
\xi_{\mathbf{e}_{rj}}^{\rho_{n}},
\xi_{\mathbf{e}_{rj}}^{\rho_{n}}
\right)\stackrel{\mbox{\eqref{eqn: fields of covariances on faithful fNCP 2}}}{=}\lim_{n\rightarrow\infty}(p_{j})_{n}
F\left(
\frac{(p_{r})_{n}}{(p_{j})_{n}}
\right),
\end{equation}
where the rank-one term in $\mathfrak{C}_{\rho_{n}}
\left(
\xi_{\mathbf{e}_{rj}}^{\rho_{n}},
\xi_{\mathbf{e}_{rj}}^{\rho_{n}}
\right)$ vanishes because $\rho_{n}(\mathbf{e}_{rj})=0$.
%%%%%%%%%%%%%
Since $(p_{j})_{n}\to p_{j}$ and $(p_{r})_{n}\to 0$, it follows $\frac{(p_{r})_{n}}{(p_{j})_{n}}\to 0$.
%%%%%%%%%%%%%%%
Moreover, since $F$ is operator monotone on $(0,+\infty)$ according to proposition \ref{prop: fields of covariances on faithful states in fNCP have operator monotone functions}, it is increasing as a
scalar function, and since it is also strictly positive, the limit
\begin{equation}
F(0+)
:=
\lim_{t\rightarrow0^{+}}F(t)
=
\inf_{t>0}F(t)
\end{equation}
exists in $[0,+\infty)$. 
%%%%%%%%%%%%%%%
It then follows from equation \eqref{eqn: continuity coefficient qj} that
\begin{equation}\label{eqn: gamma equals F zero}
0<\gamma_{j}^{\rho}
=
F(0+)
\end{equation}
for every $j=1,\ldots,K$ because $p_{j}>0$. 
%%%%%%%%%%%%%%%%
In particular, the coefficient is independent of both $j$ and $\rho$.
%%%%%%%%%%%%%%%%%%
We may therefore define a continuous extension of $F$ to $[0,\infty)$ setting $F(0):=F(0+)$.
%%%%%%%%%%%%%%%%%%%
By definition, the supported modular operator vanishes on the  subspace $\mathcal{H}_{\rho}^{qp}$.
%%%%%%%%%%%%%%
Moreover, it is $\mathcal{H}_{\rho}^{qp}\perp \xi_{\mathbb{I}}^{\rho}$, so that equation \eqref{eqn: gamma equals F zero} implies
\begin{equation}
\mathbf{T}_{\rho}
\big|_{\mathcal{H}_{\rho}^{qp}}
=
F(0)
\mathbb{I}_{\mathcal{H}_{\rho}^{qp}}
=
F(\Delta_{\rho})
\big|_{\mathcal{H}_{\rho}^{qp}}.
\end{equation}
%%%%%%%%%%%%%%%%%%%
Together with equation
\eqref{eqn: non-faithful covariance on pp}, with the invariance of the
orthogonal decomposition
\begin{equation}
\mathcal{H}_{\rho}
=
\mathcal{H}_{\rho}^{pp}
\oplus
\mathcal{H}_{\rho}^{qp}
\end{equation}
under $\mathbf{T}_{\rho}$, and the polarization identity, we thus conclude  that
\begin{equation}
\mathfrak{C}_{\rho}(\xi,\eta)
=
\left\langle
\xi
\mid
F(\Delta_{\rho})(\eta)
\right\rangle_{\rho}
\quad+
\bigl(\alpha-F(1)\bigr)
\left\langle
\xi
\mid
\xi_{\mathbb{I}}^{\rho}
\right\rangle_{\rho}
\left\langle
\xi_{\mathbb{I}}^{\rho}
\mid
\eta
\right\rangle_{\rho}
\end{equation}
for all $\xi,\eta\in\mathcal{H}_{\rho}$ when
$\mathscr{A}=\mathcal{B}(\mathcal{H})$ with $N=\mathrm{dim}(\mathcal{H})<+\infty$.
%%%%%%%%%%%%%%%%%%%%%%%%

Finally, let $\mathscr{A}$ be an arbitrary finite-dimensional
$C^*$-algebra that, without loss of generality, can be written as
\begin{equation}
\mathscr{A}
=
\bigoplus_{\ell=1}^{L}
\mathcal{B}(\mathcal{H}_{\ell}).
\end{equation}
%%%%%%%%%%%%%%%%
The compression onto the supported corner
$\mathbf{p}\mathscr{A}\mathbf{p}$, together with the CPU section
\begin{equation}
\mathbf{x}
\longmapsto
\mathbf{x}
+
\widetilde{\rho}(\mathbf{x})\mathbf{q},
\end{equation}
determines the covariance on
$\mathbf{p}\mathscr{A}\mathbf{p}$ as above.
%%%%%%%%%%%%%%
On each matrix summand, the same unitary-symmetry argument used above
shows that the restriction of $\mathbf{T}_{\rho}$ to every nonzero
subspace of the form $\mathbf{q}_{\ell}
\mathcal{B}(\mathcal{H}_{\ell})
\mathbf{e}_{\ell,j}$ is scalar. 
%%%%%%%%%%%%%%%
Choosing a faithful tracial state $\tau$ on $\mathscr{A}$
and the commuting sequence $\rho_{n}
=
(1-\varepsilon_{n})\rho
+
\varepsilon_{n}\tau$, the continuity argument above shows that this scalar is $F(0)$ on
every such subspace. 
%%%%%%%%%%%
Since $\Delta_{\rho}=0$ on $\mathbf{q}\mathscr{A}\mathbf{p}$, equation \eqref{eqn: field of covariances on non-faithful fNCP} follows.
%%%%%%%%%%%%%%%%%%%
\end{proof}

\subsection{Characterization of continuous fields of covariances in finite dimensions}\label{subsec: zoo}

We are finally in the position to present a complete characterization of \textit{continuous fields of covariances} on $\mathsf{fNCP}$ in the sense of definitions \ref{defn: field of covariances} and \ref{defn: continuity for fields of covariances}.
%%%%%%%%%%%%%%%%%%%%%%%%%%%%%%%%%%%

\begin{theorem}\label{thm: full classification of fields of covariances on fNCP}		
A functor $\mathfrak{C}\colon \mathsf{fNCP}\rightsquigarrow\mathsf{Hilb}$ is a continuous field of covariances as in definitions \ref{defn: field of covariances} and \ref{defn: continuity for fields of covariances} \textbf{if and only if} the Hilbert space 
$$
\mathfrak{C}_{0}(\mathscr{A},\rho)\equiv \mathcal{H}_{\rho}^{\mathfrak{C}}
$$ 
is the \textup{GNS} Hilbert space of $\rho$ endowed with the alternative Hilbert product determined by the sesquilinear form 
\begin{equation}\label{eqn: covariance form on fNCP}
\mathfrak{C}_{\rho}(\xi,\eta) = \langle\xi \mid F(\Delta_{ \rho })(\eta)\rangle_{\rho} + (\alpha -F(1))\langle\xi\mid\xi^{\rho}_{\mathbb{I}}\rangle_{\rho}\langle\xi^{\rho}_{\mathbb{I}}\mid\eta\rangle_{\rho},
\end{equation}
for a unique choice of a strictly positive number $\alpha$ and of a continuous function  $F: [0,\infty) \to (0,\infty)$ that   is operator monotone  on $(0,\infty)$. 
%%%%%%%%%%%%%%%%%%%%%%%%%%%%

\end{theorem}

\begin{proof}
The  \textbf{only if} part follows from proposition
\ref{prop: continuous fields of covariances on fNCP}.
%%%%%%%%%%%%%%%%%%%%%%%%%%%%
Concerning the \textbf{if} part, from equation \eqref{eqn: covariance form on fNCP} it follows that the covariance operator uniquely associated with $\mathfrak{C}$ according to equation \eqref{eqn: covariance operator} is
\begin{equation}\label{eqn: F-covariance operator}
\mathbf{T}_{\rho}= F(\Delta_{ \rho })  + (\alpha-F(1))|\xi^{\rho}_{\mathbb{I}}\rangle_{\rho}\langle\xi^{\rho}_{\mathbb{I}}| ,
\end{equation}
so that  strict positivity of $\mathbf{T}_{\rho}$ follows immediately from $F(t)>0$ on $[0,\infty)$ and $\alpha>0$.
%%%%%%%%%%%%%%%%%%%%%%%%%%%%%%%%%%%%%%

Given a morphism $\hat{\Phi}: (\mathscr{A},\rho) \to (\mathscr{B},\sigma)$ in $\mathsf{fNCP}$, the functoriality of $\mathfrak{C}$ is encoded in the monotonicity property \eqref{eqn: categorical monotonicity 3}, which is equivalent to
 { \begin{equation}\label{eqn: monotonicity of F-covariance operator}
\tilde{\Phi}^{*}\,\mathbf{T}_{\rho}\,\tilde{\Phi}\leq \mathbf{T}_{\sigma},
\end{equation}} 
where $\tilde{\Phi}$ is as in equation \eqref{eqn: from CPU to linear contractions}.
%%%%%%%%%%%%%%%%%%%%%%%%%%%
Taking into account equation \eqref{eqn: F-covariance operator} and the fact that $\tilde{\Phi}$ is a contraction, equation \eqref{eqn: monotonicity of F-covariance operator} is equivalent to
 { \begin{equation} 
\tilde{\Phi}^{*}\, F(\Delta_{\rho})\, \tilde{\Phi}\leq  F(\Delta_{\sigma}),
\end{equation}} 
which we have to prove.
%%%%%%%%%%%%%%%%%%%%%%%%%%%%%%%%

Since $F$ is operator monotone on $(0,\infty)$, it is operator concave there. 
%%%%%%%%%%%%
Moreover, because $F$ admits a continuous extension to
$[0,\infty)$, this extension is operator concave and operator monotone on $[0,\infty)$. 
%%%%%%%%%%%%%%%%%%%%%%%
Indeed, both properties follow by applying them to $A+\varepsilon I$ and $B+\varepsilon I$ and then letting
$\varepsilon\to0^{+}$.
%%%%%%%%%%%%%%%%%%%%%
Consequently, the function $G=-F$ is operator convex on $[0,\infty)$ and
satisfies $G(0)=-F(0)<0$. 
%%%%%%%%%%%%%%%%%%%%%
Jensen's operator inequality for contractions \cite{HP1982, HP2003} therefore gives
\begin{equation}
G\left(
\widetilde{\Phi}^{*}\Delta_{\rho}\widetilde{\Phi}
\right)
\leq
\widetilde{\Phi}^{*}G(\Delta_{\rho})\widetilde{\Phi},
\end{equation}
and hence
\begin{equation}
\widetilde{\Phi}^{*}F(\Delta_{\rho})\widetilde{\Phi}
\leq
F\left(
\widetilde{\Phi}^{*}\Delta_{\rho}\widetilde{\Phi}
\right).
\end{equation}
Using lemma \ref{lem: monotonicity of modular operator} and the operator
monotonicity of the extension of $F$ to $[0,\infty)$, we conclude that
\begin{equation}
\widetilde{\Phi}^{*}F(\Delta_{\rho})\widetilde{\Phi}
\leq
F\left(
\widetilde{\Phi}^{*}\Delta_{\rho}\widetilde{\Phi}
\right)
\leq
F(\Delta_{\sigma}).
\end{equation}

Consider a support-compatible sequence $\{\rho_{n}\}_{n\in\mathbb{N}}$ for $\rho$ according to definition \ref{defn: commuting sequence}.
%%%%%%%%%%%%%%%%%%%%
This implies that the density operator of $\rho_{n}$ commutes with the support projection of $\rho$ as in the hypothesis of lemma \ref{lem: convergence of modular operator}.
%%%%%%%%%%%%%%%%%%%%%%%%%%%%%%%%%%
To discuss the continuity properties of $\mathfrak{C}$, we express the covariance at $\rho$ and at $\rho_{n}$ on the reference Hilbert space $\mathcal{H}_{\tau}$ associated with a faithful tracial state $\tau$ on $\mathscr{A}$, in analogy to what is done before lemma \ref{lem: convergence of modular operator} for the modular operator.
%%%%%%%%%%%%%%%%%%%
From equation \eqref{eqn: modular operator wrt tracial state} and equation \eqref{eqn: covariance form on fNCP} it follows that
\begin{equation} 
\begin{split}
\mathfrak{C}_{\rho}(\xi_{\mathbf{a}}^{\rho},\xi_{\mathbf{a}}^{\rho})&=\langle\xi_{\mathbf{a}}^{\tau}|F(\tilde{\Delta}_{\rho})R_{\rho}(\xi_{\mathbf{a}}^{\tau})\rangle_{\tau} +(\alpha -\beta)\,|\rho(\mathbf{a})|^{2} \\ & \\
\mathfrak{C}_{\rho_{n}}(\xi_{\mathbf{ap}}^{\rho_{n}},\xi_{\mathbf{ap}}^{\rho_{n}})&=\langle\xi_{\mathbf{ap}}^{\tau}|F(\tilde{\Delta}_{\rho_{n}})R_{\rho_{n}}(\xi_{\mathbf{ap}}^{\tau})\rangle_{\tau} +(\alpha -\beta)\,|\rho_{n}(\mathbf{ap})|^{2}.
\end{split}
\end{equation}
%%%%%%%%%%%%%%%%%%
Since $\xi_{\mathbf{ap}}^{\tau}$ lives in $\mathcal{H}_{\rho}\equiv\mathscr{A}\mathbf{p}\subseteq\mathscr{A}\equiv\mathcal{H}_{\tau}$, lemma \ref{lem: convergence of modular operator} together with the continuity of $F$ and the continuity of functional calculus ensure that equation \eqref{eqn: definition of continuity} holds and thus $\mathfrak{C}$ is \textbf{continuous} as in definition \ref{defn: continuity for fields of covariances}. 
%%%%%%%%%%%%%%%%%%%%%%%%%%%%%
\end{proof}

\begin{remark}[On the role and consequences of the continuity condition]
\label{rem: continuity holds for all sequences converging to faithful states}
The \textbf{only if} part of theorem
\ref{thm: full classification of fields of covariances on fNCP}
uses the continuity requirement in definition
\ref{defn: continuity for fields of covariances}
at two stages.
%%%%%%%%%%%%%%%%%%%%%%%%%
First, continuity is used in proposition
\ref{prop: classification of invariant covariances for any algebra and tracial states}
to pass from faithful rational tracial states to arbitrary tracial states.
%%%%%%%%%%%%%
This also determines the covariance on tracial states and, by restriction to the centralizer, fixes its form on the centralizer $\mathscr{M}_{\rho}$ of an
arbitrary faithful state.
%%%%%%%%%%%%%%%%%%%%%%%%%
Second, continuity is used in proposition
\ref{prop: continuous fields of covariances on fNCP}
to determine the covariance at non-faithful, non-tracial states by proving that
the coefficient on the subspace
$\mathcal{H}_{\rho}^{\mathbf{qp}}$ is equal to $F(0)$.
%%%%%%%%%%%%%%%%%%%%%%%%%%%%%%%%%%%%%%%%%%%%%%%%%%

The support-compatibility condition in definition \ref{defn: commuting sequence} imposes no restriction when the limiting state $\rho$ is faithful, since in this case
$\mathbf{p}_{\rho}=\mathbb{I}$.
%%%%%%%%%%%%%%%%%%%%%%%%%%%
Consequently, definition \ref{defn: continuity for fields of covariances} requires continuity along every sequence of faithful states converging to a
faithful state.
%%%%%%%%%%%%%%%%%%%%
In particular, when $\mathscr{A}\cong\mathbb{C}^{N}$, the continuity condition on the faithful state space agrees with the one used by Čencov to prove the uniqueness of the Fisher--Rao metric tensor\cite{C1981a}, and when $\mathscr{A}=\mathcal{B}(\mathcal{H})$ with
$\mathcal{H}$ finite-dimensional, it agrees with the continuity condition imposed on the contravariant inverses of quantum monotone metric tensors in \cite{P1996,GHP2009}.
%%%%%%%%%%%%%%%%%%%%%%%%%

When $\rho$ is non-faithful, the \textbf{if} part of theorem \ref{thm: full classification of fields of covariances on fNCP} proves continuity along every sequence of faithful states that is support-compatible with $\rho$.
%%%%%%%%%%%%%%%%%%
We do not claim continuity along arbitrary sequences converging to a non-faithful state.
%%%%%%%%%%%%%%%%%%%%%%%
The support-compatibility hypothesis is precisely the condition used in lemma \ref{lem: convergence of modular operator} to obtain convergence of the supported modular operators on the limiting
\textup{GNS} space.
%%%%%%%%%%%%%%%%%%%%%%%%%

The resulting extension to non-faithful states provides a generalization of
the radial procedure of Petz and Sudár \cite{PS1996} that is not restricted to pure limiting states.
%%%%%%%%%%%%%%%%%%%%%
\end{remark}

\begin{remark}[On the necessity of the support projection in the continuity condition]
\label{rem: continuity on reduced directions is necessary}
Let $\mathscr{A}=\mathcal{B}(\mathcal{H})$, with $\mathcal{H}$ finite-dimensional, and let $\rho$ be a non-faithful state.
%%%%%%%%%%%%%%%%%%%%%%%%%
Choose an orthonormal basis $\{|j\rangle\}_{j=1}^{N}$ in which the density operator of $\rho$ is diagonal, and write 
\begin{equation}
\varrho
=
\sum_{j=1}^{N}p_{j}|j\rangle\langle j|,
\end{equation}
where $p_{j}>0$ for every $j\neq N$ and $p_{N}=0$.
%%%%%%%%%%%%%%
Let $\mathbf{e}_{jk}=|j\rangle\langle k|$, and let $\mathbf{p}_{\rho}$ be the support projection of $\rho$.
%%%%%%%%%%%%%%%%%%%%%%%%%

Consider the sequence $\{\rho_{n}\}_{n\in\mathbb{N}}$ of faithful states given by
\begin{equation}
\varrho_{n}
=
\sum_{j=1}^{N}p_{j}^{n}|j\rangle\langle j|,
\end{equation}
where $p_{j}^{n}>0$ and $p_{j}^{n}\rightarrow p_{j}$ for every $j$.
%%%%%%%%%%%%%%%%%%%%%%%%
This sequence is support-compatible with $\rho$ according to definition \ref{defn: commuting sequence}, since $\mathbf{p}_{\rho}$ is fixed by the modular automorphism group of every $\rho_{n}$.
%%%%%%%%%%%%%%%%%%%%%%%%%

For $j\neq N$, one has
$\mathbf{e}_{jN}\mathbf{p}_{\rho}=0$, and therefore
$\mathbf{e}_{jN}$ belongs to the Gelfand ideal of $\rho$, so that
\begin{equation}
\mathfrak{C}_{\rho}
\left(
\xi_{\mathbf{e}_{jN}}^{\rho},
\xi_{\mathbf{e}_{jN}}^{\rho}
\right)
=
0.
\end{equation}
%%%%%%%%%%%%%%%%%%%%
On the other hand, equation
\eqref{eqn: covariance form on fNCP}
gives
\begin{equation}
\label{eqn: why continuity with support projection is necessary 1}
\mathfrak{C}_{\rho_{n}}
\left(
\xi_{\mathbf{e}_{jN}}^{\rho_{n}},
\xi_{\mathbf{e}_{jN}}^{\rho_{n}}
\right)
=
p_{N}^{n}
F\left(
\frac{p_{j}^{n}}{p_{N}^{n}}
\right).
\end{equation}
%%%%%%%%%%%%%%%%%%%%%%%%%

Suppose that the continuity condition in definition
\ref{defn: continuity for fields of covariances}
were formulated without multiplying the tested elements on the right by $\mathbf{p}_{\rho}$.
%%%%%%%%%%%%%%%%%%%%%%%%%
Applied to the sequence above and to
$\mathbf{a}=\mathbf{e}_{jN}$, it would imply
\begin{equation}
0
=
\lim_{n\rightarrow\infty}
\mathfrak{C}_{\rho_{n}}
\left(
\xi_{\mathbf{e}_{jN}}^{\rho_{n}},
\xi_{\mathbf{e}_{jN}}^{\rho_{n}}
\right)
=
\lim_{n\rightarrow\infty}
p_{j}^{n}
\frac{p_{N}^{n}}{p_{j}^{n}}
F\left(
\frac{p_{j}^{n}}{p_{N}^{n}}
\right)
=
p_{j}
\lim_{t\rightarrow\infty}
\frac{F(t)}{t}.
\end{equation}
Since $p_{j}>0$ for $j\neq N$, this would force
\begin{equation}
\label{eqn: why continuity with support projection is necessary 2}
\lim_{t\rightarrow\infty}
\frac{F(t)}{t}
=
0.
\end{equation}
%%%%%%%%%%%%%%%%%%%%%%%%%
However, if $F$ satisfies the Petz symmetry condition
$F(t)=tF(t^{-1})$ \cite{P1996}, then
\begin{equation}
\lim_{t\rightarrow\infty}
\frac{F(t)}{t}
=
\lim_{t\rightarrow\infty}
F(t^{-1})
=
F(0).
\end{equation}
%%%%%%%%%%%%%%%
For the functions considered in proposition
\ref{prop: continuous fields of covariances on fNCP},
one has $F(0)>0$ and equation
\eqref{eqn: why continuity with support projection is necessary 2} would contradict the Petz symmetry condition.
%%%%%%%%%%%%%%%%%%%%%%%%%
Moreover, $F(0)=0$ would also violate the strict positivity of the field of covariances.
%%%%%%%%%%%%%%%%%%%%%%%%%%%%%%%%%

The multiplication by $\mathbf{p}_{\rho}$ in definition
\ref{defn: continuity for fields of covariances}
is therefore necessary if we consider fields of covariances that are strictly positive and we want  to include the Petz-symmetric covariances associated with quantum monotone metric tensors.
%%%%%%%%%%%%%%%%%%%%%%%%%%%%%%%

\end{remark}

\section{Geometric tensors induced by fields of covariances}
\label{sec:applications}

The characterization obtained in theorem \ref{thm: full classification of fields of covariances on fNCP} contains, within a single operator-algebraic construction, the contravariant counterparts of
several fundamental metric tensors in classical and quantum information geometry. 
%%%%%%%%%%%
More generally, it determines positive contravariant symmetric tensors on the homogeneous manifolds of states generated by the natural action of the group of invertible elements of a finite-dimensional
$C^*$-algebra.
%%%%%%%%%%%%%

Let $\mathfrak{C}$ be a continuous field of covariances, determined by a strictly
positive constant $\alpha$ and by a continuous function
$F\colon[0,+\infty)\rightarrow(0,+\infty)$ that is operator monotone on
$(0,+\infty)$. 
%%%%%%%%%%
Thus, for every state $\rho$ on a finite-dimensional
$C^*$-algebra $\mathscr A$, the covariance is given by
\begin{equation}
\mathfrak{C}_\rho(\xi,\eta)
=
\left\langle
\xi\,\middle|\,F(\Delta_\rho)(\eta)
\right\rangle_\rho
+
\bigl(\alpha-F(1)\bigr)
\left\langle
\xi\,\middle|\,\xi^\rho_{\mathbb I}
\right\rangle_\rho
\left\langle
\xi^\rho_{\mathbb I}\,\middle|\,\eta
\right\rangle_\rho .
\label{eq:covariance-classification-geometric}
\end{equation}
The rank-one term in equation~\eqref{eq:covariance-classification-geometric}
only affects the direction generated by the identity element. Since the
tangent spaces to manifolds of normalized states are contained in the affine
hyperplane annihilating $\mathbf I$, this term disappears from all the
geometric tensors considered below. Consequently, the geometry induced on
normalized state spaces depends on $F$, but not on $\alpha$.

\subsection{Faithful states on finite-dimensional $C^*$-algebras}
\label{subsec:faithful-geometric-tensors}

Let $\mathcal S_f(\mathscr A)$ denote the manifold of faithful states on the
finite-dimensional $C^*$-algebra $\mathscr A$. Since every state satisfies
$\rho(\mathbf I)=1$, the tangent space at $\rho\in\mathcal S_f(\mathscr A)$
is naturally identified with
\begin{equation}
T_\rho\mathcal S_f(\mathscr A)
=
\left\{
\zeta\in\mathscr A_{\mathrm{sa}}^*
\;\middle|\;
\zeta(\mathbf I)=0
\right\}.
\label{eq:tangent-faithful-state-space}
\end{equation}
The cotangent space may be identified with the quotient
$\mathscr A_{\mathrm{sa}}/\mathbb R\mathbf I$. We represent every equivalence
class by its unique $\rho$-centred element, obtaining
\begin{equation}
T_\rho^*\mathcal S_f(\mathscr A)
\cong
\mathscr A_{\rho,0}^{\mathrm{sa}}
:=
\left\{
\mathbf a\in\mathscr A_{\mathrm{sa}}
\;\middle|\;
\rho(\mathbf a)=0
\right\}.
\label{eq:cotangent-faithful-state-space}
\end{equation}
%%%%%%%%%%%%%%
The real part of the covariance therefore defines a positive contravariant symmetric tensor on $\mathcal S_f(\mathscr A)$ by
\begin{equation}
\mathcal R_\rho^F(\mathbf a,\mathbf b)
:=
\operatorname{Re}
\mathfrak{C}_\rho
\left(
\xi^\rho_{\mathbf a},
\xi^\rho_{\mathbf b}
\right),
\qquad
\mathbf a,\mathbf b\in
\mathscr A_{\rho,0}^{\mathrm{sa}}.
\label{eq:contravariant-faithful-definition}
\end{equation}
Because $\rho(\mathbf a)=\rho(\mathbf b)=0$, the rank-one contribution in
equation~\eqref{eq:covariance-classification-geometric} vanishes, and hence
\begin{equation}
\mathcal R_\rho^F(\mathbf a,\mathbf b)
=
\operatorname{Re}
\left\langle
\xi^\rho_{\mathbf a}
\,\middle|\,
F(\Delta_\rho)
\left(
\xi^\rho_{\mathbf b}
\right)
\right\rangle_\rho .
\label{eq:contravariant-faithful-modular}
\end{equation}

Fix a faithful tracial state $\tau$ on $\mathscr A$, and let
$\boldsymbol{\varrho}_\rho\in\mathscr A$ be the invertible density associated
with $\rho$ as in equation \eqref{eqn: density operator of a state}.
%%%%%%%%%%%%%
As in section \ref{sec: ncp}, let $L_\rho$ and $R_\rho$ denote left and right
multiplication by $\boldsymbol{\varrho}_\rho$ on the GNS Hilbert space of $\tau$. 
%%%%%%%%%%
Introduce the positive invertible operator
\begin{equation}
\mathcal K_\rho^F
:=
F\left(
L_\rho R_\rho^{-1}
\right)R_\rho .
\label{eq:K-F-faithful}
\end{equation}
Equation~\eqref{eq:contravariant-faithful-modular} then becomes
\begin{equation}
\mathcal R_\rho^F(\mathbf a,\mathbf b)
=
\operatorname{Re}
\tau\left(
\mathbf a\,
\mathcal K_\rho^F(\mathbf b)
\right).
\label{eq:contravariant-faithful-K}
\end{equation}
%%%%%%%%%%%%%%%
The tensor $\mathcal R^F$ is defined for every function $F$ appearing in the
classification theorem. To recover the usual information-geometric metric
tensors, we additionally impose the Petz symmetry condition
\begin{equation}
F(t)
=
tF\left(t^{-1}\right),
\qquad
t>0.
\label{eq:Petz-symmetry-section-five}
\end{equation}
Under this condition, $\mathcal K_\rho^F$ preserves the self-adjoint part of
$\mathscr A$, and equation~\eqref{eq:contravariant-faithful-K} reduces to
\begin{equation}
\mathcal R_\rho^F(\mathbf a,\mathbf b)
=
\tau\left(
\mathbf a\,
\mathcal K_\rho^F(\mathbf b)
\right).
\label{eq:contravariant-faithful-symmetric}
\end{equation}
Moreover,
\begin{equation}
\tau\left(
\mathcal K_\rho^F(\mathbf a)
\right)
=
F(1)\rho(\mathbf a),
\label{eq:K-centering}
\end{equation}
so that $\mathcal K_\rho^F$ restricts to an isomorphism
\begin{equation}
\mathcal K_\rho^F
\colon
\left\{
\mathbf a\in\mathscr A_{\mathrm{sa}}
\;\middle|\;
\rho(\mathbf a)=0
\right\}
\longrightarrow
\left\{
\mathbf x\in\mathscr A_{\mathrm{sa}}
\;\middle|\;
\tau(\mathbf x)=0
\right\}.
\label{eq:K-hyperplane-isomorphism}
\end{equation}
%%%%%%%%%%%%%%%%%%
Every tangent vector $\zeta\in T_\rho\mathcal S_f(\mathscr A)$ is represented
by a unique element $\mathbf x_\zeta\in\mathscr A_{\mathrm{sa}}$ satisfying
$\tau(\mathbf x_\zeta)=0$ and
\begin{equation}
\zeta(\mathbf b)
=
\tau\left(
\mathbf x_\zeta\mathbf b
\right)
\end{equation}
for every $\mathbf b\in\mathscr A_{\mathrm{sa}}$. Let
\begin{equation}
\mathbf a_\zeta
:=
\left(
\mathcal K_\rho^F
\right)^{-1}
\left(
\mathbf x_\zeta
\right).
\label{eq:cotangent-associated-zeta}
\end{equation}
Then $\mathbf a_\zeta$ is the cotangent vector associated with $\zeta$ by
the contravariant tensor $\mathcal R_\rho^F$. Indeed, for every
$\mathbf b\in\mathscr A_{\mathrm{sa}}$ such that $\rho(\mathbf{b})=0$, it holds
\begin{equation}
\mathcal R_\rho^F
\left(
\mathbf b,\mathbf a_\zeta
\right)
=
\tau\left(
\mathbf b\mathbf x_\zeta
\right)
=
\zeta(\mathbf b).
\end{equation}
The inverse of $\mathcal R_\rho^F$ therefore defines the Riemannian metric
tensor
\begin{equation}
\mathcal G_\rho^F(\zeta,\eta)
=
\tau\left(
\mathbf x_\zeta
\left(
\mathcal K_\rho^F
\right)^{-1}
\left(
\mathbf x_\eta
\right)
\right).
\label{eq:faithful-inverse-metric-general-algebra}
\end{equation}
%%%%%%%%%%%
The normalization $F(1)=1$ gives the conventional normalization of the
Fisher--Rao and Morozova--Čencov--Petz metric tensors. For arbitrary
$F(1)>0$, the corresponding metric tensors are multiplied by the overall
factor $F(1)^{-1}$.

\paragraph{The Fisher--Rao metric tensor.}

Let $\mathscr A=\mathbb C^n\cong L^\infty(X_n,\#)$, where $X_n$ is a set with
$n$ elements, and let
\begin{equation}
\tau(\mathbf a)
=
\frac{1}{n}
\sum_{j=1}^n a_j.
\end{equation}
A faithful state $\rho$ is identified with a strictly positive probability
vector $p=(p_1,\ldots,p_n)$, with $p_j>0$ and
$\sum_{j=1}^n p_j=1$. Its density with respect to $\tau$ is
\begin{equation}
\boldsymbol{\varrho}_\rho
=
(np_1,\ldots,np_n).
\end{equation}
Since $\mathscr A$ is Abelian, left and right multiplication agree. Hence
the modular operator is the identity and
\begin{equation}
\mathcal K_\rho^F
=
F(1)R_\rho.
\end{equation}
Explicitly,
\begin{equation}
\mathcal K_\rho^F(\mathbf a)_j
=
nF(1)p_ja_j.
\label{eq:classical-K}
\end{equation}

Let $u=(u_1,\ldots,u_n)$ and $v=(v_1,\ldots,v_n)$ be tangent vectors at $p$,
so that
\begin{equation}
\sum_{j=1}^n u_j
=
\sum_{j=1}^n v_j
=
0.
\end{equation}
Their representatives relative to $\tau$ are
$\mathbf x_u=(nu_1,\ldots,nu_n)$ and
$\mathbf x_v=(nv_1,\ldots,nv_n)$. Substitution into
equation~\eqref{eq:faithful-inverse-metric-general-algebra} gives
\begin{equation}
\mathcal G_p^F(u,v)
=
\frac{1}{F(1)}
\sum_{j=1}^n
\frac{u_jv_j}{p_j}.
\label{eq:Fisher-Rao-general-normalization}
\end{equation}
After imposing $F(1)=1$, this is precisely the Fisher--Rao metric tensor on
the open probability simplex, while $\mathcal R_p^F$ is its contravariant
inverse, namely the classical covariance tensor
\cite{AJLS2017,C1981a,N2024b}.

This calculation also clarifies the origin of the classical uniqueness
phenomenon. The dependence on the function $F$ disappears because the
modular operator is trivial. The same collapse occurs for faithful tracial
states on arbitrary finite-dimensional, possibly noncommutative,
$C^*$-algebras. Thus, the uniqueness of the resulting covariance on centred
elements is controlled by the triviality of the modular operator rather than
by commutativity alone.

\paragraph{The regular Morozova--Čencov--Petz metric tensors.}

Let $\mathscr A=\mathcal B(\mathcal H)$, with
$N=\dim(\mathcal H)<+\infty$, and choose
\begin{equation}
\tau(\mathbf a)
=
\frac{1}{N}
\operatorname{Tr}(\mathbf a).
\end{equation}
Write the density matrix of a faithful state $\rho$ as
\begin{equation}
\boldsymbol{\varrho}_\rho
=
\sum_{j=1}^N
p_j
\lvert j\rangle\langle j\rvert,
\qquad
p_j>0,
\qquad
\sum_{j=1}^N p_j=1.
\label{eq:faithful-density-spectral}
\end{equation}
The density of $\rho$ relative to $\tau$ is
$N\boldsymbol{\varrho}_\rho$. Consequently,
\begin{equation}
\mathcal K_\rho^F
=
NF\left(
L_{\boldsymbol{\varrho}_\rho}
R_{\boldsymbol{\varrho}_\rho}^{-1}
\right)
R_{\boldsymbol{\varrho}_\rho}.
\end{equation}
%%%%%%%%%%
Let $\zeta$ and $\eta$ be tangent vectors represented by traceless self-adjoint matrices $\mathbf X$ and $\mathbf Y$ according to
\begin{equation}
\zeta(\mathbf a)
=
\operatorname{Tr}(\mathbf X\mathbf a),
\qquad
\eta(\mathbf a)
=
\operatorname{Tr}(\mathbf Y\mathbf a).
\end{equation}
Equation~\eqref{eq:faithful-inverse-metric-general-algebra} gives
\begin{equation}
\mathcal G_\rho^F(\zeta,\eta)
=
\operatorname{Tr}
\left(
\mathbf X
\left[
F\left(
L_{\boldsymbol{\varrho}_\rho}
R_{\boldsymbol{\varrho}_\rho}^{-1}
\right)
R_{\boldsymbol{\varrho}_\rho}
\right]^{-1}
(\mathbf Y)
\right).
\label{eq:MCP-operator-form}
\end{equation}
In the eigenbasis of $\boldsymbol{\varrho}_\rho$, this becomes
\begin{equation}
\mathcal G_\rho^F(\zeta,\eta)
=
\sum_{j,k=1}^N
\frac{
\overline{X_{jk}}Y_{jk}
}{
p_kF(p_j/p_k)
}.
\label{eq:MCP-component-form}
\end{equation}
The Petz symmetry condition implies
\begin{equation}
p_kF(p_j/p_k)
=
p_jF(p_k/p_j),
\end{equation}
so that the coefficients in equation~\eqref{eq:MCP-component-form} are
symmetric under the exchange of $j$ and $k$. As $F$ ranges over the
normalized symmetric operator monotone functions admitting a strictly
positive value at zero, equation~\eqref{eq:MCP-component-form} gives the
regular Morozova--Čencov--Petz family of quantum monotone metric tensors
\cite{MC1991,P1996,GHP2009}.
%%%%%%%%%%%%%%%%%%

A distinguished member of this family is determined by
\begin{equation}
F_{\mathrm{SLD}}(t)
=
\frac{1+t}{2}.
\label{eq:SLD-function-section-five}
\end{equation}
In this case,
\begin{equation}
F_{\mathrm{SLD}}
\left(
L_{\boldsymbol{\varrho}_\rho}
R_{\boldsymbol{\varrho}_\rho}^{-1}
\right)
R_{\boldsymbol{\varrho}_\rho}
=
\frac{
L_{\boldsymbol{\varrho}_\rho}
+
R_{\boldsymbol{\varrho}_\rho}
}{2},
\end{equation}
and hence
\begin{equation}
\mathcal G_\rho^{\mathrm{SLD}}(\zeta,\eta)
=
\sum_{j,k=1}^N
\frac{
2\overline{X_{jk}}Y_{jk}
}{
p_j+p_k
}.
\label{eq:SLD-faithful}
\end{equation}
This is the symmetric logarithmic derivative quantum Fisher metric.
Under a common normalization convention, the Bures metric is one quarter
of this tensor; alternative conventions absorb this factor into the name
Bures--Helstrom metric \cite{C2020a,SK2020,SASD2025}.
%%%%%%%%%%%%%%%%%%%
It also coincides with the Riemannian metric tensor obtained from the Jordan product of self-adjoint elements in \cite{CJS2020,CJS2020a,CJS2023}.
%%%%%%%%%%%%%%%%%%
Among monotone Riemannian metric tensors, the SLD metric is minimal with
respect to the conventional pointwise ordering \cite{P1996,PG2011}.
Operationally, for every one-parameter tangent direction, the corresponding
SLD quantum Fisher information is the supremum of the classical Fisher
information over all quantum measurements
\cite{P2009a,BC1994,BG2000}.
%%%%%%%

\subsection{Non-faithful states and homogeneous state manifolds}
\label{subsec:nonfaithful-geometric-tensors}

We now pass  to the smooth
manifolds of states determined by the natural action of the group of invertible elements of $\mathscr A$. This provides an intrinsic geometric
interpretation of the covariance field at arbitrary non-faithful states.

Let $\mathscr G(\mathscr A)$ denote the Lie group of invertible elements of
$\mathscr A$. It acts smoothly on the state space $\mathcal S(\mathscr A)$
according to
\begin{equation}
\bigl(\mathbf g\cdot\rho\bigr)(\mathbf a)
=
\frac{
\rho\left(
\mathbf g^\dagger\mathbf a\mathbf g
\right)
}{
\rho\left(
\mathbf g^\dagger\mathbf g
\right)
},
\qquad
\mathbf g\in\mathscr G(\mathscr A),
\quad
\mathbf a\in\mathscr A.
\label{eq:invertible-group-action}
\end{equation}
For a fixed state $\rho$, we denote its orbit by
\begin{equation}
\mathcal O_\rho
:=
\mathscr G(\mathscr A)\cdot\rho.
\label{eq:state-orbit}
\end{equation}
These orbits are precisely the homogeneous manifolds considered in
\cite{CJS2020,CJS2020a,CJS2023}. 
%%%%%%%%%%
With the identification
\begin{equation}
\mathscr A
\cong
\bigoplus_{\ell=1}^N
\mathcal B(\mathcal H_\ell),
\end{equation}
the orbit is determined by the ranks of the density components of $\rho$ in
the simple summands. In the Abelian case, the orbits are the relative
interiors of the faces of the probability simplex. When
$\mathscr A=\mathcal B(\mathcal H)$, they are the manifolds of density
operators of fixed rank.
%%%%%%%%%%%%%%%%%%%%%%%

Let $\rho$ be a state on the finite-dimensional $C^*$-algebra $\mathcal A$, let $\mathbf p$ be its support projection, and set $\mathbf q=\mathbb{I}-\mathbf p$. 
%%%%%%%%%%%
For $\mathbf a\in\mathscr A_{\mathrm{sa}}$, let
$f_{\mathbf a}\colon\mathcal O_\rho\rightarrow\mathbb R$ be the
expectation-value function
\begin{equation}
f_{\mathbf a}(\sigma)
=
\sigma(\mathbf a).
\end{equation}
The elements of
\begin{equation}
\mathbb R\mathbf p
\oplus
\mathbf q\mathscr A_{\mathrm{sa}}\mathbf q
\label{eq:cotangent-annihilator-orbit}
\end{equation}
determine expectation-value functions with vanishing differential at $\rho$ along $\mathcal O_\rho$, and their complement in $\mathscr{A}_{sa}$ provides an identification of the cotangent space at $\rho$.
%%%%%%%%%

\begin{lemma}\label{lem:cotangent-homogeneous-orbit}
Let $\rho$ be as above.
%%%%%%%%%%%
Fix a faithful tracial state $\tau$ on $\mathcal A$, and let $\varrho$ be the density operator of $\rho$ with respect to $\tau$ as in equation \eqref{eqn: density operator of a state}.
%%%%%%%%%%%%%%%%%%%%%%%%%%%
Then the tangent space to the orbit $\mathcal O_\rho=\mathcal G(\mathcal A)\cdot\rho$ can be identified with
\begin{equation}
    T_\rho\mathcal O_\rho
    \cong
    \left\{
        \mathbf X\in\mathcal A_{\mathrm{sa}}
        \,\middle|\,
        \mathbf q\mathbf X\mathbf q=0,\ 
        \tau(\mathbf X)=0
    \right\},
\end{equation}
where $\mathbf X$ represents the tangent functional
\begin{equation}
    \zeta_{\mathbf X}(\mathbf a)
    =
    \tau(\mathbf X\mathbf a).
\end{equation}
Moreover, the kernel of the map from $\mathscr{A}_{sa}$ to $T_\rho^*\mathcal O_\rho$ given by $\mathbf a\longmapsto (d f_{\mathbf a})_\rho$ is
\begin{equation}
  \mathbb R\mathbf p\oplus
    \mathbf q\mathcal A_{\mathrm{sa}}\mathbf q.
\end{equation} 
Consequently, it holds
\begin{equation}
    T_\rho^*\mathcal O_\rho
    \cong
    \mathcal A_{\mathrm{sa}}
    \Big/
    \left(
        \mathbb R\mathbf p
        \oplus
        \mathbf q\mathcal A_{\mathrm{sa}}\mathbf q
    \right),
\end{equation}
and every equivalence class admits a unique representative in
\begin{equation}
    \mathfrak m_\rho
    :=
    \left\{
        \mathbf a\in\mathcal A_{\mathrm{sa}}
        \,\middle|\,
        \mathbf q\mathbf a\mathbf q=0,\ 
        \rho(\mathbf a)=0
    \right\}.
\end{equation}
\end{lemma}

\begin{proof}
Let $\mathbf x\in\mathcal A$, and consider a smooth curve $g_t$ in the group $\mathcal G(\mathcal A)$ of invertible elements of $\mathscr{A}$ such that
\begin{equation}
    g_0=\mathbb I,
    \qquad
    \dot g_0=\mathbf x.
\end{equation}
With respect to the reference trace $\tau$, the density operator of
$g_t\cdot\rho$ is
\begin{equation}
    \varrho_t
    =
    \frac{g_t\varrho g_t^\dagger}
    {\tau(g_t\varrho g_t^\dagger)}.
\end{equation}
Differentiating at $t=0$ gives
\begin{equation}
    \dot\varrho_0
    =
    \mathbf x\varrho+\varrho\mathbf x^\dagger
    -
    \tau\left(
        \mathbf x\varrho+\varrho\mathbf x^\dagger
    \right)\varrho.
\end{equation}
Therefore, $\dot\varrho_0$ is self-adjoint, satisfies
$\tau(\dot\varrho_0)=0$, and, since
$\mathbf q\varrho=\varrho\mathbf q=0$, satisfies
\begin{equation}
    \mathbf q\dot\varrho_0\mathbf q=0.
\end{equation}

Conversely, let $\mathbf X\in\mathcal A_{\mathrm{sa}}$ satisfy
$\mathbf q\mathbf X\mathbf q=0$ and $\tau(\mathbf X)=0$. Let
$\varrho^+$ denote the inverse of $\varrho$ in the supported algebra
$\mathbf p\mathcal A\mathbf p$, extended by zero on
$\mathbf q\mathcal A\mathbf q$, and define
\begin{equation}
    \mathbf x
    :=
    \frac{1}{2}
    \mathbf p\mathbf X\mathbf p\,\varrho^+
    +
    \mathbf q\mathbf X\mathbf p\,\varrho^+.
\end{equation}
A direct computation gives
\begin{equation}
    \mathbf x\varrho+\varrho\mathbf x^\dagger
 =
    \mathbf p\mathbf X\mathbf p
    +
    \mathbf q\mathbf X\mathbf p
    +
    \mathbf p\mathbf X\mathbf q =
    \mathbf X,
\end{equation}
where the last equality follows from
$\mathbf q\mathbf X\mathbf q=0$. Since $\tau(\mathbf X)=0$, the
normalization term vanishes, and hence $\mathbf X$ is tangent to
$\mathcal O_\rho$. This proves the claimed description of the tangent
space.

For every $\mathbf a\in\mathcal A_{\mathrm{sa}}$ and every tangent
vector represented by $\mathbf X$, one has
\begin{equation}
    (d f_{\mathbf a})_\rho(\zeta_{\mathbf X})
    =
    \tau(\mathbf X\mathbf a).
\end{equation}
If
$\mathbf a=c\mathbf p+\mathbf b$, with $c\in\mathbb R$ and
$\mathbf b\in\mathbf q\mathcal A_{\mathrm{sa}}\mathbf q$, then
\begin{equation}
    \tau(\mathbf X\mathbf a)
    =
    c\,\tau(\mathbf X\mathbf p)
    +
    \tau(\mathbf X\mathbf b)
    =
    0
\end{equation}
for every tangent $\mathbf X$. Thus,
\begin{equation}
    \mathbb R\mathbf p
    \oplus
    \mathbf q\mathcal A_{\mathrm{sa}}\mathbf q
    \subseteq
    \ker\left(
        \mathbf a\longmapsto(d f_{\mathbf a})_\rho
    \right).
\end{equation}

Conversely, suppose that
\begin{equation}
    \tau(\mathbf X\mathbf a)=0
\end{equation}
for every self-adjoint $\mathbf X$ satisfying
$\mathbf q\mathbf X\mathbf q=0$ and $\tau(\mathbf X)=0$. Testing
against arbitrary self-adjoint off-diagonal elements in
\begin{equation}
    \mathbf p\mathcal A\mathbf q
    \oplus
    \mathbf q\mathcal A\mathbf p
\end{equation}
and using the nondegeneracy of the trace pairing gives
\begin{equation}
    \mathbf p\mathbf a\mathbf q
    =
    \mathbf q\mathbf a\mathbf p
    =
    0.
\end{equation}
Testing against all $\tau$-traceless self-adjoint elements of
$\mathbf p\mathcal A\mathbf p$ then gives
\begin{equation}
    \mathbf p\mathbf a\mathbf p
    =
    c\mathbf p
\end{equation}
for some $c\in\mathbb R$. Therefore,
\begin{equation}
    \mathbf a
    =
    c\mathbf p+\mathbf q\mathbf a\mathbf q,
\end{equation}
and hence
\begin{equation}
    \ker\left(
        \mathbf a\longmapsto(d f_{\mathbf a})_\rho
    \right)
    =
    \mathbb R\mathbf p
    \oplus
    \mathbf q\mathcal A_{\mathrm{sa}}\mathbf q.
\end{equation}
Since the trace pairing is nondegenerate and all the spaces involved
are finite-dimensional, the map
$\mathbf a\mapsto(d f_{\mathbf a})_\rho$ is surjective onto
$T_\rho^*\mathcal O_\rho$. The quotient identification follows.
%%%%%%%%%%%%%%%%%
Finally, the class of $\mathbf b\in\mathcal A_{\mathrm{sa}}$ is
represented by
\begin{equation}
    \mathbf a
    =
    \mathbf p\mathbf b\mathbf p
    +
    \mathbf p\mathbf b\mathbf q
    +
    \mathbf q\mathbf b\mathbf p
    -
    \rho(\mathbf p\mathbf b\mathbf p)\mathbf p.
\end{equation}
This element satisfies
$\mathbf q\mathbf a\mathbf q=0$ and $\rho(\mathbf a)=0$, and uniqueness follows because
\begin{equation}
    \mathfrak m_\rho
    \cap
    \left(
        \mathbb R\mathbf p
        \oplus
        \mathbf q\mathcal A_{\mathrm{sa}}\mathbf q
    \right)
    =
    \{0\}.
\end{equation}
\end{proof}

The GNS Hilbert space of $\rho$ admits the orthogonal decomposition
\begin{equation}
\mathcal H_\rho
\cong
\mathbf p\mathscr A\mathbf p
\oplus
\mathbf q\mathscr A\mathbf p.
\label{eq:GNS-orbit-decomposition}
\end{equation}
The identification of the real cotangent space with a real subspace of
$\mathcal H_\rho$ is not canonical at the level of normalization. Indeed,
for every constant $c>0$, one may define the real-linear injective map
\begin{equation}
\Lambda_\rho^{(c)}(\mathbf a)
:=
\xi^\rho_{
\mathbf p\mathbf a\mathbf p
+
c\,\mathbf q\mathbf a\mathbf p
},
\qquad
\mathbf a\in\mathfrak m_\rho.
\label{eq:Lambda-orbit-general}
\end{equation}
Different values of $c$ rescale only the contribution associated with
support-changing directions. The covariance field itself determines this
contribution through the value $F(0)$, whereas its overall numerical
normalization depends on the choice of $\Lambda_\rho^{(c)}$.

Throughout this section, we adopt the convention $c=\sqrt{2}$ and write
\begin{equation}
\Lambda_\rho(\mathbf a)
:=
\Lambda_\rho^{(\sqrt{2})}(\mathbf a)
=
\xi^\rho_{
\mathbf p\mathbf a\mathbf p
+
\sqrt{2}\,\mathbf q\mathbf a\mathbf p
}.
\label{eq:Lambda-orbit}
\end{equation}
This convention will be seen below to agree with the standard
normalizations of the Jordan-product and Fubini--Study tensors.
%%%%%%%%%%%%%%%

We define a real bilinear form on $T_\rho^*\mathcal O_\rho$ by
\begin{equation}
\mathcal R_\rho^F(\mathbf a,\mathbf b)
:=
\operatorname{Re}
\mathfrak{C}_\rho
\left(
\Lambda_\rho(\mathbf a),
\Lambda_\rho(\mathbf b)
\right),
\qquad
\mathbf a,\mathbf b\in\mathfrak m_\rho.
\label{eq:orbit-contravariant-definition}
\end{equation}
%%%%%%%%%%%%%%%%%
Let $\widetilde{\rho}$ denote the faithful restriction of $\rho$ to the
supported algebra $\mathbf p\mathscr A\mathbf p$, whose identity element is
$\mathbf p$. The modular operator $\Delta_\rho$ restricts to the modular
operator $\Delta_{\widetilde{\rho}}$ on
$\mathbf p\mathscr A\mathbf p$, and vanishes on
$\mathbf q\mathscr A\mathbf p$. Using
equation~\eqref{eq:covariance-classification-geometric} and the orthogonality
of the decomposition~\eqref{eq:GNS-orbit-decomposition}, we obtain
\begin{equation}
\mathcal R_\rho^F(\mathbf a,\mathbf b)
=
\operatorname{Re}
\left\langle
\xi^\rho_{\mathbf p\mathbf a\mathbf p}
\,\middle|\,
F\left(
\Delta_{\widetilde{\rho}}
\right)
\left(
\xi^\rho_{\mathbf p\mathbf b\mathbf p}
\right)
\right\rangle_\rho
+
2F(0)
\operatorname{Re}
\rho\left(
\mathbf p\mathbf a\mathbf q\mathbf b\mathbf p
\right).
\label{eq:orbit-tensor-general-formula}
\end{equation}
The rank-one contribution involving $\alpha-F(1)$ vanishes because
$\rho(\mathbf a)=\rho(\mathbf b)=0$.
%%%%%%%%%%%%
The coefficient $2F(0)$ in the second term of
equation~\eqref{eq:orbit-tensor-general-formula} is a consequence of the
normalization chosen in equation~\eqref{eq:Lambda-orbit}. 
%%%%%%%%%%%
More generally, had we used the map $\Lambda_\rho^{(c)}$ in
equation~\eqref{eq:Lambda-orbit-general}, the second term in
equation~\eqref{eq:orbit-tensor-general-formula} would be multiplied by
$c^2F(0)$ instead of $2F(0)$. Thus, the dependence on $F(0)$ is intrinsic
to the covariance field, while the additional numerical factor is fixed by
the normalization chosen in the identification of cotangent vectors with
GNS vectors.
%%%%%%%%%%%%%%

The map $\Lambda_\rho$ is injective. Indeed, if
$\Lambda_\rho(\mathbf a)=0$, then
$\mathbf p\mathbf a\mathbf p=0$ and
$\mathbf q\mathbf a\mathbf p=0$. Since $\mathbf a$ is self-adjoint and
$\mathbf q\mathbf a\mathbf q=0$, it follows that $\mathbf a=0$. The strict
positivity of the covariance therefore implies that
$\mathcal R_\rho^F$ is positive definite. Moreover, operator monotone
functions are real analytic on $(0,+\infty)$, while the support projection
and the inverse of the density on its support depend smoothly on a
constant-rank orbit. It follows that the family
$\rho\mapsto\mathcal R_\rho^F$ defines a smooth positive contravariant
symmetric tensor on every orbit $\mathcal O_\rho$. Its inverse defines a
Riemannian metric tensor, which we denote by $\mathcal G^F$.

\paragraph{Non-faithful states on Abelian algebras.}

Let $\mathscr A=\mathbb C^n$ and  $\rho$ corresponding to a probability vector whose support contains $r<n$ points. 
%%%%%%%%%%%%
Since $\mathbf p$
and $\mathbf q$ are central, it holds $\mathbf q\mathscr A\mathbf p= \{0\}$.
%%%%%%%%%%%%%%%%%
There are therefore no support-changing directions inside the orbit. The
orbit $\mathcal O_\rho$ is the relative interior of the face of the
probability simplex determined by the support of $\rho$, and
equation~\eqref{eq:orbit-tensor-general-formula} reduces to the covariance
tensor of the faithful state $\widetilde{\rho}$ on the supported Abelian
algebra $\mathbf p\mathscr A\mathbf p\cong\mathbb C^r$. Its inverse is the
Fisher--Rao metric tensor on that face:
\begin{equation}
\mathcal G_p^F(u,v)
=
\frac{1}{F(1)}
\sum_{j\in\operatorname{supp}(p)}
\frac{u_jv_j}{p_j}.
\label{eq:Fisher-Rao-face}
\end{equation}
Thus, the covariance field extends the contravariant form of Čencov's
theorem from the open probability simplex to the relative interiors of all
its faces.

\paragraph{Non-faithful, non-pure states on $\mathcal B(\mathcal H)$.}

Let $\mathscr A=\mathcal B(\mathcal H)$, with
$N=\dim(\mathcal H)<+\infty$, and let $\rho$ be represented by a density
matrix of rank $r$, where $1<r<N$. Choose an orthonormal basis in which
\begin{equation}
\boldsymbol{\varrho}_\rho
=
\sum_{j=1}^r
p_j
\lvert j\rangle\langle j\rvert,
\qquad
p_j>0,
\qquad
\sum_{j=1}^r p_j=1.
\label{eq:intermediate-rank-density}
\end{equation}
The support projection is
\begin{equation}
\mathbf p
=
\sum_{j=1}^r
\lvert j\rangle\langle j\rvert,
\qquad
\mathbf q
=
\mathbf I-\mathbf p.
\end{equation}
The orbit $\mathcal O_\rho$ is the manifold $\mathcal S_r(\mathcal H)$ of
density operators of rank $r$.

Let $\mathbf a,\mathbf b\in\mathfrak m_\rho$, and write
\begin{equation}
a_{jk}
=
\langle j\mid\mathbf a\mid k\rangle,
\qquad
b_{jk}
=
\langle j\mid\mathbf b\mid k\rangle.
\end{equation}
Equation~\eqref{eq:orbit-tensor-general-formula} becomes
\begin{equation}
\mathcal R_\rho^F(\mathbf a,\mathbf b)
=
\operatorname{Re}
\left(
\sum_{j,k=1}^r
p_kF(p_j/p_k)
\overline{a_{jk}}b_{jk}
\right)
+
2F(0)
\operatorname{Re}
\left(
\sum_{\mu=r+1}^d
\sum_{k=1}^r
p_k
\overline{a_{\mu k}}b_{\mu k}
\right).
\label{eq:intermediate-rank-F-tensor}
\end{equation}
The first term is the regular contravariant Morozova--Čencov--Petz tensor associated
with the faithful restriction of $\rho$ to
$\mathbf p\mathcal B(\mathcal H)\mathbf p$. The second term describes the
directions in which the support of the state changes and is weighted by the
single value $2F(0)$.
%%%%%%%%%%%%

Consider now the function
$F_{\mathrm{SLD}}(t)=(1+t)/2$. Since
$F_{\mathrm{SLD}}(0)=1/2$,
equation~\eqref{eq:intermediate-rank-F-tensor} gives
\begin{equation}
\mathcal R_\rho^{\mathrm{SLD}}(\mathbf a,\mathbf b)
= 
\operatorname{Re}
\left(
\sum_{j,k=1}^r
\frac{p_j+p_k}{2}
\overline{a_{jk}}b_{jk}
\right)
+
\operatorname{Re}
\left(
\sum_{\mu=r+1}^d
\sum_{k=1}^r
p_k
\overline{a_{\mu k}}b_{\mu k}
\right).
\label{eq:intermediate-rank-SLD-components}
\end{equation}
A direct block computation shows that
\begin{equation}
\mathcal R_\rho^{\mathrm{SLD}}(\mathbf a,\mathbf b)
=
\rho\left(
\mathbf a\circ\mathbf b
\right),
\qquad
\mathbf a,\mathbf b\in\mathfrak m_\rho,
\label{eq:SLD-Jordan-centred}
\end{equation}
where
\begin{equation}
\mathbf a\circ\mathbf b
:=
\frac{
\mathbf a\mathbf b+\mathbf b\mathbf a
}{2}
\label{eq:Jordan-product-section-five}
\end{equation}
is the Jordan product on
$\mathcal B(\mathcal H)_{\mathrm{sa}}$. Equivalently, without choosing
centred representatives,
\begin{equation}
\mathcal R_\rho^{\mathrm{SLD}}
\left(
(df_{\mathbf a})_\rho,
(df_{\mathbf b})_\rho
\right)
=
\rho\left(
\mathbf a\circ\mathbf b
\right)
-
\rho(\mathbf a)\rho(\mathbf b).
\label{eq:Jordan-covariance-general}
\end{equation}
%%%%%%%%%%%%%%%%%
The tensor in equation~\eqref{eq:Jordan-covariance-general} is precisely the
contravariant tensor determined by the Jordan product in
\cite{CJS2020,CJS2020a,CJS2023}. Its inverse is the Riemannian metric
tensor obtained there from the Jordan-algebraic construction. 
%%%%%%%%%
Thus, the field of covariances associated with
\begin{equation}
F(t)
=
\frac{1+t}{2}
\end{equation}
recovers the inverse of the Jordan metric on every manifold
$\mathcal S_r(\mathcal H)$ of non-faithful, non-pure quantum states. 
%%%%%%%%%%%

\paragraph{Pure states on $\mathcal B(\mathcal H)$.}

Let $\mathscr A=\mathcal B(\mathcal H)$, with
$N=\dim(\mathcal H)<+\infty$, and let $\rho$ be a pure state, which means it is represented by a density
operator of rank $r=1$  according to
\begin{equation}
\mathbf p
=
\lvert\psi\rangle\langle\psi\rvert,
\qquad
\langle\psi\mid\psi\rangle=1.
\end{equation}
The orbit $\mathcal O_{\mathbf p}$ is the manifold of pure states, naturally
identified with the complex projective space
$\mathbb{CP}(\mathcal H)$. Every tangent vector at $\mathbf p$ may be written
uniquely in the form
\begin{equation}
\mathbf X_v
=
\lvert v\rangle\langle\psi\rvert
+
\lvert\psi\rangle\langle v\rvert,
\qquad
v\in\mathbf q\mathcal H.
\label{eq:pure-tangent-vector}
\end{equation}
We adopt the normalization of the Fubini--Study metric tensor given by
\begin{equation}
\mathcal G_{\mathbf p}^{\mathrm{FS}}
\left(
\mathbf X_v,\mathbf X_w
\right)
=
4\operatorname{Re}
\langle v\mid w\rangle.
\label{eq:Fubini-Study-normalization}
\end{equation}

For every $\mathbf a\in\mathcal B(\mathcal H)_{\mathrm{sa}}$, consider the
expectation-value function
\begin{equation}
f_{\mathbf a}(\sigma)
=
\operatorname{Tr}(\sigma\mathbf a).
\end{equation}
Its differential at $\mathbf p$ satisfies
\begin{equation}
(df_{\mathbf a})_{\mathbf p}(\mathbf X_v)
=
2\operatorname{Re}
\left\langle
\mathbf q\mathbf a\psi
\,\middle|\,
v
\right\rangle.
\label{eq:differential-expectation-pure}
\end{equation}
Consequently, the inverse Fubini--Study tensor is characterized by
\begin{equation}
\left(
\mathcal G_{\mathbf p}^{\mathrm{FS}}
\right)^{-1}
\left(
(df_{\mathbf a})_{\mathbf p},
(df_{\mathbf b})_{\mathbf p}
\right)
=
\operatorname{Re}
\left\langle
\mathbf q\mathbf a\psi
\,\middle|\,
\mathbf q\mathbf b\psi
\right\rangle.
\label{eq:inverse-FS-expectation-functions}
\end{equation}

Let $\mathbf a\in\mathfrak m_{\mathbf p}$ be the canonical cotangent
representative. Since $\mathbf p$ has rank one, the centering condition
implies
\begin{equation}
\mathbf p\mathbf a\mathbf p
=
\rho(\mathbf a)\mathbf p
=
0.
\end{equation}
Therefore, it holds
\begin{equation}
\Lambda_{\mathbf p}(\mathbf a)
=
\sqrt{2}\,
\xi^{\mathbf p}_{\mathbf q\mathbf a\mathbf p}.
\end{equation}
The contribution associated with the positive spectrum of the modular
operator disappears, and
equation~\eqref{eq:orbit-tensor-general-formula} reduces to
\begin{equation}
\mathcal R_{\mathbf p}^F
\left(
(df_{\mathbf a})_{\mathbf p},
(df_{\mathbf b})_{\mathbf p}
\right)
=
2F(0)
\operatorname{Re}
\left\langle
\mathbf q\mathbf a\psi
\,\middle|\,
\mathbf q\mathbf b\psi
\right\rangle.
\label{eq:pure-F-covariance-explicit}
\end{equation}
Combining equations~\eqref{eq:inverse-FS-expectation-functions} and
\eqref{eq:pure-F-covariance-explicit}, we obtain
\begin{equation}
\mathcal R_{\mathbf p}^F
=
2F(0)
\left(
\mathcal G_{\mathbf p}^{\mathrm{FS}}
\right)^{-1}.
\label{eq:pure-contravariant-FS}
\end{equation}
Accordingly, the inverse Riemannian metric tensor is
\begin{equation}
\mathcal G_{\mathbf p}^F
=
\frac{1}{2F(0)}
\mathcal G_{\mathbf p}^{\mathrm{FS}}.
\label{eq:pure-covariant-FS}
\end{equation}

Thus, every function $F$ appearing in the characterization of continuous
fields of covariances induces a multiple of the Fubini--Study metric tensor
on the manifold of pure states. No Petz symmetry condition is needed for
this conclusion.
%%%%%%%
Indeed, the pure-state tensor induced by any field of covariances is necessarily unitarily invariant, and this singles out the Fubini--Study geometry.
%%%%%%%%%%%
Then, the explicit choice of a field of covariances  determines the proportionality coefficient to be
$2F(0)$ at the contravariant level, or $(2F(0))^{-1}$ at the covariant level.
%%%%%%%%%%%%%%%
This coefficient agrees with the radial extension procedure of Petz and Sudár \cite{PS1996} because of the choice $c=\sqrt{2}$ in defining $\Lambda_{\rho}$ in equations \eqref{eq:Lambda-orbit-general} and \eqref{eq:Lambda-orbit}. 
%%%%%%%%%%%%
More precisely, in the normalization
\eqref{eq:Fubini-Study-normalization}, the radial extension of the regular
Morozova--Čencov--Petz metric associated with $F$ restricts to
$(2F(0))^{-1}$ times the Fubini--Study metric tensor. 
%%%%%%%%%%%%%
The present construction obtains the same tensor directly from the covariance assigned to the pure state, without introducing a limiting procedure from the
faithful state manifold.
%%%%%%%%%%%%%%%%%%%%%

For the SLD function, one has
\begin{equation}
2F_{\mathrm{SLD}}(0)
=
1.
\end{equation}
Consequently,
\begin{equation}
\mathcal R_{\mathbf p}^{\mathrm{SLD}}
=
\left(
\mathcal G_{\mathbf p}^{\mathrm{FS}}
\right)^{-1},
\qquad
\mathcal G_{\mathbf p}^{\mathrm{SLD}}
=
\mathcal G_{\mathbf p}^{\mathrm{FS}}.
\label{eq:SLD-pure-FS}
\end{equation}
This is the rank-one instance of the agreement between the SLD covariance
field and the group-theoretic Jordan construction.
%%%%%%%%%%%%
\\

The preceding results in this section exhibit the common geometric content of a continuous
field of covariances. On faithful states, the positive spectrum of the
modular operator determines the regular Morozova--Čencov--Petz tensors,
while its triviality in the Abelian and tracial cases produces the
Fisher--Rao uniqueness phenomenon. On a non-faithful homogeneous state
manifold, the positive spectral ratios determine the geometry internal to
the support, whereas the zero eigenspace determines the geometry of the
support-changing directions through the coefficient $F(0)$. The particular
choice $F(t)=(1+t)/2$ recovers the Jordan-product metric on every orbit,
including the Fisher--Rao metric on classical probability simplices, the
SLD/Bures--Helstrom metric on faithful quantum states, and the Fubini--Study
metric on pure quantum states.
%%%%%%%%%%%%%%%%%%%

\section{Conclusions and future work}\label{sec: conclusions}

We investigated the problem of unifying and generalizing classical and quantum statistical covariances. 
%%%%%%%%%%%
The starting observation is that classical statistical covariance is induced by the GNS Hilbert product when a
probability measure is regarded as a state on a commutative $C^*$-algebra, and that the GNS construction defines a contravariant functor from the
category of non-commutative probability spaces to the category of Hilbert spaces and contractions. 
%%%%%%%%
This led us to introduce a field of covariances as
a contravariant functor that preserves the algebraic action of state-preserving completely positive unital maps while deforming the Hilbertian structure of the corresponding GNS spaces in definition \ref{defn: field of covariances}.
%%%%%%%%%%%%%%%

Our main result, theorem \ref{thm: full classification of fields of covariances on fNCP}, gives a complete characterization of the continuous fields of covariances on the category $\mathrm{fNCP}$ of
finite-dimensional non-commutative probability spaces.
%%%%%%%%%%%%
Every such field is uniquely determined by a strictly positive constant $\alpha$ and a continuous function
$F\colon[0,+\infty)\rightarrow(0,+\infty)$ that is operator monotone on $(0,+\infty)$. 
%%%%%%%%%%%%%%%%%%%%%%%
More explicitly, the covariance at a state $\rho$ is given by
\begin{equation}
\mathfrak{C}_\rho(\xi,\eta)
=
\left\langle
\xi\,\middle|\,F(\Delta_\rho)(\eta)
\right\rangle_\rho
+
\bigl(\alpha-F(1)\bigr)
\left\langle
\xi\,\middle|\,\xi^\rho_{\mathbb I}
\right\rangle_\rho
\left\langle
\xi^\rho_{\mathbb I}\,\middle|\,\eta
\right\rangle_\rho ,
\end{equation}
where $\Delta_\rho$ is the modular operator associated with $\rho$, including the non-faithful case. 
%%%%%%%%%%
The continuity condition for a field of covariances is formulated in terms of support-compatible sequences of faithful states in definition \ref{defn: continuity for fields of covariances}. 
%%%%%%%%%%%%
On faithful state spaces it reduces to the usual continuity assumption employed in classical
and quantum information geometry. 
%%%%%%%%%%%
At non-faithful states, the restriction to support-compatible sequences and to vectors surviving in the limiting GNS space is necessary to retain the regular Petz-symmetric covariances \cite{MC1991,P1996,GHP2009}.
%%%%%%%%%%%%%%%%%%%%%

The characterization we provide unifies several familiar geometric structures. 
%%%%%%%%%%%%
On commutative algebras, the modular operator is trivial and the covariance reduces, on centred elements, to a multiple of the ordinary statistical
covariance. 
%%%%%%%%%%%
Its inverse is the Fisher--Rao metric tensor, both on the open probability simplex and on the relative interiors of its boundary faces.
%%%%%%%%%%%%%%%%%%%%%%
The same collapse occurs for tracial states on arbitrary finite-dimensional $C^*$-algebras.
%%%%%%%%%%%%%%
Consequently, the uniqueness phenomenon underlying Čencov's theorem is governed by the triviality of the modular operator rather than by the commutativity of the algebra alone.
%%%%%%%%%%%%%%%%%%%%%%%%%%%%%%%%%%

On the manifold of faithful states of an arbitrary finite-dimensional $C^*$-algebra, the real part of a field of covariances determines a positive
contravariant symmetric tensor. 
%%%%%%%%%%%%
When $F$ satisfies the Petz symmetry condition $F(t)=tF(t^{-1})$, the inverse tensor specializes, on
$\mathcal B(\mathcal H)$, to the regular Morozova--Čencov--Petz monotone metric associated with $F$. 
%%%%%%%%%%%%%%
The parameter $\alpha$ does not contribute to
these tensors because its rank-one term lies in the direction generated by the identity, which is transverse to the affine hyperplane in which states live.
%%%%%%%%%%%%%%%%%%%

The construction also applies intrinsically to non-faithful states. 
%%%%%%%
The state space of a finite-dimensional $C^*$-algebra is partitioned into homogeneous manifolds generated by the action of the group of invertible
elements. 
%%%%%%%%%%
After identifying the cotangent space of one of these manifolds with a real subspace of the GNS Hilbert space, the covariance field induces a positive contravariant symmetric tensor on the manifold. 
%%%%%%%%
Its supported component is governed by the positive spectrum of the modular operator, while the component corresponding to variations of the support is governed
by $F(0)$. 
%%%%%%%%%
This is a direct construction at the non-faithful state and should not be confused with a radial extension from the faithful state space.
%%%%%%%%%%%%%%%%%%%%%%%%%%%
The normalization of the identification between real cotangent vectors and GNS vectors is not fixed by the covariance field. 
%%%%%%%%%%%
If the off-diagonal component is multiplied by a constant $c>0$, the support-changing part of the induced tensor is multiplied by $c^2F(0)$. 
%%%%%%%%%%%%%
Thus, the abstract dependence on $F(0)$ is intrinsic, whereas the additional numerical coefficient is
conventional. 
%%%%%%%%%%%
With the normalization $c=\sqrt{2}$ adopted in section \ref{sec:applications}, the   coefficient is $2F(0)$.
%%%%%%%%%%
This convention has the advantage that the field of covariances associated with $F(t)=\frac{1+t}{2}$ recovers exactly the contravariant tensor induced by the Jordan product as in \cite{CJS2020,CJS2020a,CJS2023}.
%%%%%%%%%%%%%%%%%%%%
For $\mathcal B(\mathcal H)$, this includes the SLD metric on faithful states, the corresponding Jordan metric on non-faithful states of intermediate rank, and the Fubini--Study metric on pure states.
%%%%%%%%%%%%%%%%%%%%%%%

More generally, on the pure-state manifold every covariance field induces a multiple of the inverse Fubini--Study tensor whose proportionality constant is $2F(0)$,  in agreement with the
coefficient obtained through the radial extension of Petz and Sudár \cite{PS1996}. 
%%%%%%%
The difference with \cite{PS1996} is however methodological: the present tensor is obtained directly from the field of covariances which is defined on all states directly, whereas the Petz--Sudár construction approaches the pure-state manifold from faithful states.
%%%%%%%%%%%%%%%%%%%%%%%%%%%%%

The extension of theorem \ref{thm: full classification of fields of covariances on fNCP} to infinite-dimensional operator algebras is
a natural direction for future work. 
%%%%%%%%%%%
A reasonable first setting is the full subcategory of $\mathrm{NCP}$ whose objects are pairs $(\mathscr A,\rho)$ with $\mathscr A$ an injective $W^*$-algebra and $\rho$ a normal state. 
%%%%%%%%%%%
In this context, support projections and
state-preserving conditional expectations remain available, but several
finite-dimensional ingredients used here require substantial analytic replacement. 
%%%%%%%%%
In particular, covariance forms need not be represented by bounded operators on the GNS Hilbert space, modular operators are generally
unbounded, and continuity must be formulated together with suitable domain and closability conditions.
%%%%%%%%%

\addcontentsline{toc}{section}{Funding}
\section*{Funding}

This work has been supported by the Madrid Government under the Multiannual Agreement with UC3M in the line of “Research Funds for Beatriz Galindo Fellowships” (C\&QIG-BG-CM-UC3M), and in the context of the V PRICIT (Regional Programme of Research and Technological Innovation), and through the project \textbf{TEC-2024/COM-84 QUITEMAD-CM}.
%%%%%%%%%%%%%%%%%%%%
This article/publication is based upon work from COST Action CaLISTA CA21109 supported by COST (European Cooperation in Science and Technology).
%%%%%%%%%%%%%%%%%%%%%%%%%%%%%%

\addcontentsline{toc}{section}{References}
{\footnotesize
	\printbibliography[title=References]
}
\end{document}